\documentclass{SciPost}

\usepackage{amsthm}

\newtheorem{theorem}{Theorem}[section]

\newcommand{\frontmattertitle}[1]{%
  \clearpage
  \thispagestyle{plain}
  \vspace*{2cm}
  \begin{center}
    {\LARGE\bfseries #1}
  \end{center}
  \vspace{1.5cm}
}

\hypersetup{
    colorlinks,
    linkcolor={red!50!black},
    citecolor={blue!50!black},
    urlcolor={blue!80!black}
}

\usepackage[bitstream-charter]{mathdesign}

\usepackage{makecell}

\setcellgapes{3pt}
\makegapedcells

\numberwithin{equation}{section}

\usepackage[most]{tcolorbox} % robust; works fine with SciPost

\tcbset{
  boxrule=0.6pt, arc=6pt, breakable,
  colframe=black!55, colback=black!3,   % light gray fill, dark-gray frame
  left=10pt, right=10pt, top=8pt, bottom=8pt
}

\newtcolorbox{scipostbox}[1][]{%
  enhanced, % enable title attachment etc.
  title={#1}, fonttitle=\bfseries, coltitle=black,
  colbacktitle=black!12,                      % slightly darker title chip
  boxed title style={boxrule=0.6pt, arc=4pt, colframe=black!55, colback=black!12},
  attach boxed title to top left={xshift=6pt, yshift*=-\tcboxedtitleheight/2}, % small tab
}

\DeclareSymbolFont{usualmathcal}{OMS}{cmsy}{m}{n}
\DeclareSymbolFontAlphabet{\mathcal}{usualmathcal}

\fancypagestyle{SPstyle}{
\fancyhf{}
\lhead{\colorbox{scipostblue}{\bf \color{white} ~Wrocław University of Science and Technology}}
\rhead{{\bf \color{scipostdeepblue} ~PhD Thesis }}

\fancyfoot[C]{\textbf{\thepage}}
}

\fancypagestyle{emptyheader}{%
  \fancyhf{}% kompletnie pusto
  
}

\usepackage[main=english,polish]{babel}

\newcommand{\be}{\begin{equation}}
\newcommand{\ee}{\end{equation}}
\newcommand{\bs}{\begin{equation}\begin{aligned}}
\newcommand{\es}{\end{aligned}\end{equation}}
\newcommand{\bg}{\begin{gathered}}
\newcommand{\eg}{\end{gathered}}

\begin{document}

\pagestyle{SPstyle}

\begin{center}{\huge \textbf{\color{scipostdeepblue}{
%%%%%%%%%% TODO: Write your article's title here
 Effective theories for many-body systems with nonuniform symmetries\\
}}}\end{center}

%%%%%%%%%No phage number %%%%%%%%%
\thispagestyle{empty}
%%%%%%%%%

\begin{center}\LARGE \textbf{
% Use (full) first name (+ middle name initials) + surname format.
% Separate subsequent authors by a comma, omit comma and use "and" for the last author.
% Mark the corresponding author(s) with a superscript symbol in this order
% \star, \dagger, \ddagger, \circ, \S, \P, \parallel, ...
Aleksander Głódkowski
}\end{center}

\noindent\rule{\textwidth}{1pt}
\vspace{50pt}

\begin{center} \Large PhD thesis prepared under supervision of \end{center}

\begin{center} \Large
\textbf{Prof. Piotr Surówka} \end{center}

\vspace{50pt}

\vspace{50pt}

\begin{figure}[h!]
\centering    \includegraphics[width=0.25\linewidth]{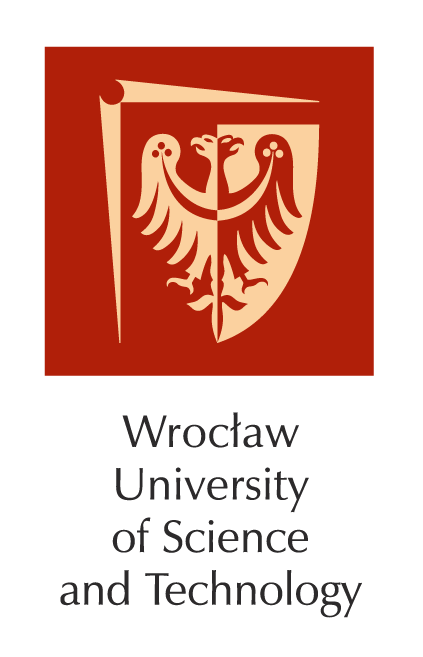}
\end{figure}

\vspace{50pt}

%\begin{center} \LARGE
%\textit{God used beautiful mathematics in creating the world.}
%\end{center}

\vspace{\baselineskip}

% For convenience during refereeing we turn on line numbers:
%\linenumbers

\clearpage
\thispagestyle{empty}
\mbox{}
\clearpage

\clearpage
\thispagestyle{empty}   % no page number on the quote page

\vspace*{\fill}

\begin{center}
{\LARGE\itshape
Nothing in life is to be feared, it is only to be understood.\\
Now is the time to understand more, so that we may fear less.
}
\end{center}

\vspace{0.8em}

\begin{flushright}
{\Large \textit{Maria Skłodowska-Curie}}
\end{flushright}

\vspace*{\fill}

\clearpage
\thispagestyle{empty}
\mbox{}
\clearpage

\clearpage
\pagenumbering{roman}
\setcounter{page}{1}

\frontmattertitle{Abstract}

\addcontentsline{toc}{section}{Abstract}
\vspace{-25pt}

\noindent The low-energy dynamics of many-body systems is governed by gapless modes whose properties are dictated by symmetry. Their existence follows from Goldstone’s theorem, while their effective description at zero temperature is determined by the pattern of symmetry breaking. At finite temperature, an analogous role is played by hydrodynamics, which describes the universal collective behavior of many-body systems over long times and large distances.

These principles are well understood for \textit{uniform} symmetries, which act homogeneously in spacetime and lead to a correspondence between massless Goldstone modes and broken generators, as well as gapless hydrodynamic modes and conserved charges. However, this simple picture changes in the presence of \textit{nonuniform} symmetries, whose generators do not commute with spacetime translations. The low-energy implications of these symmetries remain less understood, as they do not introduce additional gapless modes but instead constrain the dynamics of the existing degrees of freedom.

In this thesis, we develop a unified framework for many-body systems with nonuniform symmetries and show that their effects can be understood in terms of additional fields that are not independent at low energies and can be eliminated, leading to kinematic constraints that reshape the infrared dynamics. In systems with spontaneous symmetry breaking, this mechanism modifies the effective theory and often softens the dispersion relations of the remaining modes. At finite temperature, it manifests in hydrodynamics as constraints on macroscopic currents. As a result, nonuniform symmetries give rise to qualitatively new physical phenomena, including modified spectra of collective excitations, exemplified by transverse Tkachenko oscillations in quantum vortex crystals, and unconventional transport phenomena, such as anomalously slow diffusion and softened sound modes in dipole-conserving systems.

\clearpage
\thispagestyle{empty}
\mbox{}
\clearpage

\newpage
\vspace{10pt}
\noindent

%\noindent\rule{\textwidth}{1pt}
%\frontmattertitle{Streszczenie}
 
\frontmattertitle{Streszczenie}
\addcontentsline{toc}{section}{Streszczenie}
\vspace{-25pt}

{\selectlanguage{polish}
\noindent  W niskich energiach dynamika układów wielu ciał jest zdominowana przez bezmasowe stopnie swobody, których własności są wyznaczone przez symetrie układu. Ich istnienie wynika ze spontanicznego złamania symetrii, zgodnie z twierdzeniem Goldstone’a, natomiast ich efektywny opis w zerowej temperaturze jest wyznaczony przez schemat tego złamania. Analogiczną rolę w skończonej temperaturze pełni hydrodynamika, która opisuje uniwersalne zachowanie układów wielu ciał w granicy długich czasów i dużych odległości.

Powyższy obraz jest dobrze rozumiany dla symetrii \textit{jednorodnych}, których działanie nie zależy jawnie od punktu w czasoprzestrzeni. Wówczas każdemu złamanemu generatorowi odpowiada bezmasowy bozon Goldstone’a, a każdemu zachowanemu ładunkowi odpowiada hydrodynamiczne wzbudzenie kolektywne. Obraz ten ulega jednak istotnej zmianie w obecności symetrii \emph{niejednorodnych}, których generatory nie komutują z translacjami w czasoprzestrzeni. Konsekwencje tych symetrii w niskich energiach pozostają słabiej poznane, ponieważ nie prowadzą one do powstania dodatkowych bezmasowych wzbudzeń, lecz ograniczają dynamikę pozostałych stopni swobody.

W niniejszej rozprawie rozwijamy ujednolicony formalizm układów wielu ciał z niejednorodnymi symetriami. Pokazujemy, że ich efekty mają swoje źródło w obecności dodatkowych pól, które nie stanowią niezależnych stopni swobody w granicy niskich energii i po ich eliminacji narzucają kinematyczne ograniczenia na dynamikę układu. W układach ze spontanicznym złamaniem symetrii mechanizm ten modyfikuje efektywną teorię i często prowadzi do zmiany dyspersji, co przekłada się na wolniejszą propagację bozonów Goldstone’a. Natomiast w hydrodynamice przejawia się w postaci ograniczeń nakładanych na makroskopowe prądy. W konsekwencji niejednorodne symetrie prowadzą do jakościowo nowych zjawisk fizycznych. Obejmują one zmiany w widmie wzbudzeń kolektywnych, takie jak poprzeczne oscylacje Tkaczenki w kwantowych kryształach wirów, a także nietypowe zjawiska transportowe, w tym anomalnie wolną dyfuzję oraz fale dźwiękowe o kwadratowej dyspersji w układach zachowujących moment dipolowy.
}

\vspace{10pt}
\newpage

\clearpage
\thispagestyle{empty}
\mbox{}
\clearpage

\newpage

\section*{List of Publications}
During the course of this PhD, the author has published the following articles.

\begin{enumerate}

\item[A1] \textbf{A.~Głódkowski}, 
\textit{A complex scalar field theory for charged fluids, superfluids, and fracton fluids}, 
\href{https://doi.org/10.1007/JHEP03(2026)134}{JHEP \textbf{03} (2026) 134}.

\item[A2] \textbf{A.~Głódkowski}, S.~Moroz, F.~Peña-Benítez, P.~Surówka, 
\textit{Effective field theory for the superfluid vortex lattice from coset construction}, 
\href{https://doi.org/10.1103/m1b1-94f7}{Phys.\ Rev.\ B \textbf{113}, 024506 (2026)}.

\item[A3] \textbf{A.~Głódkowski}, P.~Matus, F.~Peña-Benítez, L.~Tsaloukidis, 
\textit{Quadrupole gauge theory: Anti-Higgs mechanism and elastic dual}, 
\href{https://doi.org/10.1103/wnvt-gmwv}{Phys.\ Rev.\ D \textbf{112}, L121702 (2025)}.

\item[A4] \textbf{A.~Głódkowski}, F.~Peña-Benítez, P.~Surówka, 
\textit{Dissipative fracton superfluids}, 
\href{https://doi.org/10.1007/JHEP07(2024)285}{JHEP \textbf{07} (2024) 285}.

\item[A5] S.~Nandy, J.~Herbrych, Z.~Lenarčič, \textbf{A.~Głódkowski}, P.~Prelovšek, M.~Mierzejewski, 
\textit{Emergent dipole moment conservation and subdiffusion in tilted chains}, 
\href{https://doi.org/10.1103/PhysRevB.109.115120}{Phys.\ Rev.\ B \textbf{109}, 115120 (2024)}.

\item[A6] \textbf{A.~Głódkowski}, F.~Peña-Benítez, P.~Surówka, 
\textit{Hydrodynamics of dipole-conserving fluids}, 
\href{https://doi.org/10.1103/PhysRevE.107.034142}{Phys.\ Rev.\ E \textbf{107}, 034142 (2023)}.

\end{enumerate}

\subsection*{Statement of originality}

Parts of this thesis are based on the author's previously published work [A2, A4--A6], as indicated in the manuscript. The remaining results represent original contributions by the author.


\clearpage
\thispagestyle{empty}
\mbox{}
\clearpage

\newpage
\vspace{10pt}
\noindent

%\noindent\rule{\textwidth}{1pt}
%\frontmattertitle{Streszczenie}
 
\frontmattertitle{Acknowledgements}
\vspace{-25pt}

\noindent I have become who I am through the many interactions I have had over the years. Some led to lasting connections, while others to instructive lessons. With this in mind, I would like to thank all those I have encountered throughout my doctoral studies and in life more broadly.

First, I would like to thank my supervisor and friend, Piotr Surówka, for his invaluable guidance throughout this journey, both scientific and personal, as well as for his constant support. I am also grateful for his help in improving the clarity of this thesis. His sense of humor and relaxed attitude made my PhD considerably less stressful and fostered a more balanced and resilient perspective. Second, I am indebted to Francisco Peña-Benítez for his patience over the years during countless hours at the board, for introducing me to some fancy mathematics, and for shaping the way I think about physical problems.

I extend my thanks to my collaborators, in particular to Sergej Moroz for hosting me at Nordita, teaching me about quantum vortex crystals, and for his continued support, as well as to Paweł Matus for many insightful conversations about physics. I am also grateful to the people at Wrocław University of Science and Technology with whom I shared many lunches and coffee breaks, especially Marcin Mierzejewski, Jacek Herbrych, Maciej Maśka, Maksymilian Kliczkowski, and Jakub Pawłowski. Our numerous discussions on physics and life made my PhD especially enjoyable, and I could not have hoped for a better environment in which to pursue my doctoral studies.

I also wish to acknowledge those who are often overlooked, namely my high school mathematics teacher, Nina Filipowicz, and my physics teacher, Artur Kolincio. I am deeply grateful for the strong foundation in mathematics and physics that you provided, as well as for your guidance and encouragement in the early stages of my development.

I would also like to thank my parents, Wioletta and Jarosław Głódkowski, for giving me exactly the upbringing I needed to flourish in today’s world. Although I have not always been the most expressive child, I am truly grateful for everything you have done for me. I also thank my brothers, Łukasz and Miłosz. I know I can always count on you.

Finally, I am deeply grateful to Natalia for weathering the storm with me during the most challenging stages of this thesis.
\newline 

\hfill A.G.

\vspace{10pt}
\newpage

\clearpage
\thispagestyle{empty}
\mbox{}
\clearpage

\newpage

\vspace{10pt}
%\noindent\rule{\textwidth}{1pt}
\tableofcontents
\noindent\rule{\textwidth}{1pt}
\vspace{10pt}
\thispagestyle{emptyheader}
\newpage

% Write your article contents here, starting from first \section.
% An example structure is given below.
\clearpage
\pagenumbering{arabic}
\setcounter{page}{1}

\section{Introduction}
Symmetries are the organizing principles of nature, underlying both the geometric fabric of spacetime and the interactions between the fundamental particles that inhabit it. In many-body systems, symmetry provides structure, classifying phases of matter and constraining their low-energy excitations despite the large number of interacting constituents. 

This universality stems from the dominance of gapless degrees of freedom at low energies. In the long-wavelength limit, these modes can be excited with arbitrarily small energy,
\be
\omega(k \to 0) \to 0\,,
\ee
and therefore remain relevant over long timescales. On the other hand, excitations separated by a finite energy gap are not accessed at energies below the gap, and their effect is reduced to a renormalization of the effective properties of the gapless modes. Over the past century, this perspective has been shaped by powerful general principles, including Noether’s theorem \cite{Noether1918}, Goldstone’s theorem \cite{Goldstone:1961eq,PhysRev.127.965}, Landau’s theory of symmetry breaking \cite{Landau:1980mil}, and the Wilsonian renormalization group framework \cite{PhysRevB.4.3174,Wilson:1971dh}.

These developments culminate in the modern framework of effective field theory \cite{Weinberg:1978kz}, where symmetry largely determines the structure of the low-energy description. Both the relevant degrees of freedom and their interactions follow from symmetry considerations, independently of microscopic details. The short-distance physics is encoded in a finite set of phenomenological coefficients, fixed either experimentally or by matching to an underlying microscopic theory. This perspective extends naturally to finite temperature, where symmetry constrains hydrodynamic equations, organizes transport, and governs hydrodynamic modes \cite{Son:2009tf,Kovtun:2012rj}. Hydrodynamics can thus be understood as an effective field theory describing the universal long-wavelength behavior of many-body systems near equilibrium \cite{Dubovsky:2011sj,Crossley:2015evo}. Not all symmetries, however, stand on an equal footing. %In this thesis, we focus on \textit{continuous symmetries}, which play a central role in effective field theory and underlie the emergence of gapless modes.

In the conventional classification, continuous symmetries are divided into \textit{internal} and \textit{spacetime} symmetries. Internal symmetries act solely on the field space, whereas spacetime symmetries act nontrivially on the spacetime coordinates. For instance, the generators of the Poincaré algebra are spacetime symmetries, while a global $\rm U(1)$ phase shift symmetry is internal. While this distinction is useful, it does not capture the different ways in which symmetries affect low-energy physics. From the perspective of effective field theory, a more refined classification distinguishes symmetries according to whether their generators act uniformly in spacetime, leading to a distinction between \textit{uniform} and \textit{nonuniform} symmetries, whose consequences for low-energy dynamics are fundamentally different.

For uniform symmetries, conserved charges give rise to conserved densities obeying continuity equations, leading to hydrodynamic modes whose gaplessness follows from conservation laws and that govern long-wavelength dynamics at finite temperature. On the other hand, spontaneous symmetry breaking leads to Goldstone fields whose interactions are derivatively coupled, with their gaplessness protected by symmetry.

However, this simple framework breaks down in a number of physically relevant systems. For example, rotating superfluids can form vortex lattice phases whose low-energy spectrum contains only a single gapless mode, despite the spontaneous breaking of multiple continuous symmetries \cite{Watanabe:2013iia,Glodkowski:2025krf}. At the same time, systems with conserved multipole moments of charge exhibit excitations with restricted mobility leading to anomalous transport that lies beyond the scope of conventional hydrodynamic descriptions \cite{Guardado_Sanchez_2020,Gromov_hydro_2020,PhysRevResearch.3.043186,Glorioso:2022,Glodkowski:2022xje}. These examples appear to challenge the conventional correspondence between symmetries and gapless modes and raise the question of how to systematically construct effective theories for such systems. In this thesis, we address this question and show that these phenomena have a common origin in symmetries that act nonuniformly in spacetime, i.e., whose generators do not commute with spacetime translations.

Such \textit{nonuniform symmetries} are in fact ubiquitous in many-body systems. A simple and physically transparent example is provided by spatial rotations. For a single particle, the generator of rotations is the angular momentum,
\be
\boldsymbol{L} = \boldsymbol{x} \times \boldsymbol{p} \,,
\ee
which depends explicitly on the position of the particle. Under a spatial translation,
\be 
\boldsymbol{x} \to \boldsymbol{x} + \boldsymbol{a}\,,
\ee
the angular momentum transforms as
\be
\boldsymbol{L} \to \boldsymbol{L} + \boldsymbol{a} \times \boldsymbol{p} \,,
\ee
and is therefore not invariant. This reflects the fact that rotations act nonuniformly in space and do not commute with spatial translations. The physical consequences of rotational symmetry are well understood, even if often treated implicitly. In particular, in a translationally invariant system, rotations do not give rise to independent conservation laws in the long-wavelength limit, but instead impose kinematic constraints, enforcing the symmetry of the stress tensor. Similarly, in crystals, both translations and rotations are spontaneously broken, yet the low-energy spectrum contains only phonons associated with the broken translations, with no additional gapless modes arising from the broken rotations.

However, more exotic classes of nonuniform symmetries, such as multipole and subsystem symmetries, lead to more dramatic consequences, qualitatively reorganizing the low-energy dynamics. In these systems, nonuniform symmetries do not merely constrain the dynamics but reshape it at a fundamental level, giving rise to novel phases of matter characterized by new forms of topological order \cite{Vijay_2015,PhysRevB.94.235157}, as well as distinct universality classes with qualitatively altered transport properties \cite{Gromov_hydro_2020,Glorioso:2022}. These phenomena have attracted significant interest, including in quantum information, where such phases may offer new mechanisms for quantum error correction. Despite their physical relevance, no unified theoretical framework exists for nonuniform symmetries comparable to that available for their uniform counterparts.

\subsection{Nonuniform symmetries}
We now formalize the notion of \emph{nonuniform symmetries}. To make this precise, we work in the Hamiltonian formulation of classical field theory, where symmetries act via Poisson brackets. Since the arguments are largely algebraic, they carry over directly to quantum theories, with Poisson brackets replaced by equal-time commutators.

The phase space of a classical field theory consists of maps from physical space into a finite-dimensional target manifold, $\phi^a: \mathbb{R}^3 \to M$, where $\phi^a$ collectively denotes all independent canonical variables. Local observables, such as conserved currents, are expressed as local functionals of these variables evaluated at the spatial point $\boldsymbol{x}$. The Poisson structure on phase space defines the action of global symmetry generators on the fields. In particular, the generators of spatial translations act on the canonical fields as,\footnote{Here we
denote the generator of time translations as $P_0 \equiv H$, where
$H$ is the Hamiltonian. Translational invariance is a symmetry of the
theory provided that the Hamiltonian is a functional of the canonical fields,
$ H\equiv H[\phi^a]$, with no explicit dependence on spacetime coordinates.}
\be \label{eq:translationCanonical}
 \{P_\mu,  \phi^a(\boldsymbol x) \} = - \partial_\mu \phi^a(\boldsymbol x) \,,
\ee
and hence act on any local functional of the fields by the induced spacetime derivative.\footnote{In writing Eq.~\eqref{eq:translationCanonical}, we have left the time dependence of the fields implicit.} Note, however, that contrary to finite-dimensional classical mechanics, the spacetime coordinates themselves do not belong to the phase space and therefore do not enter the Poisson algebra, which is defined with respect to the canonical fields. In particular, one has $\{ P_\mu , x^\nu \} = 0$. Therefore, a local observable does not in general inherit the transformation property of the canonical fields, Eq.~\eqref{eq:translationCanonical}, when it exhibits explicit coordinate dependence, i.e., when it is \textit{nonuniform}.

To define uniform symmetry, we consider a global continuous symmetry with conserved current $j = j_\mu dx^\mu$ satisfying
\be
d \star j = 0 \,,
\ee
and associated conserved charge
\be 
Q = \int \star j \,,
\ee
where the integration is carried over a spatial hypersurface of fixed time.\footnote{One can also consider integrating the current over more general hypersurfaces. The resulting operators remain topological, but no longer define conserved charges. This perspective has recently shaken our understanding of global symmetries \cite{Gaiotto:2014kfa}.}

Such a symmetry is said to be \emph{uniform} if the conserved current carries no explicit dependence on spacetime coordinates. Equivalently, under spacetime translations generated by $P_\mu$, the current transforms homogeneously,
\be
\{ P_\mu, j(\boldsymbol{x}) \} = - \partial_\mu j(\boldsymbol{x}) \,,
\ee
which implies that the associated charge commutes with translations,
\be
\{ P_\mu, Q \} = 0 \,.
\ee
Conversely, nonuniform symmetries fail to commute with spacetime translations, and their associated currents carry explicit coordinate dependence. 

The central claim of this thesis is that nonuniform symmetries do not give rise to additional independent gapless modes, but instead impose constraints on the low-energy theory. As a result, nonuniform symmetry generators are realized on the same degrees of freedom as the uniform ones. At zero temperature, this leads to a nontrivial counting of Goldstone modes, in which not all broken generators correspond to independent gapless excitations. The remaining would-be Goldstone fields are either auxiliary variables reflecting a redundancy in the parameterization of the vacuum manifold, or correspond to gapped excitations. These would-be Goldstone fields can be eliminated, leaving an imprint on the remaining gapless modes and constraining their dynamics in a manner consistent with the nonuniform symmetry. At finite temperature, nonuniform symmetries similarly give rise to gapped modes which, upon integrating out, impose kinematic constraints on the remaining hydrodynamic currents.

In this thesis, we develop a general framework for many-body systems with nonuniform symmetries that makes these mechanisms precise. We establish general counting rules for both gapless Goldstone excitations and hydrodynamic modes. We then analyze concrete physical systems, including quantum vortex lattices and dipole-conserving fluids, to elucidate their nontrivial dynamics and trace their origin to nonuniform symmetries.
%%%%%%%

%%%%%%%%%
%%%%%%%%
%%%%%%%%%
%%%%%%%%
%%%%%%%%%%
%%%%%%%%%%%

%%%%%%%%%
%%%%%%%%%%%
%%%%%%%%%%%

%%%%%%%%%
%%%%%%%%%%%
%%%%%%%%%%%

%%%%%%%%%
%%%%%%%%%%%
%%%%%%%%%%%

%%%%%%%%%
%%%%%%%%%%%
%%%%%%%%%%%

%%%%%%%%%
%%%%%%%%%%%
%%%%%%%%%%%

%%%%%%%%%
%%%%%%%%%%%
%%%%%%%%%%%

%%%%%%%%%
%%%%%%%%%%%
%%%%%%%%%%%
\subsection{Organization of the thesis}

This thesis is organized in two parts. 

Part \ref{part:1} is devoted to the study of nonuniform symmetry breaking and its implications for the low-energy spectrum. In Sec.~\ref{sec:nonuniformGoldstone}, we develop the framework for nonlinearly realized nonuniform symmetries and establish a counting rule for the number of gapless modes. We demonstrate that the counting formula correctly reproduces the number of phonons when applied to elastic media. In Sec.~\ref{sec:vortex}, we study quantum vortex crystals, which provide a particularly nontrivial realization of nonuniform symmetry breaking with numerous such symmetries spontaneously broken in the ground state. In this setting, we systematically construct an effective field theory based on symmetry principles, verify the general counting of gapless modes, and analyze the resulting collective excitations.

In Part \ref{part:2}, we study many-body dynamics in systems with nonuniform symmetries. In Sec.~\ref{sec:nonuniformHydrodynamics}, we adopt the canonical paradigm of hydrodynamics to formulate the hydrodynamic theory governing the diffusion of charges in the presence of nonuniform symmetries. In doing so, we establish a counting rule for gapless hydrodynamic modes. We then turn to Aristotelian fluids, which are homogeneous and isotropic systems without boost symmetry, and construct their hydrodynamic theory from first principles, emphasizing the role of nonuniform symmetries. In Sec.~\ref{sec:fracton}, we introduce fracton phases of matter, focusing on their realization in systems with nonuniform multipole symmetries that constrain the microscopic dynamics of elementary charges. We develop a dipole sigma model, which provides a nonuniform generalization of the ordinary $ \mathrm U(1)$ superfluid and establish a charge-dipole-vortex duality, mapping the theory to an emergent gauge theory with tensor gauge fields. Then, we analyze how the presence of nonuniform dipole symmetry qualitatively modifies charge diffusion, leading to an anomalous slowdown of transport. We also provide a physical realization in terms of a lattice model with emergent dipole conservation that captures this universality class. Finally, in Sec.~\ref{sec:fractonFluids}, we construct a hydrodynamic theory for fracton fluids, which are phases exhibiting a nonuniform dipole symmetry. We identify two distinct fluid phases and systematically classify their hydrodynamic transport properties. We demonstrate how the constraints imposed by nonuniform dipole symmetry on microscopic kinematics give rise to unconventional collective dynamics.

\part{Nonuniform symmetry breaking}\label{part:1}

  \section{Counting gapless modes with nonuniform symmetries} \label{sec:nonuniformGoldstone}

In this section, we develop a general counting rule for gapless Nambu--Goldstone (NG) modes in systems with spontaneously broken nonuniform symmetries.\footnote{ Throughout this thesis, we use the term \emph{Goldstone fields} to denote coordinates on the coset space appearing in the nonlinear realization of broken symmetries, while the term \emph{Nambu--Goldstone (NG) modes} is reserved for the genuinely gapless excitations in the physical spectrum. Elucidating the subtle relation between these two notions in the presence of nonuniform symmetries is a central focus of Part~\ref{part:1} of this thesis.} The result applies to both relativistic and nonrelativistic systems, as well as to abelian and nonabelian charges, and relies only on the assumption of unbroken translational invariance. The construction is purely algebraic, based on the nonlinear realization of the underlying symmetry algebra, and is therefore agnostic to the microscopic details of the symmetry breaking mechanism. To the best of our knowledge, the counting formula provided here has not been previously reported in the literature. Finally, we apply the general formula to crystalline solids and show that it correctly accounts for the number of acoustic phonons.

\subsection{Counting gapless modes}
The celebrated Goldstone theorem states that for each spontaneously broken generator there exists a gapless NG mode in the spectrum \cite{PhysRev.127.965},
\be
n_{\mathrm{NG}} = n_{\mathrm{BG}} \,.
\ee
In general, however, this one-to-one relation holds in relativistic systems with Lorentz invariance and, crucially, only when the spontaneously broken generators are \textit{uniform}. In fact, most condensed matter systems tend to violate both of these assumptions. A canonical example is provided by crystalline solids, which spontaneously break spatial translations and rotations, as well as Galilean boosts. The total number of broken generators is then
\be 
n_{\mathrm{BG}} = d + d + \frac{d(d-1)}{2}\,,
\ee 
where $d$ denotes the spatial dimension.
However, the low-energy spectrum contains only the acoustic phonons, and therefore 
\be 
n_{\mathrm{NG}} = d < n_{\mathrm{BG}} \,.
\ee 
This simple example demonstrates that the one-to-one correspondence between broken generators and gapless modes does not hold in general.

In the absence of Lorentz invariance, this correspondence is modified due to the emergence of a nontrivial symplectic structure. This induces a pairing mechanism among the Goldstone modes, which become canonically conjugate to each other, leading to the Nielsen--Chadha classification of type-I and type-II Goldstone modes \cite{Nielsen:1975hm}. For uniform broken generators, the number of gapless modes is determined by the expectation values of the commutators of conserved charges, as captured by the Watanabe--Murayama theorem \cite{Watanabe:2012hr,Watanabe:2014fva} (the counting was first proposed without proof by Watanabe and Brauner in \cite{Watanabe:2011ec}). However, this counting rule does not resolve the discrepancy exhibited by the phonon example because no such pairing takes place in elastic crystals. Indeed, even relativistic solids support only acoustic phonons \cite{Endlich:2012pz,PhysRevD.89.045002}, indicating that the reduction in the number of gapless modes must occur via a different mechanism.

This mechanism is rooted in the presence of spontaneously broken nonuniform symmetries. While the reduction in the number of gapless excitations in systems with nonuniform symmetry breaking has been discussed in various contexts \cite{PhysRevLett.88.101602,Watanabe:2011ec,Watanabe:2013iia,Brauner:2014aha}, a general counting rule is currently lacking. Within the coset construction (see Appendix~\ref{sec:coset} for a review), this reduction is typically implemented via the inverse Higgs mechanism \cite{Ivanov:1975zq}, which provides a systematic procedure for eliminating inessential Goldstone fields (see Appendix~\ref{sec:inverseHiggsMechanism}). However, this procedure does not determine the number of independent gapless modes. In particular, the number of massless excitations cannot, in general, be obtained by simply subtracting the number of imposed inverse Higgs constraints from the number of broken generators, i.e.,\footnote{Even after accounting for possible pairings between the remaining Goldstone modes, this relation does not hold in general.}
\be
n_{\mathrm{NG}} \neq n_{\mathrm{BG}} - n_{\mathrm{IH}}\,,
\ee
where $n_{\mathrm{BG}}$ denotes the total number of broken generators and $n_{\mathrm{IH}}$ is the number of inverse Higgs constraints.

In the remainder of this section, we show that the number of gapless modes in the presence of nonuniform symmetries can nevertheless be determined by a simple algebraic argument, and is encoded in the kernel of the action of translations on the space of charges. We also demonstrate that this counting rule correctly accounts for the number of gapless modes in crystalline solids.

%%%%%%%%%

\subsection{A minimal example}
Before developing the general formalism, we consider a minimal example of the nonlinear realization of nonuniform symmetries and illustrate its consequences for the low-energy spectrum. To this aim, we consider a translationally invariant system in one spatial dimension which, in addition to conserved momentum $P$, carries two conserved charges, a uniform (monopole) charge $Q$ and a nonuniform (dipole) charge $D$ satisfying the following algebra 
\be \label{eq:dipoleAlgebra1d}
\{P,D\}=Q\,.
\ee
We assume that both symmetries $Q$ and $D$ are spontaneously broken in the ground state and introduce the Goldstone fields $\varphi$ and $\psi$, corresponding to broken $Q$ and $D$, respectively, so that\footnote{
These relations encode the shift symmetry of the Goldstone fields under the action of the broken charges.
}
\bs
\{Q, \varphi(  x) \} &= 1\,, \quad &\{D, \varphi(  x) \} &= 0\,, \\
\{Q, \psi(  x) \} &= 0\,, \quad &\{D, \psi(  x) \} &= 1\,.
\es 
%%%
Notice that the symmetry algebra implies 
\be 
1 =\{ Q, \varphi(x) \} =\{ \{P,D\}, \varphi(x) \} = -\{ D,  \{ P,  \varphi(x) \} \} \,,
\ee 
where in the second equality we have invoked the Jacobi identity together with $\{D, \varphi( x) \} = 0$. Therefore, we conclude that the $\varphi$ transforms inhomogeneously under translations 
\be 
\{ P,  \varphi( x) \} = - \partial_x \varphi( x) - \psi( x)\,.
\ee 
Since the Goldstone field $\varphi$ does not transform homogeneously under translations, it is useful to decompose it into a homogeneous mode $\tilde\varphi$ and a piece that captures the inhomogeneous contribution,
\be 
 \varphi = \tilde \varphi - x\psi \,,
\ee
so that $\tilde\varphi$ transforms homogeneously under translations
\be 
\{ P,  \tilde \varphi(x) \} = - \partial_x \tilde \varphi(x)\,.
\ee 
However, $\tilde \varphi$ is charged under $D$,
\be 
\{ D, \tilde \varphi(x)\} =  x \,.
\ee 
Therefore, the homogeneous mode $\tilde \varphi$ realizes nonlinearly both the uniform symmetry $Q$ and the nonuniform symmetry $D$. It is then straightforward to verify that the combinations $d\psi$ and 
\bs
D\tilde\varphi \equiv d\tilde\varphi - \psi dx
\es
are invariant under the symmetries generated by $Q$ and $D$ and therefore provide the basic invariant building block for the effective theory. 

In order to elucidate the low-energy content of the effective theory, we consider a quadratic action 
\begin{equation}
S=\frac12\int dt dx
\Big[
\frac{1}{v^2}(\partial_t\tilde\varphi)^2
- (D_x\tilde\varphi)^2
+ \frac{1}{\Omega^2}(\partial_t\psi)^2
- \frac{1}{\Lambda^2} (\partial_x\psi)^2
\Big] \,,
\end{equation}
written in the $\hbar  = 1$ units so that
\bs 
[\tilde \varphi] &= L^{1/2} T^{-1/2}\,, \\ 
[\psi] &= L^{-1/2} T^{-1/2}\,, \\ 
[v] &= L T^{-1}\,, \\ 
[\Omega] &= T^{-1}\,, \\ 
[\Lambda] &= L^{-1}\,.
\es  
%%%%%%%%%
%%%%%%%%
%%%%%%%%%
%%%%%%%%
%%%%%%%%%
%%%%%%%%
%
Going to Fourier space, $\partial_t\to - i\omega$ and $\partial_x\to i k$, the equations of motion can be written as
\begin{equation}
\begin{pmatrix}
\omega^2 - v^2 k^2 & -i v^2 k \\
-i k & 1 - \omega^2/\Omega^2 + k^2/\Lambda^2
\end{pmatrix}
\begin{pmatrix}
\tilde\varphi \\ \psi
\end{pmatrix}
=0 \,.
\end{equation}
Nontrivial solutions exist when the determinant vanishes, yielding the dispersion relation
\begin{equation}\label{eq:dispersionNonuniformToy}
(\omega^2-v^2 k^2)\Big(1-\frac{\omega^2}{\Omega^2}+\frac{k^2}{\Lambda^2}\Big)+v^2 k^2=0,
\end{equation}
which is quadratic in $\omega^2$ and therefore admits two branches $\omega_\pm(k)$. Expanding at small momentum, $k\ll \Lambda$, one finds a gapless NG mode
\be \label{eq:quadratic}
\omega_-(k)=\pm \frac{v}{\Lambda} k^2 + O(k^4)\,,
\ee
and a gapped mode
\be
\omega_+(k)=\pm \left( \Omega+\frac{v^2 \Lambda^2 + \Omega^2}{2 \Lambda^2 \Omega} k^2 \right) + O(k^4)\,.
\ee
In particular, we see that the theory contains a massive excitation with gap $\Omega$, which at small $k \ll \Lambda$ is dominated by the $\psi$ field. At low energies $\omega\ll\Omega$ and $k \ll \Lambda$, the massive field $\psi$ can be integrated out from its equation of motion
\be
\frac{1}{\Omega^2}\partial_t^2\psi - \frac{1}{\Lambda^2}\partial_x^2\psi  - D_x \tilde \varphi  = 0\,.
\ee 
This gives
\be 
\psi = \left(1 + \frac{\partial_t^2 }{\Omega^2} - \frac{\partial_x^2}{\Lambda^2} \right)^{-1} \partial_x \tilde \varphi \simeq \partial_x \tilde \varphi + \mathcal O\left(\frac{\omega^2}{\Omega^2}, \frac{k^2}{\Lambda^2} \right)\,.
\ee 
Therefore, in the low-energy, long-wavelength limit $\omega\ll\Omega$ and $k\ll\Lambda$ the $\psi$ field does not correspond to an independent degree of freedom but rather is soldered to $\tilde \varphi$ via 
\be 
\psi \simeq \partial_x \tilde \varphi\,.
\ee  
Plugging this back into the Lagrangian we obtain the following low-energy theory at the leading order\footnote{Since the remaining Goldstone exhibits a quadratic dispersion we infer scaling $\partial_t \sim \partial^2_x$.}
\begin{equation} \label{eq:lowenergyTheory}
S=\frac12\int dt dx
\Big[
\frac{1}{v^2}(\partial_t\tilde\varphi)^2
- \frac{1}{\Lambda^2} (\partial^2_x\tilde \varphi)^2
\Big] \,.
\end{equation}
We find that the low-energy theory contains a single massless NG mode $\tilde \varphi$ with a quadratic dispersion relation Eq.~\eqref{eq:quadratic}, which nonlinearly realizes both spontaneously broken symmetries. The reduction in the number of massless Goldstone modes and the softening of the remaining mode are direct consequences of the nonuniform symmetry breaking pattern. In particular, nonuniform symmetries do not give rise to additional gapless excitations, as the associated mode is not independent and, in this case, manifests as a gapped excitation. Once integrated out, this mode qualitatively modifies the low-energy spectrum by softening the dispersion of the remaining mode, reflecting the constraints imposed by the nonuniform dipole symmetry.

% --- Box ---
 \begin{scipostbox}[Dual description]
Interestingly, the resulting low-energy theory Eq.~\eqref{eq:lowenergyTheory} admits a dual description in terms of an emergent gauge theory. To see this, we first rescale $\tilde \varphi \rightarrow \sqrt{v \Lambda} \tilde \varphi$ such that the theory takes the following form
\be \label{eq:dipoleLagrangian1D}
\mathcal L = \frac{g}{2}
(\partial_t\tilde\varphi)^2
- \frac{1}{2g} (\partial^2_x\tilde \varphi)^2
 \,, 
\ee
where $g =\frac{\Lambda}{v}$. We will assume that $\tilde \varphi$ is smooth, so that integrations by parts can be performed without generating boundary or defect terms.\footnote{In the compact theory the $\tilde\varphi$ field may contain singular configurations corresponding to vortices. In this case the theory is no longer self-dual, although it still admits a dual gauge theory description in which vortices play the role of gauge charges.} 
 The equation of motion is 
\be \label{eq:dipoleEquation1D}
\partial_t j_t + \partial^2_x j_{xx} = 0 \,,
\ee 
where $j_t = g \partial_t \tilde \varphi$ and $j_{xx} = \frac{1}{g} \partial^2_x \tilde \varphi$. Now we perform a Legendre transformation to express the Lagrangian in terms of the current
\be
\mathcal L = j_t \partial_t \tilde \varphi - \frac{1}{2g} j_t^2 - j_{xx} \partial^2_x \tilde \varphi + \frac{g}{2} j_{xx}^2
 \,.
\ee
Integrating out $\tilde \varphi$ then imposes the conservation equation Eq.~\eqref{eq:dipoleEquation1D}, which can be resolved by introducing a scalar gauge potential $\phi$,
\be 
j_t = \partial^2_x \phi \,, \quad j_{xx} = - \partial_t \phi\,,
\ee 
with gauge freedom 
\be 
\phi \rightarrow \phi + a(t) + b(t) x \,,
\ee 
which leaves Eq.~\eqref{eq:dipoleEquation1D} invariant. Then we arrive at the following description
\be
\mathcal L =  \frac{g}{2} \left( \partial_t \phi \right)^2  - \frac{1}{2g} \left( \partial^2_x \phi \right) ^2  
 \,.
\ee

Therefore, we find that the theory Eq.~\eqref{eq:dipoleLagrangian1D} is self-dual
\be \boxed{
g \partial_t\tilde\varphi \leftrightarrow \partial_x^2\phi\,, 
\quad
\frac{1}{g} \partial_x^2\tilde\varphi \leftrightarrow -\partial_t\phi \,.}
\ee

 \end{scipostbox}
% --- Box ---
%%%%%%%%%

The example above may suggest that the number of gapless modes is determined simply by counting the number of uniform broken generators. While correct in spirit, this picture is too naive, since the algebra need not be expressed in a basis adapted to the decomposition into uniform and nonuniform directions.

To illustrate this, consider the following nonuniform algebra
\bs \label{eq:nonuniformAlgebra12}
\{P,Q^1\} = \{P,Q^2\} = Q^2 - Q^1\,,
\es
where $Q^1$ and $Q^2$ denote broken generators. In particular, neither $Q^1$ nor $Q^2$ commutes with translations, and both are nonuniform symmetries. One might then naively conclude that no uniform broken generators are present and therefore that the system is gapped.

However, the algebra Eq.~\eqref{eq:nonuniformAlgebra12} is simply the dipole algebra Eq.~\eqref{eq:dipoleAlgebra1d} written in a different basis of the broken charge space,
\be
Q^1 \equiv D\,, \qquad Q^2 \equiv D+Q\,.
\ee
This example shows that the existence of gapless modes is not determined by whether the individual generators in a given presentation of the algebra are uniform, but instead requires a basis-independent classification. This classification is controlled by the kernel of the translation action on the space of broken charges.

To make this explicit, introduce the charge vector
\be
Q^A =
\begin{pmatrix}
Q^1 \\
Q^2
\end{pmatrix}\,.
\ee
Then the action of translations can be written as
\be
\{ P, Q^A \} = \lambda^A{}_B \, Q^B \,,
\ee
with
\be
\lambda =
\begin{pmatrix}
-1 & -1 \\
1 & 1
\end{pmatrix}.
\ee
The kernel of $\lambda$ is nontrivial,
\be
\ker \lambda = \mathrm{span}\!\left\{
\begin{pmatrix}
-1 \\
1
\end{pmatrix}
\right\},
\ee
corresponding to the linear combination
\be
 Q^2 - Q^1 = Q\,,
\ee
which is precisely the uniform broken generator.

In the following, we formalize this observation and establish a general correspondence between massless NG modes and the kernel of the translation action, leading to a simple counting formula for the number of gapless excitations.

%%%%%%%%%%%
\subsection{Nonuniform abelian symmetries}
We consider a system invariant under spacetime translations generated by $P_\mu$ and carrying a set of nonuniform charges $\{ Q^A \}$ obeying the algebra\footnote{In general, the right-hand side of the nonuniform algebra Eq.~\eqref{eq:nonuniformAlgebra} may also contain unbroken generators. If present, these may impose symmetry constraints on the effective action but do not influence the counting of the massless NG modes, which depends only on the projection of the algebra onto the broken charge sector.}
\be \label{eq:nonuniformAlgebra}
\left\{ P_\mu, Q^A \right\} = {(\lambda_\mu)}^A{}_B Q^B\,,
\ee 
where $\lambda_\mu$ are constant matrices furnishing a representation of the translation algebra on the charge space.

Throughout this section we assume that the spacetime translations $P_\mu$ and the
charges $\{ Q^A \}$ commute among themselves, i.e.,
\be 
\left\{ P_\mu, P_\nu \right\} = 0\,, \quad \left\{ Q^A, Q^B \right\} = 0\,,
\ee 
and that all charges $Q^A$ are
spontaneously broken, while the translations $P_\mu$ remain unbroken. Since spacetime translations commute, the Jacobi identity for charges
$(P_\mu,P_\nu,Q^A)$ implies
\be \label{eq:lambdaCommutator}
[\lambda_\mu,\lambda_\nu]=0 \,,
\ee
so that the matrices $\lambda_\mu$ are mutually commuting.

\subsubsection{Nonuniform Goldstone fields}
Let us introduce a set of Goldstone fields $\{ \varphi_A \}$ associated with the broken generators $\{ Q^A \}$, together with the canonical Poisson bracket structure
\be \label{eq:abelianCanonicalCommutation}
\left\{ Q^A, \varphi_B( \mathbf x)  \right \} = \delta^A{}_B\,.
\ee 
Invoking the nonuniform algebra Eq.~\eqref{eq:nonuniformAlgebra} and the canonical Poisson brackets of the Goldstone fields Eq.~\eqref{eq:abelianCanonicalCommutation}, we find
\be 
\left \{ \{ P_\mu, Q^A\}, \varphi_B( \mathbf x) \right \} = {(\lambda_\mu)}^A{}_C \left \{  Q^C, \varphi_B( \mathbf x) \right \} = {(\lambda_\mu)}^A{}_B\,. 
\ee 
On the other hand, by the Jacobi identity,
\bs 
\left \{ \{ P_\mu, Q^A\}, \varphi_B( \mathbf x) \right \} &= \left  \{ P_\mu, \{Q^A, \varphi_B (\mathbf x) \} \right \} - \left \{ Q^A, \{P_\mu, \varphi_B (\mathbf x)  \} \right \} \,,  \\
&= - \left \{ Q^A, \{P_\mu, \varphi_B (\mathbf x)  \} \right \}\,,
\es 
where in the second line we have used the canonical Poisson brackets Eq.~\eqref{eq:abelianCanonicalCommutation}.

 We therefore conclude that the Goldstone fields $\varphi_A$ transform inhomogeneously under spacetime translations,
\be \label{eq:Goldstonetransformation}
\{P_\mu, \varphi_A (\mathbf x)  \} = - \partial_\mu  \varphi_A (\mathbf x) - (\lambda_\mu^T)_A{}^B \varphi_B(\mathbf x)\,.
\ee 
It is then possible to factor out the inhomogeneous part of the transformation by writing
\be \label{eq:homogeneousGoldstone}
\varphi_A = {(e^{-x^\mu \lambda_\mu^T})}_A{}^B \tilde \varphi_B \,.
\ee 
It is straightforward to verify that the $\tilde \varphi_A$ fields transform homogeneously under spacetime translations,
\be 
\{P_\mu, \tilde \varphi_A (\mathbf x)  \} = - \partial_\mu  \tilde \varphi_A (\mathbf x)\,,
\ee 
but no longer satisfy the canonical Poisson brackets \eqref{eq:abelianCanonicalCommutation}. Instead, we find that
\be  \label{eq:nonAbelianHomogenousGoldstone}
 \{Q^A, \tilde \varphi_B(\mathbf x) \} = (e^{x^\mu \lambda_\mu^T})_B{}^C  \{Q^A, \varphi_C(\mathbf x) \} = (e^{x^\mu \lambda_\mu^T})_B{}^A\,.
\ee 
We can then construct the covariant derivatives 
\be \label{eq:covariantDerivative}
D \tilde \varphi_A = d \tilde \varphi_A - (\lambda_\mu^T)_A{}^B \tilde \varphi_B dx^\mu \,,
\ee 
which are invariant under the action of the broken charges,
\be \label{eq:covariantDeriveHomogeneousGoldsotne}
\{ Q^A, D \tilde \varphi_B(\mathbf x) \} = 0\,.
\ee 
%%%%%
%%%%%
%%%%%
\subsubsection{Kernel decomposition}\label{sec:kernelDecomposition}
Let $V$ denote the Goldstone vector space dual to the charge space $V^\star$. 
Define the common kernel of the matrices $\lambda_\mu^T$ as the subspace
\be \label{eq:kernel}
K \equiv  \bigcap_\mu \ker \lambda_\mu^T \subset V \,,
\ee
together with a complementary subspace $K^\perp$ such that
$V = K \oplus K^\perp$.
Accordingly, any vector $\tilde\varphi \in  V$ admits a unique decomposition
\be \label{eq:GoldstoneDecompositon}
\tilde\varphi = \tilde\varphi^{\parallel} + \tilde\varphi^{\perp} \,, 
\ee
where $\tilde\varphi^{\parallel}\in K$ and $\tilde\varphi^{\perp}\in K^\perp\,.$

Choose a basis $\{e^a\}_{a=1}^{\dim K}$ of $K$ and a basis
$\{e^\alpha\}_{\alpha=1}^{\dim K^\perp}$ of a complementary subspace $K^\perp$ such that 
\be 
\tilde\varphi^{\parallel} = e^a \tilde\varphi^{\parallel}_a\,, \quad \tilde\varphi^{\perp} = e^\alpha \tilde\varphi^{\perp}_\alpha \,.
\ee 
By construction, the basis vectors spanning $K$ satisfy
\be \label{eq:kernelBasisCondition}
\lambda_\mu^T e^a = 0 \,.
\ee
The combined set
\be
\{e^A\}\equiv\{e^a,e^\alpha\}
\ee
forms a basis of $V$ adapted to the decomposition
$V = K \oplus K^\perp$.

Let us introduce the dual basis in the charge space $E_A = \{ E_a, E_\alpha \}$, defined by  
\be 
\langle E_A,  e^B \rangle  = \delta_A{}^B\,.
\ee 
Any charge $Q \in V^\star$ can then be decomposed as
\be 
Q =  E_A Q^A = E_a Q^a + E_\alpha Q^\alpha \,. 
\ee 
%%%%%%%%%%
%break%
%%%%%%%
The kernel decomposition makes manifest a one-to-one correspondence between the Goldstone modes in the kernel and the uniform generators. In particular, in a basis adapted to the kernel decomposition
Eq.~\eqref{eq:GoldstoneDecompositon}, to each kernel Goldstone mode
$\tilde\varphi^\parallel_a$ one can associate a uniform broken generator
$Q^a$ satisfying
\be \label{eq:canonicalUniform}
\{Q^a, \tilde\varphi^\parallel_b \} = \delta^a{}_b \,.
\ee
To see this, let us first show that the $\{ Q^a \}$ charges are uniform. 
By Eq.~\eqref{eq:kernelBasisCondition}, one has
\be
\lambda_\mu^T e^a = 0 \quad \Longrightarrow \quad \langle E_A , \lambda_\mu^T e^a \rangle
= \langle \lambda_\mu E_A , e^a \rangle = 0\,.
\ee
Expanding the action on the dual basis,
\be
\lambda_\mu E_A = E_B (\lambda_\mu)^B{}_A \,,
\ee
it follows that
\be \label{eq:lambdaCondition}
 (\lambda_\mu)^B{}_A \langle E_B,e^a\rangle
= (\lambda_\mu)^a{}_A = 0\,.
\ee
Therefore, in the basis adapted to the kernel decomposition of the Goldstone vector space, the charges $\{ Q^a \}$ are uniform 
\be 
\left \{ P_\mu, Q^a \right \} = (\lambda_\mu)^a{}_A Q^A = 0\,. 
\ee 
To show canonical conjugacy of the kernel modes, we start from Eq.~\eqref{eq:nonAbelianHomogenousGoldstone}, which implies 
    \be  
 \{Q^a, \tilde \varphi_b(\mathbf x) \}  = (e^{x^\mu \lambda_\mu^T})_b{}^a\,.
\ee 
On the other hand, by Eq.~\eqref{eq:kernelBasisCondition} we have
\be 
\lambda_\mu^T e^a = 0 \quad \Longrightarrow \quad e^{x^\mu \lambda_\mu^T} e^a = e^a \,. 
\ee 
Then it follows that
\be 
(e^{x^\mu \lambda_\mu^T})_b{}^a = \langle E_b, e^{x^\mu \lambda_\mu^T} e^a \rangle = \langle E_b,  e^a \rangle = \delta_b{}^a\,.
\ee 
Thus, each Goldstone mode in the kernel $K$ is canonically conjugate to a uniform broken generator.
%
%%%%%%
%%%%%

\subsubsection{Counting formula}

We now establish a general counting rule for the number of gapless NG modes in systems with spontaneously broken nonuniform symmetries. The key observation is that only those Goldstone modes lying in the kernel of the translation action remain derivatively coupled, while all others generically acquire a mass.

Consider a system with broken charges $\{Q^A\}$ obeying the algebra Eq.~\eqref{eq:nonuniformAlgebra}, and let $V$ denote the Goldstone vector space of homogeneous fields. The number of gapless NG modes is given by
\be
\boxed{
n_{\rm NG} = \dim K \,,
\quad
K \equiv \bigcap_\mu \ker \lambda_\mu^T \subset V \,.
}
\ee

To derive this result, we note that the covariant derivatives Eq.~\eqref{eq:covariantDerivative} are invariant under the action of the broken charges. As a consequence, any local Hamiltonian consistent with the symmetry can depend on the Goldstone fields $\tilde\varphi_A$ only through $D_\mu \tilde\varphi_A$, i.e.
\be
\mathcal H \equiv \mathcal H \left[D_\mu \tilde\varphi_A\right] \,.
\ee
Employing a basis adapted to the kernel decomposition, the covariant derivatives take the form
\bs \label{eq:covariantDevsKernelDecompositoon}
D_\mu \tilde\varphi^{\parallel}_a &= \partial_\mu \tilde\varphi^{\parallel}_a - (\lambda_\mu^T)_a{}^\alpha \tilde \varphi^\perp_\alpha\,, \\
D_\mu  \tilde\varphi^{\perp}_\alpha  &= \partial_\mu \tilde\varphi^{\perp}_\alpha - (\lambda_\mu^T)_\alpha{}^\beta \tilde \varphi^\perp_\beta \,.
\es 
where we used Eq.~\eqref{eq:lambdaCondition}, implying
\be 
(\lambda_\mu)^a{}_A =  (\lambda_\mu^T)_A{}^a = 0\,.
\ee 
The Hamiltonian can therefore be written as
\be
\mathcal H \equiv \mathcal H \left[D_\mu \tilde\varphi_a^\parallel, D_\mu \tilde\varphi_\alpha^\perp \right] \,.
\ee 

From Eq.~\eqref{eq:covariantDevsKernelDecompositoon} it follows that the fields $\tilde\varphi^{\parallel}_a$ are derivatively coupled and therefore remain gapless at low energies. Since the broken charges are abelian, no presymplectic structure is generated and no pairing mechanism is available (see Sec.~\ref{sec:nonabelianGoldsotne}). Consequently, each $\tilde\varphi^{\parallel}_a$ corresponds to an independent massless NG mode.

To determine the fate of the remaining fields $\tilde\varphi^{\perp}_\alpha$, consider the most general quadratic contribution to the Hamiltonian consistent with the symmetries,
\be \label{eq:quadraticHamiltonianAllowed}
\mathcal H
=
\frac12 (\Sigma^{\mu\nu})^{A B}\,
D_\mu \tilde\varphi_A\, D_\nu \tilde\varphi_B \,,
\ee
with arbitrary constant coefficients $(\Sigma^{\mu\nu})^{A B }$.

Expanding the covariant derivatives, one finds a contribution of the form
\be
\mathcal H_{\rm mass}
=
\frac12 \tilde\varphi^\perp_\alpha\,
M^{\alpha \beta}\,
\tilde\varphi^\perp_\beta \,,
\ee
where
\be \label{eq:massMatrix}
M^{\alpha \beta} \equiv 
(\lambda_\mu^T)_A{}^\alpha 
(\Sigma^{\mu\nu})^{AB}
(\lambda_\nu^T)_B{}^\beta \,.
\ee

For generic values of $(\Sigma^{\mu\nu})^{AB}$, the matrix $M^{\alpha\beta}$ is non-degenerate, and the fields $\tilde\varphi^\perp_\alpha$ are therefore gapped.\footnote{\label{fn:auxiliary} Strictly speaking, the fields $\tilde\varphi^\perp_\alpha$ need not correspond to independent propagating degrees of freedom. Depending on the realization, they may either appear as massive modes or as auxiliary variables that can be eliminated by a suitable parametrization \cite{PhysRevLett.88.101602}. In either case, they do not represent independent low-energy degrees of freedom.}

We thus conclude that only the modes spanning the kernel $K$ remain gapless, leading to
\be
n_{\rm NG} = \dim K \,.
\ee
The above result admits a natural interpretation directly in charge space. Defining
\be 
K^\star \equiv \bigcap_\mu \ker \lambda_\mu \subset V^\star \,,
\ee 
we identify $K^\star$ as the subspace of uniform generators.

Indeed, by definition of the representation $\lambda_\mu$ in
Eq.~\eqref{eq:nonuniformAlgebra}, a charge $Q\in V^\star$
is uniform if and only if
\be
\{P_\mu,Q\}=\lambda_\mu Q =0
\ee
for all $\mu$, which is equivalent to $Q \in K^\star$.

Furthermore, in a basis adapted to the kernel decomposition of the Goldstone space, each massless NG mode
$\tilde\varphi^\parallel_a \in K$ is canonically conjugate to a uniform generator $Q^a$, see Eq.~\eqref{eq:canonicalUniform}. Since
$\dim K^\star=\dim K$, this establishes a one-to-one correspondence between the elements of $K^\star$ and massless NG modes.

The counting of gapless modes may therefore be performed equivalently in the charge representation as
\be \boxed{
n_{\rm NG} = \dim K^\star\,, \quad K^\star = \bigcap_\mu \ker\lambda_\mu \,.}
\ee
Thus, determining the number of massless NG modes reduces to computing the kernel $K^\star$. Its elements are precisely the uniform generators, whose canonically conjugate Goldstone fields exhaust the set of gapless excitations. This provides a simple and general counting rule for systems with nonuniform symmetries. Notably, the result does not rely on Lorentz or Galilean invariance, nor on rotational symmetry. 

In the following section, we generalize this analysis to nonabelian algebras, where additional pairing mechanisms between Goldstone modes may arise.

%%%%%%%
%%%%%

%%%%%%%
%%%%%%%
%RANDOM%
%%%%%%%
%%%%%%%
%%%%%%%
%RANDOM%

%%%%%%%
%%%%%%%%%%
%%%%%%%%
\subsection{Nonuniform nonabelian symmetries}\label{sec:nonabelianGoldsotne}
 We now extend the analysis to the case in which the broken nonuniform charges $\{Q^A\}$ form a
nonabelian algebra. In particular, we consider the following algebra\footnote{Consistency of the algebra further requires the Jacobi identity for
$(P_\mu,Q^A,Q^B)$, which implies that $\lambda_\mu$ acts as a derivation of the
Lie algebra generated by $Q^A$, i.e.
\be
(\lambda_\mu)^A{}_D f^{DB}{}_C
+(\lambda_\mu)^B{}_D f^{AD}{}_C
-f^{AB}{}_D (\lambda_\mu)^D{}_C = 0 \,.
\ee
}
\be \label{eq:nonabelianAlgebra}
\left\{ P_\mu, Q^A \right\} = {(\lambda_\mu)}^A{}_B Q^B\,, \quad
\left\{ Q^A, Q^B \right\} = f^{AB}{}_C Q^C \,.
\ee
This algebra is very general in that it relies only on the existence of unbroken, mutually commuting spacetime translations, and therefore applies to a large class of physical systems.

Applying the argument from the previous section, we conclude that only those
Goldstone fields lying in the common kernel of the matrices $\lambda_\mu^T$ can
possibly give rise to gapless excitations. Accordingly, using the decomposition
Eq.~\eqref{eq:GoldstoneDecompositon}, the fields $\tilde\varphi^\perp_\alpha \in K^\perp$ are generically gapped.\footnote{Not all these fields correspond to independent dynamical variables and may in some cases be eliminated from the outset by a suitable choice of parametrization (see footnote~\ref{fn:auxiliary}).} They can therefore be
integrated out in the low-energy regime $\omega,|\mathbf{k}|\ll \Delta$, where
$\Delta$ denotes the corresponding gap.\footnote{When multiple gapped modes are
present, $\Delta$ should be chosen as the smallest of the corresponding gaps.}
The resulting low-energy effective theory can then be formulated purely in terms
of the remaining gapless NG fields $\tilde\varphi^{\parallel}_a \in K$.

In particular, at low energies the generating functional can be expressed in the
path integral formalism as
\be
\mathcal Z_{\omega,|\mathbf{k}|\ll \Delta}
=
\int \mathcal D \tilde \varphi^{\parallel}_a \,
e^{-S_{\rm eff}[\tilde \varphi^{\parallel}_a]}\,,
\ee
where the effective action $S_{\rm eff}[\tilde \varphi^{\parallel}_a]$ is defined by formally integrating out the massive fields,
\be
e^{-S_{\rm eff}[\tilde \varphi^{\parallel}_a]}
=
\int \mathcal D\tilde\varphi^\perp_\alpha\,
e^{-S[\tilde \varphi^{\parallel}_a,\tilde\varphi^\perp_\alpha]}\,.
\ee
%%%%%%
%%%%%%%
%break%
%%%%%%
Therefore, the low-energy dynamics is governed by a set of homogeneous
Goldstone fields $\tilde \varphi^\parallel_a \in K$, which furnish a nonlinear realization of the spontaneously broken nonuniform symmetries after integrating out the gapped sector.

Moreover, note that the uniform generators
$Q^a$ form a closed Lie subalgebra. To see this, let $Q^a$ and $Q^b$ be uniform, the Jacobi identity for $(P_\mu,Q^a,Q^b)$ then implies
\be 
\{P_\mu,\{Q^a,Q^b\}\}=0 \,, 
\ee 
and hence $\{Q^a,Q^b\}$ is uniform. Thus, the subspace of broken generators commuting with spacetime
translations is closed under the Lie bracket so that the uniform generators form a Lie subalgebra,
\be\label{eq:closedAlgebra}
\{Q^a,Q^b\}=f^{ab}{}_{c}Q^c \,.
\ee

The low-energy sector is thus associated with spontaneously broken uniform
nonabelian symmetries forming the closed Lie algebra
Eq.~\eqref{eq:closedAlgebra}. Nonvanishing commutators of the generators
$Q^a$ can induce a presymplectic structure on the Goldstone field space,
leading to pairing and a further reduction in the number of gapless modes. The corresponding counting rule was established by Watanabe and Murayama in \cite{Watanabe:2012hr,Watanabe:2014fva}, building on earlier work by Brauner and Watanabe \cite{Watanabe:2011ec}.
\begin{theorem}[Watanabe--Murayama counting theorem]
\label{thm:WM}
Consider a translationally invariant system with spontaneously broken continuous uniform
symmetries generated by $\{Q^a\}$. For a broken ground state with charge
expectation values $\langle Q^a \rangle$, the total number of NG
modes is
\be
n_{\rm NG}
=
n_{\rm B}
-
\frac12\,\mathrm{rank}\,\rho,
\qquad
\rho^{ab} \equiv \langle \{ Q^a, Q^b \} \rangle \,,
\ee
where $n_{\rm B}$ denotes the number of broken generators.
\end{theorem}

We are now in a position to combine this result with the previous analysis. As shown above, the gapless sector is governed by the generators in the kernel $K^\star$, which form a closed Lie subalgebra Eq.~\eqref{eq:closedAlgebra}. The problem thus reduces to a system with spontaneously broken uniform symmetries, to which the Watanabe--Murayama counting applies.

We thus arrive at the general counting formula for gapless NG modes in systems with spontaneously broken nonuniform symmetries. For a system with nonuniform charges $\{Q^A\}$ obeying the algebra Eq.~\eqref{eq:nonabelianAlgebra}, and a ground state with expectation values $\langle Q^A \rangle$, the number of gapless NG modes is
\be \boxed{
n_{\rm NG}
=
\dim K^\star
-
\frac12\,\mathrm{rank}\,\rho\Big|_{K^\star}\,, }
\ee
where $\rho\big|_{K^\star}$ denotes the restriction of the matrix
\be
\rho^{AB} \equiv \langle \{Q^A,Q^B\}\rangle
\ee
to the kernel subspace $K^\star \subset V^\star$. This result follows from applying the Watanabe--Murayama counting to the restricted symmetry sector defined by $K^\star$.

Operationally, one proceeds in two steps. First, one restricts the algebra to the kernel $K^\star$, thereby isolating the gapless sector and reducing the problem to an effective uniform symmetry sector. Then, one applies the Watanabe--Murayama result to account for possible pairing among the remaining generators.

%%%%%%%%%%
%%%%%%%%%
%break%
%%%%%%%
%%%%%%%
%%%%%%%%%%
%%%%%%%%%
%break%
%%%%%%%
%%%%%%%
\subsection{Crystalline solids}\label{sec:solids}
In this section, we apply the counting formula to elastic crystals and demonstrate that it reproduces the correct number of acoustic phonons. We then construct an explicit effective field theory in which all spontaneously broken symmetries are nonlinearly realized on the phononic degrees of freedom, and show that it reduces to the standard theory of elasticity.

\paragraph{Symmetries of solids.} We consider a nonrelativistic many-body system whose microscopic dynamics is invariant under the Bargmann algebra
\bs
\{ L,B_i \} &= \epsilon_{ij} B_j \,, \qquad &
\{L,P_i\} &= \epsilon_{ij} P_j \,, \\
\{H,B_i\} &= -P_i \,, \qquad &
\{B_i,P_j\} &= \delta_{ij} M \,.
\es
We assume that the system crystallizes and develops long-range translational order. 
At long wavelengths the crystal is described by material coordinates $\phi^I \equiv \phi^I(x)$, which encode the dynamics of the lattice points. 
These coordinates possess internal shift symmetries generated by the crystal translation operators $T_I$,
\be \label{eq:material}
\{T_I,\phi^J\} = \delta_I{}^J \,,
\ee
reflecting the periodicity of the underlying lattice. We further assume that the lattice exhibits a discrete rotational symmetry $C_n$.

\paragraph{Symmetry breaking and mode counting.} There is a freedom in choosing the internal labeling of the material coordinates $\phi^I$. 
We fix this choice by selecting a reference configuration in which the material coordinates coincide with spatial coordinates in equilibrium. 
In other words, in the crystalline ground state we choose
\be\label{eq:equilibriumConfig}
\langle \phi^I \rangle_{\rm eq} = \sqrt{\rho_0}\delta^I{}_i x^i \,.
\ee
This configuration spontaneously breaks spacetime boosts $B_i$, rotations $L$ and translations $P_i$, as well as crystal translations $T_I$.
However, one can still define the diagonal combination
\be
\bar P_i \equiv P_i + \sqrt{\rho_0} \delta_i{}^I T_I\,,
\ee
which remains unbroken, providing a notion of unbroken translations. The symmetry breaking pattern is\footnote{Notice that the generator of spacetime translations $P_i$ is not independent from $T_I$ and $\bar P_i$ and hence including it would be redundant.}
\be 
\text{Broken: }  B_i, T_I, L \,, \quad \text{Unbroken: } \bar P_i, H, M\,.
\ee 
In this basis, the relevant nonuniform commutators are 
\be
\{ H,B_i \} = \delta_i{}^I T_I - \bar P_i\,, \quad \{ \bar P_i, L \} = \epsilon_{ij} \left(  \delta_j{}^I T_I - \bar P_j \right) \,.
\ee
Furthermore, in this case, the spontaneously broken charges are nonabelian
\be 
\{ L,B_i \} = \epsilon_{ij} B_j\,.
\ee 
However, as we will see, the algebra restricted to the kernel is abelian, and no pairing mechanism arises.

To identify the kernel, we organize the broken generators as
\be 
Q^A = \begin{pmatrix}
    B_i\\
    T_I \\
    L
\end{pmatrix}\,.
\ee 
In this basis, the matrices $\lambda_\mu$, defined through the action of the unbroken translations $\bar P_\mu$ on the broken generators as
\be
\{ \bar P_\mu, Q^A \} = (\lambda_\mu)^A{}_B Q^B \,,
\ee
take the form
\be 
\lambda_0 =
\begin{pmatrix}
0 & 0 & 0\\
\delta_j{}^{I} & 0 & 0\\
0 & 0 & 0
\end{pmatrix} \,, \quad \lambda_i =
\begin{pmatrix}
0 & 0 & 0 \\
0 & 0 & \epsilon_{ij}\delta_j{}^{J} \\
0 & 0 & 0
\end{pmatrix}\,.
\ee 
The common kernel of the matrices $\lambda_\mu$ is spanned by the internal translation generators $T_I$, i.e.
\be 
K^\star = \mathrm{span}\{ T_I \}\,.
\ee 
Since these generators form an abelian subalgebra, the counting formula derived above implies that each of them gives rise to an independent gapless mode.
We therefore conclude that the system supports $d=2$ gapless NG modes, identified with the acoustic phonons. 
This reproduces the expected spectrum of crystalline solids.

We now turn to the explicit construction of the effective field theory describing the dynamics of these modes. 

\paragraph{Effective field theory.}
The material coordinates $\phi^I(x)$ possess a shift symmetry
Eq.~\eqref{eq:material}, while transforming as spacetime scalar fields under
time translations, spatial translations, rotations and Galilean boosts,
\bs \label{eq:materialCoordinateTransformation}
\{ H, \phi^I(x)\} &= - \partial_0 \phi^I(x) \,, \\
\{ P_i, \phi^I(x)\} &= - \partial_i \phi^I(x) \,, \\
\{ L, \phi^I(x)\} &= - \epsilon^{ij} x_i \partial_j \phi^I(x)\,, \\
\{ B_i, \phi^I(x)\} &= - t \partial_i \phi^I(x)\,.
\es
%%%
Since the crystal symmetry acts by constant shifts Eq.~\eqref{eq:material}, invariant operators
must be constructed from derivatives of $\phi^I$.
It is straightforward to verify that the spatial gradients $\partial_i \phi^I$ transform
homogeneously under all spacetime symmetries and thus constitute the basic building blocks of the effective theory, with rotationally invariant combinations constructed using tensors preserved by the  point group.

The situation is more subtle for time derivatives. Acting with a Galilean boost
on $\partial_0\phi^I$ yields
\be
\{ B_i, \partial_0 \phi^I \}
= - t \partial_i \partial_0 \phi^I
- \partial_i \phi^I \,,
\ee
showing that $\partial_0\phi^I$ transforms inhomogeneously under Galilean boosts.

To construct boost--invariant structures involving time derivatives it is convenient to introduce
the velocity field $v^\mu=(1,v^i)$ by requiring that the material
coordinates are comoving,\footnote{For a detailed discussion of the comoving formalism,
see Ref.~\cite{Glodkowski:2025tnv}.}
\be \label{eq:comovingCondition}
v^\mu \partial_\mu \phi^I
= \partial_0 \phi^I + v^i \partial_i \phi^I = 0 \,,
\ee
which defines the velocity of the medium. 
The most general solution to this condition is
\be
v^\mu
= \frac{1}{\rho}
\,\epsilon_{IJ}\epsilon^{\mu\nu\rho}
\partial_\nu \phi^I \partial_\rho \phi^J \,,
\ee
where
\be \label{eq:densityLattice}
\rho
= \epsilon^{ij}\epsilon_{IJ}
\partial_i \phi^I \partial_j \phi^J \,,
\ee
represents the density of the lattice sites. 
This allows one to construct the conserved current
\be
j^\mu \equiv \rho v^\mu
= \epsilon_{IJ}\epsilon^{\mu\nu\rho}
\partial_\nu \phi^I \partial_\rho \phi^J \,,
\ee
which identically satisfies
\be
\partial_\mu j^\mu = 0 \,.
\ee
From Eq.~\eqref{eq:comovingCondition}, it follows that the velocity field transforms inhomogeneously under Galilean boosts
\be 
\{ B_i, v^j \} = -t\partial_i v^j + \delta_i{}^j\,.
\ee 
Then, it follows that the kinetic energy, $\tfrac{1}{2} \rho v^2$,
transforms under Galilean boosts into a total derivative 
\be 
\{ B_i, \tfrac{1}{2} \rho v^2 \} = - \frac{t}{2}   \partial_i \left( \rho v^2 \right) + \rho v^i = \partial_\mu \left( x^i j^\mu - \delta^\mu{}_i \tfrac{t}{2} \rho v^2 \right) \,.
\ee 
%%%%%%%%%%
%EFT%
%%%%%%%%%
Therefore, at the leading order in derivative expansion the effective theory takes the form 
\be  \label{eq:elasticLagrangian}
\mathcal L = \frac12 \rho v^2  - U_{\rm{el}}\,,
\ee 
where the first term describes the kinetic energy of the elastic
degrees of freedom, while the second term represents the elastic strain energy 
\be 
U_{\rm{el}} \equiv \frac{1}{2} C^{ij}{}_{IJ} \partial_i \phi^I \partial_j \phi^J
\ee 
parameterized with the elastic moduli tensor $C^{ij}{}_{IJ}$ whose explicit form is constrained by the point group of the
underlying lattice.

To make contact with conventional formulations of elasticity theory,
we parametrize fluctuations around the equilibrium configuration Eq.~\eqref{eq:equilibriumConfig} as
\be \label{eq:expansionPhi}
\phi^I
=
\sqrt{\rho_0}\,\delta^I{}_i\big(x^i + u^i\big),
\ee
where $u^i \equiv u^i(x)$ is the displacement field describing phonon
excitations of the lattice and $\rho_0$ denotes the equilibrium density of the system. These fields correspond to the Goldstone modes associated with the uniform generators $T_I$, which realize all spontaneously broken symmetries nonlinearly. The nonlinear transformation laws follow directly from Eq.~\eqref{eq:materialCoordinateTransformation} upon substituting Eq.~\eqref{eq:expansionPhi}.

Up to a total derivative, we can then express the elastic energy in terms of the displacement fields as\footnote{In passing from $\phi^I$ to the displacement fields $u^i$, an overall factor of $\rho_0$ has been absorbed into a redefinition of the elastic moduli tensor $C^{ij}{}_{kl}$.} 
\be \label{eq:elasticEnergy}
U_{\rm{el}} = C^{ij}{}_{kl} \partial_i u^k \partial_j u^l \,.
\ee 
Plugging Eq.~\eqref{eq:expansionPhi} into the comoving condition Eq.~\eqref{eq:comovingCondition} we arrive at 
\be 
\partial_0 u^i + v^i + v^j \partial_j u^i = 0 \,.
\ee 
The solution is 
\be \label{eq:velocityElastic}
v^i = - \partial_0 u^i + \mathcal{O}(u^2) \,,
\ee 
so that the lattice velocity coincides with the time derivative of the displacement field at leading order.
Moreover, using Eq.~\eqref{eq:densityLattice} we find that the density $\rho$ admits the following expansion 
\be \label{eq:elasticDensity}
\rho = \rho_0 \left( 1 + \partial_i u^i \right)  + \mathcal{O}(u^2)\,.
\ee 
%%%%%
Using Eqs.~\eqref{eq:elasticDensity} and \eqref{eq:velocityElastic},
and expanding the effective Lagrangian
Eq.~\eqref{eq:elasticLagrangian} around the equilibrium configuration
to quadratic order in the displacement fields, we obtain
\be
\mathcal L
= \frac12 \rho_0 (\partial_0 u^i)^2
- U_{\rm el} \,.
\ee
If the lattice has a sufficiently large point group (e.g.\ a triangular lattice with $C_6$ symmetry),
the elastic moduli take the isotropic Lam\'e form
\be
C^{ij}{}_{kl}
=\lambda\,\delta^{ij}\delta_{kl}
+\mu\big(\delta^i{}_k\delta^j{}_l+\delta^i{}_l\delta^j{}_k\big)\,.
\ee
In this case the quadratic elastic Lagrangian can be brought, up to total derivatives, to the standard form (see e.g. \cite{Landau:1986aog})
\be
\mathcal L
=\frac12\rho_0 (\partial_0 u^i)^2
-\frac{\lambda}{2}\,(\partial_i u^i)^2
-\mu\,u_{ij}u_{ij}\,,
\quad 
u_{ij}\equiv \frac12(\partial_i u_j+\partial_j u_i)\,.
\ee
Decomposing $u^i$ into longitudinal and transverse parts in momentum space,
$u^i=u_L^i+u_T^i$ with $k_i u_T^i=0$, one obtains two acoustic
branches 
\be
\omega_L^2=\frac{\lambda+2\mu}{\rho_0} k^2 \,,
\quad 
\omega_T^2=\frac{\mu}{\rho_0} k^2 \,,
\ee
corresponding to longitudinal and transverse phonons, respectively.
%%%%%%%%%%%%%

%%%%%%%%%%%%
\section{Superfluid vortex crystals}\label{sec:vortex}
\textit{ This chapter is based on Ref.~\cite{Glodkowski:2025krf}, although the presentation differs in several respects, including the identification of the microscopic symmetries Sec.~\ref{sec:microscopicSymmetries}.} \newline 

\noindent Having developed the general framework for nonlinear realizations of nonuniform symmetries, we now turn to a concrete physical realization in the context of rotating superfluids. Rotating Bose--Einstein condensates of ultracold atoms provide a particularly nontrivial example of a system in which several nonuniform symmetries are spontaneously broken, leading to the formation of an emergent vortex crystal. The resulting vortex crystal supports a single low-energy collective excitation, the transverse Tkachenko mode, whose properties are dictated by symmetry considerations. The goal of this chapter is to construct an effective field theory for the long-wavelength oscillations in vortex crystals based purely on symmetry principles, with particular emphasis on the subtleties arising from spontaneously broken nonuniform symmetries.

This chapter thus serves both as a detailed application of the general formalism developed earlier in the thesis, particularly the application and explicit verification of the counting formula in a highly nontrivial example, and as a complementary perspective on the effective theory of vortex crystals.
%%%%%%%%%%%%

%%%%%%%%%%%%
\subsection{Emergence of vortex lattices in rotating superfluids}
In this section, we briefly review the formation of vortex lattices in rotating superfluids and discuss their experimental realization in ultracold atomic condensates.

\paragraph{Rotating bucket experiment.} The physics of rotating superfluids is most clearly understood by recalling the seminal 
``rotating bucket'' experiment. In this setup, a container 
filled with liquid helium-4 is set into rotation with angular velocity $\boldsymbol{\Omega}$.

\begin{figure}[h!]
    \centering
    \includegraphics[width=0.2\linewidth]{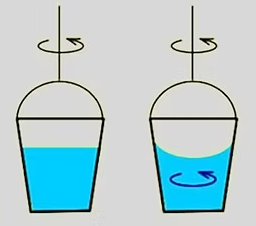}
    \caption{Schematic illustration of the rotating bucket experiment. In the normal phase, fluid equilibrates into rigid-body rotation with nonzero vorticity.}
    \label{fig:bucket}
\end{figure}

Due to viscous coupling to the walls, which transmits angular momentum throughout the fluid, a normal fluid eventually equilibrates and undergoes rigid-body rotation with vorticity determined by the rotation frequency (see Fig.~\ref{fig:bucket})
\be \label{eq:rigidRotation}
\mathbf{v} = \boldsymbol{\Omega} \times \mathbf{r}\,, \quad 
\nabla \times \mathbf{v} = 2\boldsymbol{\Omega} \,.
\ee
The system is then gradually cooled below the superfluid transition temperature 
($T \simeq 2.14\,\mathrm{K}$), at which point it enters the superfluid phase. 
In this phase, the fluid is described by a macroscopic wavefunction\footnote{At finite temperature, the system contains both superfluid and normal components \cite{Landau_Theory,Tisza}.}
\be 
\psi(\mathbf{r}) = \sqrt{n_0 + \delta n(\mathbf{r})}\, e^{i\theta(\mathbf{r})}\,,
\ee
whose phase $\theta(\mathbf{r})$ encodes the superfluid velocity,
\be 
\mathbf{v}_s = \frac{\hbar}{m} \nabla \theta\,.
\ee
%%%%%%%%%%%
%%%%%%%%
Therefore, the superfluid flow is locally irrotational,
\be
\nabla \times \mathbf{v}_s = 0\,,
\ee
and cannot sustain rigid-body rotation.

Instead, superfluids accommodate angular momentum through the formation of topological defects, 
known as \textit{quantum vortices}, corresponding to singularities of the phase. At the location of a vortex, the phase winds by an integer multiple of $2\pi$,\footnote{This quantization follows from the nontrivial topology of the order parameter manifold, $\pi_1(\mathrm U(1)) = \mathbb{Z}$.} leading to quantized circulation with vorticity localized at the vortex cores,
\be 
\nabla \times \nabla \theta = 2\pi \delta^{(2)}(\mathbf{r}-\mathbf{r}_0)\,.
\ee

\begin{figure}[h!]
    \centering
    \includegraphics[width=0.3\linewidth]{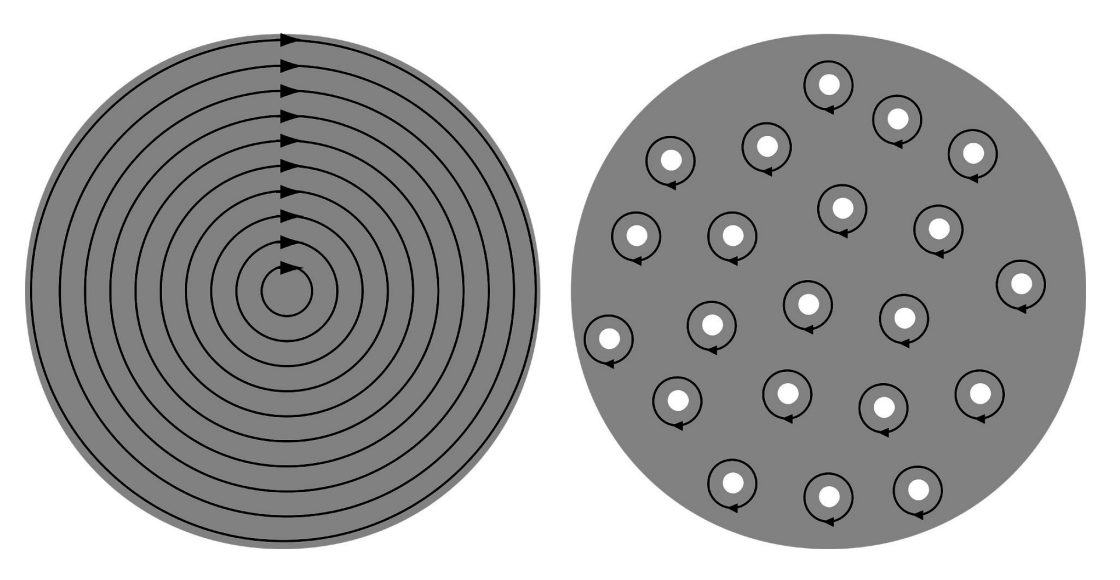}
    \caption{A normal fluid realizes angular momentum via rigid body rotation (left) whereas the superfluid state rotates by forming a distribution of quantum vortices, which mimic the rigid body rotation when viewed on large scales (right).}
    \label{fig:superfluidRotation}
\end{figure}

At finite rotation frequency $\Omega$, many such vortices are nucleated. On length scales 
large compared to the inter-vortex spacing, the coarse-grained vorticity reproduces the rigid-body result Eq.~\eqref{eq:rigidRotation} (see Fig.~\ref{fig:superfluidRotation}), leading to the Feynman relation \cite{FEYNMAN195517}
\be
n_v = \frac{2m\Omega}{h}\,,
\ee
which fixes the density of vortices in terms of the rotation rate. 

At sufficiently large $\Omega$, interactions between vortices lead to the formation of a 
regular triangular lattice, as demonstrated by Tkachenko \cite{Tkachenko:1965}. In liquid helium, however, these vortex structures are difficult to observe directly, owing to their microscopic core size and the absence of direct imaging probes.
%%%%%%%%%%%%
%%%%%%%%%%

\paragraph{Rotating condensates of ultracold atoms.} 
The experimental realization of Bose--Einstein condensation in dilute atomic gases \cite{doi:10.1126/science.269.5221.198,PhysRevLett.75.3969} has provided a clean and highly tunable platform for studying rotating superfluids. In these systems, atoms are confined in an optical trap that generates an approximately harmonic potential.\footnote{In practice, weak anharmonic corrections are often present, and different trapping geometries can be realized.} 

The atomic cloud is first cooled to temperatures slightly above the condensation temperature, of order $100\,\mathrm{nK}$. Angular momentum is then imparted by introducing a rotating (time-dependent) anisotropy in the trapping potential. As the system is further cooled below the transition temperature, it undergoes Bose--Einstein condensation, and quantized vortices are nucleated. After removing the trap anisotropy, the resulting vortex lattice persists in the harmonic potential and can be observed after time-of-flight expansion.

\begin{figure}[h!]
    \centering
    \includegraphics[width=0.7\linewidth]{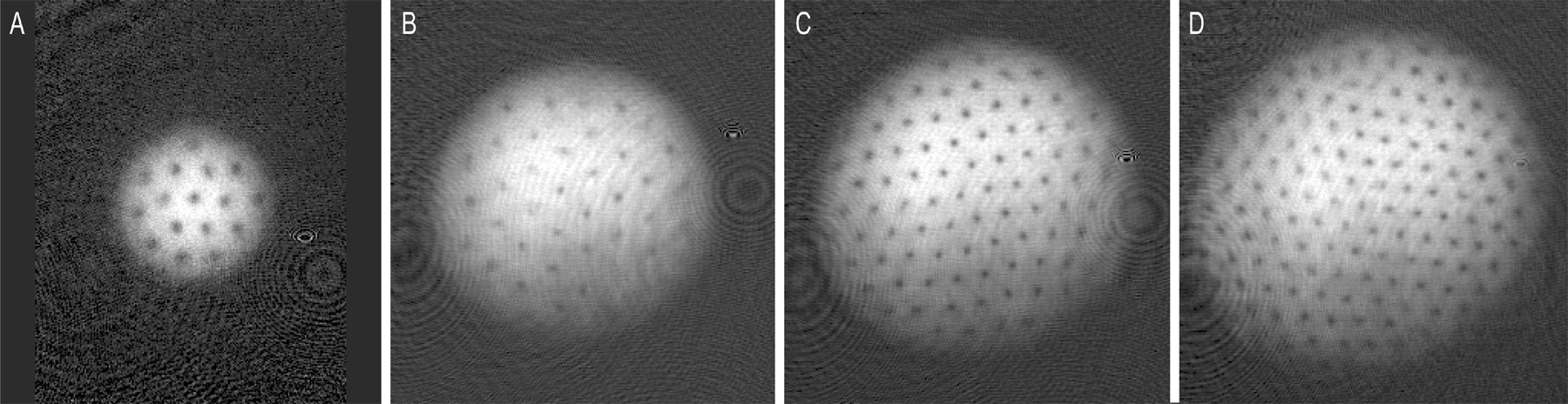}
\caption{Triangular vortex lattice observed in a rotating Bose--Einstein condensate of ultracold rubidium-87 atoms. The vortices, visible as density depletions, carry quantized circulation. Figure from \cite{doi:10.1126/science.1060182}.}
    \label{fig:quantumVortexLattice}
\end{figure}

In contrast to liquid helium, where vortex structures are difficult to observe directly, ultracold atomic systems allow for real-space imaging of vortex configurations. This provides a striking macroscopic manifestation of quantum mechanics, arising from the coherent occupation of a single quantum state. Numerous experiments have directly observed the formation of highly regular triangular vortex lattices \cite{PhysRevLett.84.806,PhysRevLett.85.2223,doi:10.1126/science.1060182,PhysRevLett.87.210403,PhysRevLett.89.100403,PhysRevLett.90.170405,PhysRevLett.88.010405,PhysRevLett.91.100402,PhysRevLett.92.040404}, see Fig.~\ref{fig:quantumVortexLattice}.\footnote{More recently, rotating condensates in the lowest Landau level (LLL) regime have been realized \cite{doi:10.1126/science.aba7202, mukherjee2022crystallization}, allowing for the study of strongly correlated vortex states.} 

In the low-energy limit, such phases behave as an elastic medium and support collective excitations associated with transverse oscillations of the lattice, known as Tkachenko waves \cite{Tkachenko:1965, sonin2014tkachenko}. In the following, we construct the effective field theory for these modes by identifying the underlying nonuniform symmetry structure and formulating its nonlinear realization. Within this framework, the gapless Tkachenko mode arises upon integrating out massive modes associated with the nonuniform symmetry generators, thereby modifying its polarization and dispersion relation.

%%%%%%%%%%%
\subsection{Global symmetries of vortex crystals}

To construct an effective field theory, we begin by identifying the global symmetries of the system and the associated pattern of symmetry breaking. We distinguish between two sets of symmetries, the microscopic symmetries of the underlying model and the emergent symmetries associated with the vortex lattice. An equilibrium embedding of the vortex lattice in physical spacetime leads to the spontaneous breaking of both symmetry groups down to a diagonal subgroup corresponding to unbroken translations. After elucidating the symmetry breaking pattern, we apply the nonuniform counting formula, which implies the existence of a single massless excitation.

\subsubsection{Microscopic model and its global symmetries}\label{sec:microscopicSymmetries} 
The microscopic dynamics of interacting bosons is governed by the Gross-Pitaevskii field theory in a two-dimensional harmonic trap,
\be\label{eq:actionHarmonic}
\mathcal L =  \frac{i}{2}  \Phi^\dag \overset{\leftrightarrow}{\partial_0} \Phi - \frac{1}{2m}\partial_i \Phi \partial_i \Phi^\dag  -\frac{m}{2} \omega^2 x^2 |\Phi|^2-V(|\Phi|) \,,
\ee 
where 
\be 
 \Phi^\dag \overset{\leftrightarrow}{\partial_0} \Phi =  \Phi^\dag \partial_0 \Phi - \Phi\partial_0 \Phi^\dag \,. 
\ee 
%%%%%%%%%%%%%%%
Furthermore, we assume that the system is prepared at finite chemical potentials $\mu$ and $\Omega$, conjugate to particle number and angular momentum, respectively. The corresponding equilibrium state is described by the density matrix
\be 
\rho = e^{-\beta (H - \mu N - \Omega L)}\,, \quad \mathcal Z = \text{Tr} \, e^{-\beta (H - \mu N - \Omega L)}\,.
\ee  
It is then convenient to analyze the system in the co-rotating frame, where the equilibrium state becomes time independent. This is achieved by the change of coordinates, 
\be 
x_i^\prime = R(\Omega t)_{ij} x_j\,,  \quad R(\Omega t)_{ij} = \delta_{ij}  \cos{(\Omega t)} - \epsilon_{ij} \sin{(\Omega t)} \,,
\ee 
which is equivalent to sending 
\be 
\partial_t \rightarrow \partial_t - \Omega \epsilon_{ij} x_j \partial_i 
\ee 
in the Lagrangian Eq.~\eqref{eq:actionHarmonic}. In the co-rotating frame the system becomes stationary, but the Lagrangian acquires additional terms associated with non-inertial effects. Explicitly, the Lagrangian density takes the form
\be\label{eq:action1.5}
\mathcal L
= \frac{i}{2}\,\Phi^\dag \overset{\leftrightarrow}{\partial_0} \Phi
- \frac{1}{2m}\left|
\partial_i \Phi - i m \Omega \epsilon_{ij} x_j \Phi
\right|^2
- \frac{m}{2}\big(\omega^2-\Omega^2\big) x^2 |\Phi|^2
- V(|\Phi|) \, ,
\ee
where the minimal coupling to the effective vector potential encodes the Coriolis force, while the shift of the harmonic potential reflects the repulsive centrifugal force. We assume that the system rotates precisely with the frequecny of the trap, $\Omega = \omega$, such that the repulsive centrifugal force exactly cancel the harmonic trap. The resulting Lagrangian  
%%%%%%%%%
\be\label{eq:action2} \boxed{
\mathcal L =  \frac{i}{2}  \Phi^\dag \overset{\leftrightarrow}{\partial_0} \Phi - \frac{1}{2m} \left |\partial_i \Phi-i m \Omega \epsilon_{ij} x_j \Phi   \right |^2 -V(|\Phi|) \,,}
\ee 
describes nonrelativistic bosons placed in an effective homogeneous magnetic field 
\be 
B_{\rm{eff}} = -2m\Omega \,.
\ee 
To identify the global symmetries of the system notice that the Lagrangian Eq.~\eqref{eq:action2} is invariant under the following transformations 
\bs 
(t, \boldsymbol x) &\mapsto (t + a, \boldsymbol x) \,, \quad & \Phi(t,\boldsymbol x) &\mapsto \Phi(t+a,\boldsymbol x)\,, \\ 
(t, \boldsymbol x) &\mapsto (t, \boldsymbol x) \,, \quad  & \Phi(t,\boldsymbol x)  & \mapsto  e^{i \alpha} \, \Phi(t,\boldsymbol x)  \,, \\ 
(t, \boldsymbol x) &\mapsto (t,  R(\theta) \, \boldsymbol x ) \,, \quad  & \Phi(t,\boldsymbol x)  & \mapsto  \Phi\!\left(t, R^{-1}(\theta) \boldsymbol x\right)  \,, \\ 
(t, \boldsymbol x) &\mapsto  (t, \boldsymbol x + \boldsymbol c) \,, \quad  & \Phi(t,\boldsymbol x)  & \mapsto    \exp \Big[-i m \Omega \, \left( \boldsymbol c \times \boldsymbol x \right) \Big] \,  \Phi\!\left(t,\boldsymbol x - \boldsymbol c\right) \,, \\ 
(t, \boldsymbol x) &\mapsto  \left(t, \boldsymbol x +  M(\Omega t) \, \boldsymbol b \right )\,, \quad  & \Phi(t,\boldsymbol x)  & \mapsto    \exp \Big[- i \frac{m}{2}\,\boldsymbol b\cdot\big(\mathbb 1+R(-2\Omega t)\big)\boldsymbol x\Big] \, \Phi\!\left(t,\boldsymbol x -  M(\Omega t) \, \boldsymbol b\right) \,, \\ 
\es 
where 
\be 
 R(\theta) = \begin{pmatrix}
\cos\theta & -\sin\theta\\
\sin\theta & \cos\theta
\end{pmatrix}\,, \quad  M(\Omega t)
=\frac{1}{2\Omega}
\begin{pmatrix}
\sin(2\Omega t) & \cos(2\Omega t) - 1\\
1 - \cos(2\Omega t)  & \sin(2\Omega t)
\end{pmatrix}\,.
\ee
%%%%%%%%%%%
%break%
%%%%%%%%%%%%
The Noether charges associated with the above global symmetries, which generate the corresponding transformations by exponentiation, are
\bs 
Q &= \int d^2 x \, \left |\Phi \right|^2 \,, \\
H &= \int d^2x \, \left[ \frac{1}{2m} \left |\partial_i \Phi - i m \Omega \epsilon_{ij} x_j \Phi       \right |^2  + V(|\Phi|) \right] \,, \\
L &= \int d^2 x \, \left[ \epsilon_{ij} x_i \frac{i}{2}  \Phi^\dag \overset{\leftrightarrow}{\partial_j} \Phi  - m \Omega \, x^2 \left| \Phi \right|^2 \right] \,, \\ 
P_i &=     \int d^2 x \, \left[ \frac{i}{2}  \Phi^\dag \overset{\leftrightarrow}{\partial_i} \Phi  - m \Omega \epsilon_{ij} x_j \left| \Phi \right|^2   \right] \,, \\    
B_i &=     \int d^2 x \, \left[ M_{ji}(\Omega t) \left( \frac{i}{2}  \Phi^\dag \overset{\leftrightarrow}{\partial_j} \Phi \right)  - \frac{m}{2} \left( \delta_{ij} + R_{ij}(-2\Omega t)  \right) x_j \left| \Phi \right|^2 \right] \,.
\es 
Imposing the canonical commutation relations, 
\be \label{eq:canonicalComplexFields}
[ \Phi(\boldsymbol x), \Phi^\dag(\boldsymbol y)] = \delta(\boldsymbol x - \boldsymbol y) \,,
\ee 
one verifies by direct (though somewhat tedious) computation that the generators obey the \textbf{magnetic Bargmann algebra}\footnote{In writing the algebra we have used anti-Hermitian basis for the generators.}
\begin{equation}\label{eq:magneticBargmann}
\boxed{
\begin{array}{rcl@{\quad}rcl}
[L, B_i] & = &  \epsilon_{ij} B_j\,, & [L, P_i] & = &  \epsilon_{ij} P_j\,, \\[4pt]
[H, B_i] & = & - P_i - 2  \Omega \epsilon_{ij} B_j\,, & [P_i, P_j] & = & - 2m\Omega \epsilon_{ij} Q\,, \\[4pt]
[B_i, P_j] & = &  \delta_{ij} m Q\,. & & &
\end{array}
}
\end{equation}
This algebra describes the conserved generators of Galilean fields propagating in a uniform magnetic field $B_{\rm{eff}} = - 2 m \Omega$.

\subsubsection{Emergent symmetries of the vortex lattice}
Having established the microscopic symmetry structure, we now focus on the \textbf{emergent symmetries} of the vortex lattice. We remain agnostic about the microscopic mechanism driving lattice formation and instead assume that the system crystallizes, using this as a starting point to analyze the symmetry breaking pattern and the resulting low-energy effective description. 
\begin{figure}[h]
    \centering
\includegraphics[scale=0.45]{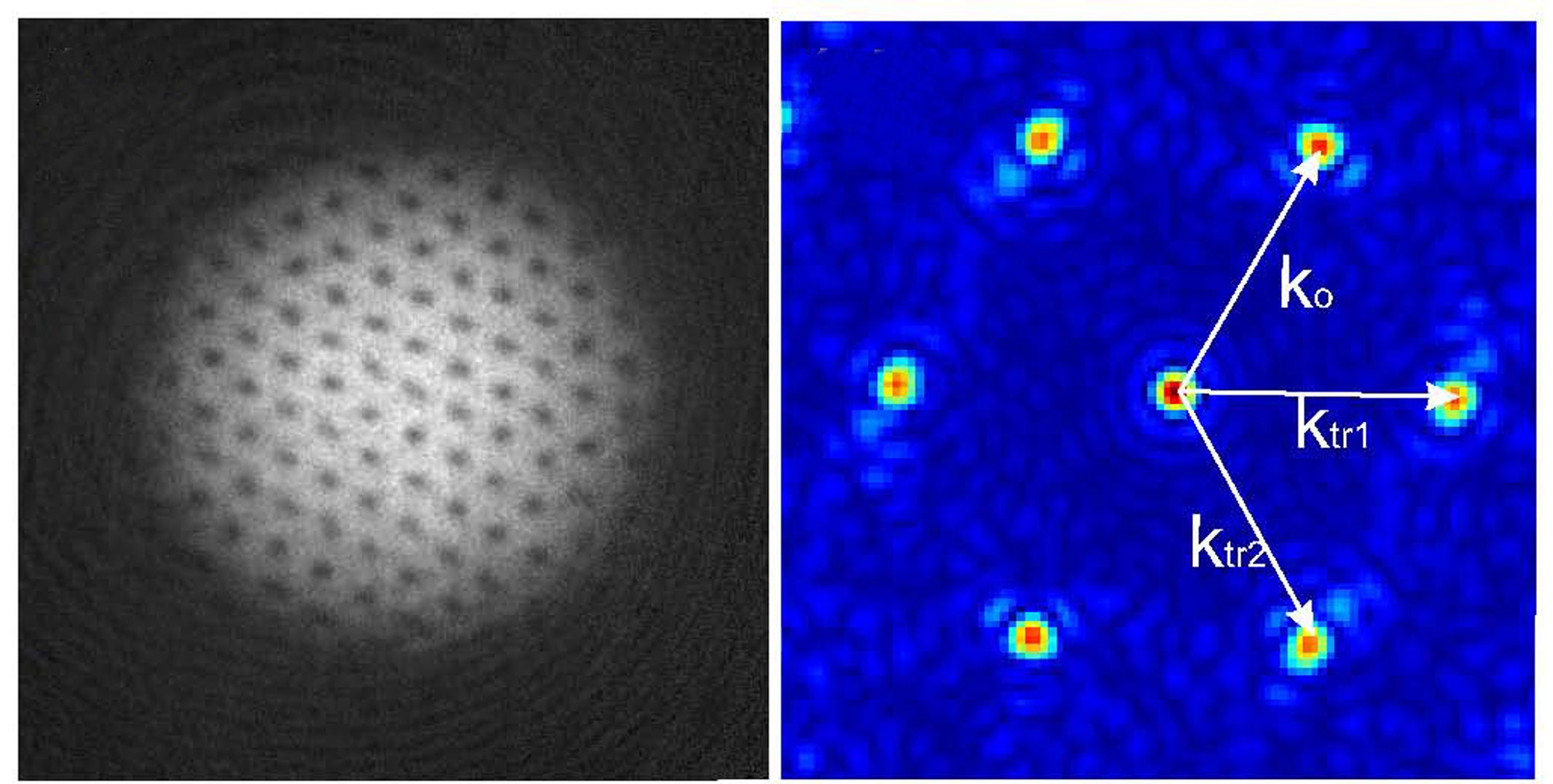}
    \caption{Quantum vortex lattice in a rotating Bose--Einstein condensate, illustrating the emergence of a crystalline structure with discrete translational and rotational symmetries. This configuration is modeled in the effective description using vortex coordinates and the superfluid phase, which together define the associated vortex space. Image adapted from the National Institute of Standards and Technology (NIST), JILA (Cornell group), public domain.}
    \label{fig:vortexLattice}
\end{figure}
Guided by the experimental observations of the crystalline order shown in Fig. \ref{fig:vortexLattice}, we now construct a coarse-grained description of the vortex lattice. To this aim, we introduce an abstract manifold, which we dub the \emph{vortex space}, equipped with coordinates $\phi^I (I=1,2)$ and $\psi$, corresponding to vortex (comoving) coordinates and $\mathrm U(1)$ phase of the superfluid, respectively. The vortex coordinates track the positions of the vortices, while the $\mathrm U(1)$ phase encodes the independent superfluid dynamics. Together, these two fields fully capture the long-wavelength dynamics.

Since we are describing a crystalline arrangement, the system is invariant under a discrete subgroup of spatial translations, reflecting the periodic structure of the underlying vortex lattice. At long wavelengths, however, this discrete symmetry is effectively promoted to a continuous one, reflecting the insensitivity of the coarse-grained theory to microscopic lattice spacings. The corresponding transformations are generated by translation operators in vortex space,
\be\label{eq:shift}
T_I : \phi^I \mapsto \phi^I + \xi^I\,,
\ee 
where $\xi^I$ are constant parameters. Furthermore, the lattice is triangular and therefore exhibits a discrete $C_6$ symmetry. This discrete rotational symmetry acts on the vortex space as a linear transformation of the comoving coordinates,
\be\label{eq:rotation}
C_6 : \phi^I \mapsto D^I{}_J \phi^J\,,
\ee 
where $D^I_J \in SO(2)$ represents a rotation in  a two-dimensional plane by a multiple of $\frac{\pi}{3}$. On the other hand, the $\psi$ field nonlinearly realizes the associated $\mathrm U(1)$ symmetry through a shift,
\be 
Q : \psi \mapsto \psi + \alpha \,. 
\ee 
Finally, we must account for the fact that vortices propagate in an effective magnetic background generated by the ambient superfluid. From the dual vortex perspective, the uniform bosonic density plays the role of a constant magnetic flux \cite{PhysRevB.39.2756}, so that the translations in vortex space no longer commute. Instead, the translation algebra closes into a central extension,
\be \label{eq:magneticTranslation}
[T_I, T_J] =   2 m \Omega \epsilon_{IJ} N \,,
\ee 
reflecting the noncommutative nature of the vortices.\footnote{This noncommutativity arises naturally in the point-vortex description, where the symplectic form treats the vortex coordinates as conjugate variables \cite{10.1063/1.2425103}.}

\subsubsection{Symmetry breaking pattern}
In this section, we construct an embedding of the vortex space into physical spacetime that represents the equilibrium vortex arrangement, and analyze the resulting pattern of spontaneous symmetry breaking. We proceed by fixing the equilibrium values of the vortex coordinates $\phi^I$ by identifying them with the vortex positions in physical spacetime at a chosen reference time. Assuming a stationary condensate of the form $\langle \Phi \rangle = e^{i\mu t} \Phi_0$, the equilibrium background is then chosen as
\be \label{eq:groundStateVortex}
\langle \phi^I \rangle_{\rm{eq}}   = \delta^I_i x^i \,, \quad   \langle \psi \rangle_{\rm{eq}}   = \mu t\,,
\ee 
where $\mu$ is the chemical potential.

The equilibrium configuration Eq.~\eqref{eq:groundStateVortex} induces spontaneous breaking of the continuous global symmetries Eq.~\eqref{eq:magneticBargmann} and Eq.~\eqref{eq:magneticTranslation}. To see this, it suffices to verify that Eq.~\eqref{eq:groundStateVortex} is not individually preserved by the action of the symmetries. For example, the action of the magnetic translations in spacetime $P_i$ and vortex translations $T_I$ is 
\bs 
 [ P_i, \phi^I] &= -  \partial_i \phi^I\,, \quad    &[  P_i, \psi] &= -\partial_i \psi - m\Omega \epsilon_{ij}  x^j \,, \\ 
  [ T_I, \phi^J] &= \delta^J_I \,, \quad    &[ T_I, \psi] &=  m\Omega \epsilon_{IJ}  \phi^J \,.
\es 
Evaluating these transformations on the background \eqref{eq:groundStateVortex} shows that it is not invariant under $P_i$ or $T_I$. Hence, both symmetries are spontaneously broken. Similarly, one can show that all generators are broken individually. 

However, one can construct specific diagonal combinations of spacetime and internal symmetries that remain unbroken. In particular, define 
\be 
\bar P_i = P_i + \delta^I_i T_I\,, \quad \bar H = H + \mu Q \,.
\ee 
Then, it is straightforward to verify that their action vanishes on the equilibrium configuration Eq.~\eqref{eq:groundStateVortex}, i.e.,  
\bs 
\langle [\bar P_i, \phi^I] \rangle_{\rm{eq}} = 0 \,, \quad  \langle [\bar P_i, \psi] \rangle_{\rm{eq}}  = 0\,, \\ 
\langle [\bar H, \phi^I] \rangle_{\rm{eq}} = 0 \,, \quad  \langle [\bar H, \psi] \rangle_{\rm{eq}}  = 0\,.
\es 
Therefore, we identify $\bar P_\mu = (\bar H, \bar P_i)$ as the generators of unbroken spacetime translations.

In addition to the continuous diagonal translations, the equilibrium configuration is invariant under a discrete subgroup of spatial rotations, generated by a simultaneous rotation in physical space and in the internal vortex space. This diagonal discrete symmetry acts identically on both sets of indices. As a result, spatial and internal coordinates transform in the same representation, and we may henceforth identify their indices. We therefore use a single label $a=1,2$ for both sectors, raised and lowered with $\delta_{ab}$, and implicitly summed over.
%%%%%
%break%
%%%%%%%%
With this in mind, the pattern of symmetry breaking can be expressed as
\be\begin{split}\label{eq:breakingPattern}
\text{Unbroken:}& \quad \bar H = H + \mu Q\,, \quad  \bar P_a = P_a + T_a\,, \\
\text{Broken:}& \quad T_a\,,  B _a\,, L\,, Q \,.
\end{split}\ee 
%%%%
Using Eq.~\eqref{eq:magneticBargmann} and Eq.~\eqref{eq:magneticAlgebra} it is straightforward to verify that the generators obey the following algebra
%%%
\be\begin{aligned}\label{eq:cosetAlgebra}
[T_a, T_b] &= 2m\Omega  \epsilon_{ab} Q\,, &\quad  [\bar P_a, T_b] &= 2m\Omega  \epsilon_{ab} Q \,, \\
[B_a, \bar P_b] &= m \delta_{ab} Q\,, &\quad  [\bar H, B_a] &= - \bar P_a +  T_a - 2\Omega \epsilon_{ab} B_b \,, \\
[L, \bar P_a] &=  \epsilon_{ab} (\bar P_b - T_b)\,, &\quad [L,  B_a] &=  \epsilon_{ab} B_b\,.
\end{aligned}
\ee 
We observe that among the broken generators ${T_a, B_a, L, Q}$ only the $\mathrm U(1)$ charge $Q$ is uniform and spans the common kernel of the translation action. Thus, applying the nonuniform counting formula of Sec.~\ref{sec:nonuniformGoldstone}, we conclude that the system supports a single gapless excitation despite the spontaneous breaking of six generators.
%%%%%%
%%%%%%%
%break%
%%%%%%%%

\subsection{Coset construction}
We now construct the low-energy effective theory dictated by the symmetry breaking pattern Eq.~\eqref{eq:breakingPattern}. Since there is only one gapless Goldstone mode, in the hydrodynamic limit all broken symmetries are necessarily realized nonlinearly on a single field. The Goldstone fields associated with nonuniform symmetries do not correspond to independent gapless excitations but instead impose kinematic constraints on the remaining mode.

To determine the invariant building blocks, we apply the coset construction, reviewed in Appendix~\ref{sec:coset}, impose the inverse Higgs constraints to eliminate redundant fields, and systematically derive the effective action governing the long-wavelength dynamics of the vortex lattice.

\subsubsection{Maurer-Cartan form}
We begin with by identifying the vacuum manifold with the coset space,
\be 
U = e^{ t \bar H} e^{ x^a \bar P_a} e^{ u^a  T_a} e^{ \varphi  N} e^{ v^a B_a} e^{ \gamma L}\,, 
\ee 
The Goldstone fields $\{u^a, v^a, \varphi, \gamma\}$ parametrize the coset and are associated with the broken vortex translations, magnetic boosts, global $\mathrm U(1)$ symmetry, and spatial rotations, respectively. In dynamical situations they are promoted to local fields defined over spacetime $u^a(x), v^a(x), \varphi(x), \gamma(x)$, which furnish a nonlinear realization of the broken symmetries. The corresponding transformation rules follow from \eqref{eq:transformation}.

Using the algebra Eq.~\eqref{eq:cosetAlgebra}, as well as the Baker--Campbell--Hausdorff formula, one can verify the following algebraic relations, 
\be\begin{split}\label{eq:identitiesCoset}
    e^{-v^a B_a} \bar  H 
    e^{v^a B_a} &= \bar H -  v^a \bar P_a + v^a T_a -  2\Omega \epsilon_{ab} v^a B_b + \frac{mv^2}{2}  N \,, \\
       e^{-\gamma L} e^{-v^a B_a} \bar  H 
    e^{v^a B_a}  e^{\gamma L}&= \bar H -  v^a R_{ab}(\gamma)\big(\bar P_b - T_b\big) -  2\Omega \epsilon_{ab} v^a R_{bc}(\gamma)B_c + \frac{mv^2}{2}  N \,, \\
      e^{-u^a T_a} \bar  P_a
    e^{u^a T_a} &= \bar P_a + 2m\Omega \epsilon_{ab} u^b N 
    \,, \\
     e^{-v^a B_a} \bar  P_a
    e^{v^a B_a} &= \bar P_a - m v^a N 
    \,, \\
      e^{-\gamma L} B_a e^{\gamma L} &= R_{ab}(\gamma) B_b\,, \\ 
          e^{-\gamma L} \bar P_a e^{\gamma L} &= R_{ab}(\gamma) \bar P_b + \big(\delta_{ab}-R_{ab}(\gamma)\big)T_b\,, \\
    e^{-u^a T_a} d e^{u^a T_a} &= du^a T_a + m\Omega \epsilon_{ab} du^a u^b N\,, \\ 
      e^{-\gamma L} B_a e^{\gamma L} &= R_{ab}(\gamma) B_b\,.
\end{split}
\ee 
With these identities, it is possible to verify that the Maurer--Cartan form 
\be
\boldsymbol{\omega} \equiv U^{-1} d U\,,
\ee 
can be expanded in the basis of generators
\be 
\boldsymbol{\omega} = \omega_H \bar H + \omega^a_P \bar P_a + \omega^a_T T_a +  \omega^a_B B_a + \omega_N N + \omega_L L\,,
\ee
with the components given by
\be \begin{split}\label{eq:components}
\omega_H &= dt \,, \\ 
\omega^a_P &= \big(dx^b - v^b dt \big) R_{ba}(\gamma) \,, \\ 
\omega^a_T &= d\phi^a - \omega^a_P  \,, \\ 
\omega^a_B &= \big( dv^b + 2\Omega  \epsilon_{bc} v^c dt \big) R_{ba}(\gamma)\,, \\ 
\omega_N &= d \varphi +2m\Omega \epsilon_{ab} u^b dx^a +  \frac{mv^2}{2} dt-  m v^a dx^a  + m\Omega \epsilon_{ab} du^a u^b  \,, \\
\omega_L &=d\gamma\,.
\end{split}
\ee 
%%%%%%%
%%%
 Employing the general decomoposition of the Maurer-Cartan form, 
\be
\boldsymbol{\omega} = e^\alpha_\mu dx^\mu \big( \bar P_\alpha + D_\alpha u^a T_a + D_\alpha v^a B_a + D_\alpha \varphi N + D_\alpha \gamma L\big)
\ee 
where $\bar P_\alpha = (\bar H, \bar P_a)$, we can identify the coset vielbein $e^\alpha_\mu$ and the covariant derivatives $D_\alpha u^a$ etc., which constitute the invariant derivative structures that we will use to construct an effective action. 

From $\omega_H$ and $\omega^a_P$ we can extract the coset vielbein and its inverse 
\be 
e^0_\mu = \delta^0_\mu \,, \quad e^a_\mu = (\delta^b_\mu - \delta^0_\mu v^b \big) R_{ba}(\gamma) \,, \quad E^\mu_0 = \delta^\mu_0+\delta^\mu_a v^a\,, \quad E^\mu_a = \delta^\mu_b R_{ba}(\gamma)\,.
\ee 
%%%%%%%%
Then, using \eqref{eq:components} we can identify the explicit expresions for the covariant derivatives
\be
\begin{split}\label{eq:covariantDerivatives}
    D_0 u^a &= E^\mu_0 \big( \partial_\mu \phi^a - e^a_\mu \big) = \big(\delta^\mu_0 + \delta^\mu_b v^b \big) \partial_\mu \phi^a = \partial_0 u^a + v^b \partial_b u^a + v^a\,, \\
        D_a u^b &= E^\mu_a \big( \partial_\mu \phi^b - e^b_\mu \big) =  R_{ca}(\gamma)\partial_c  \phi^b-\delta^b_a\,, \\
        D_0 v^a &= E^\mu_0 \big( \partial_\mu v^b + 2\Omega \epsilon_{bc} v^c \delta^0_\mu \big)R_{ba} = \big(\partial_0 v^b + v^c \partial_c v^b+ 2\Omega \epsilon_{bc} v^c\big)R_{ba}(\gamma)  \,, \\
        D_a v^b &= E^\mu_a \big( \partial_\mu v^b + 2\Omega \epsilon_{bc} v^c \delta^0_\mu \big)R_{ba} = \partial_d v^c R_{da}(\gamma) R_{cb}(\gamma)  \,, \\ 
        D_0 \varphi &= E^\mu_0 \big( \partial_\mu \varphi +2m\Omega \epsilon_{ab} u^b \delta^a_\mu +  \frac{mv^2}{2} \delta^0_\mu- m  v^a \delta^a_\mu  + m\Omega \epsilon_{ab} \partial_\mu u^a u^b \big) \,,\\
        &=  \partial_0 \varphi +  v^a\partial_a \varphi - \frac{mv^2}{2}  +  m\Omega \epsilon_{ab} v^a u^b + m\Omega \epsilon_{ab} D_0 u^a u^b \,, \\
         D_a \varphi &= E^\mu_a \big( \partial_\mu \varphi +2m\Omega \epsilon_{ab} u^b \delta^a_\mu +  \frac{mv^2}{2} \delta^0_\mu- m  v^a \delta^a_\mu  + m\Omega \epsilon_{ab} \partial_\mu u^a u^b \big) \\
         &= \big(\partial_b \varphi +2m\Omega \epsilon_{bc} u^c  - m  v^b   + m\Omega \epsilon_{cd} \partial_b u^c u^d \big) R_{ba}(\gamma) \,, \\
         D_0 \gamma &= E^\mu_0 \big( \partial_\mu \gamma \big) = \partial_0 \gamma + v^b \partial_b \gamma \,, \\
         D_a \gamma &= E^\mu_a \big( \partial_\mu \gamma \big) = \partial_b \gamma R_{ba}(\gamma) \,.
\end{split}
\ee

\subsubsection{Inverse Higgs constraints}\label{sec:inverseHiggs}

Having established the covariant derivatives in Eq.~\eqref{eq:covariantDerivatives}, one may proceed to construct the most general invariant action. Although the counting formula derived in Sec.~\ref{sec:nonuniformGoldstone} predicts the existence of a single gapless excitation, the coset construction yields additional Goldstone fields beyond those required by the counting, rendering the description unnecessarily complicated. 

In the coset construction, this redundancy can be systematically removed by imposing inverse Higgs constraints (see Appendix~\ref{sec:inverseHiggsMechanism}), which allow certain Goldstone fields to be expressed algebraically in terms of derivatives of others. In this way, the inverse Higgs constraints provide a convenient implementation of the reduction implied by the nonuniform symmetry algebra.

However, the physical role of these additional fields is not immediately transparent, as they may correspond either to pure gauge redundancies or to propagating but massive modes, depending on the representation furnished by the order parameter. The goal of this section is to identify the physical nature of the remaining Goldstone fields and to show explicitly how they can be eliminated using inverse Higgs constraints, yielding a low-energy effective theory written solely in terms of the single gapless mode.

\paragraph{Redundant Goldstone fields.}
We begin by demonstrating that both the boost and the rotational Goldstones, $v^a$ and $\gamma$, are redundant degrees of freedom. Specifically, we show that the corresponding symmetry generators do not produce independent fluctuations of the order parameter. Consequently, $v^a$ and $\gamma$ are pure gauge and can be consistently eliminated by an appropriate gauge fixing.

To see this, we consider a generic local perturbation of the order parameters, Eq.~\eqref{eq:groundStateVortex}, generated by the broken symmetry generators
\be \begin{split}
\langle \delta \phi^I \rangle_{\rm{eq}}  
&= \left \langle \big[ \gamma(x) L + u^a(x) T_a + v^a(x) B_a + \varphi(x) N \,,\, \phi^I \big] \right \rangle_{\rm{eq}}   \\
&= \delta^I_a \Big(  \gamma(x) \epsilon_{ab} x_b+ \frac{1}{2\Omega} \epsilon_{ab} v^b(x) 
-\frac{1}{2\Omega} \epsilon_{ab} R_{bc}(2\Omega t) v^c(x) + u^a(x) \Big)  \,, \\    \langle \delta \psi   \rangle_{\rm{eq}} 
&= \left \langle \big[ \gamma(x) L + u^a(x) T_a + v^a(x) B_a + \varphi(x) N \,,\, \psi \big]   \right \rangle_{\rm{eq}} \\
&= m\Omega \epsilon_{ab} u^a(x) x_b 
+ \frac{m}{2} x_a v^a(x) 
+ \frac{m}{2} R_{ab}(2\Omega t) x_a v^b(x) 
+ \varphi(x) \,,
\end{split}\ee
where Goldstone fields $\{\varphi(x), u^a(x), v^a(x), \gamma(x)\}$ are understood as local functions of space and time. 
%%%%%%%
%break%
%%%%%%
To see that the fluctuations generated by $v^a(x)$ and $\gamma(x)$ are not independent of $u^a(x)$ and $\varphi(x)$, we introduce the following change of variables
\be \begin{split}\label{eq:gaugeFix}
\varphi(x) &= \tilde \varphi(x) - m\Omega \epsilon_{ab} u^a(x) x_b - \frac{m}{2} x_a v^a(x) - \frac{m}{2} R_{ab}(2\Omega t) x_a v^b(x)\,, \\
u^a(x)&= \tilde u^a(x)-\gamma(x) \epsilon_{ab} x_b + \frac{1}{2\Omega} \epsilon_{ab} \Big(  R_{bc}(2\Omega t) v^c(x) -  v^b(x)\Big)\,,
\end{split}
\ee 
so that 
\be 
\langle \delta \phi^I \rangle_{\rm{eq}} = \delta^I_a \tilde u^a \,, \quad \langle \delta \psi \rangle_{\rm{eq}} = \tilde \varphi(x)\,.
\ee 
 Therefore, an arbitrary fluctuation is fully parameterized by the fields $\tilde u^a(x)$ and $\tilde \varphi(x)$. By fixing the gauge \eqref{eq:gaugeFix}, the redundant Goldstone fields $v^a(x)$ and $\gamma(x)$ are eliminated in favour of these independent degrees of freedom.
 
\paragraph{Massive Kohn mode.} 
After gauge fixing, we are left with the independent fluctuations $\tilde u^a(x)$ and $\tilde\varphi(x)$, which correspond to physical propagating degrees of freedom. However, as established earlier, the system supports only a single gapless NG mode. Indeed, the symmetry algebra \eqref{eq:cosetAlgebra} implies the existence of a gapped excitation with a fixed mass $2\Omega$, which is identified with the Kohn mode \cite{Glodkowski:2025krf}. The presence and gap of this mode are protected by symmetry and are independent of microscopic details.

We will not reproduce the derivation of the Kohn mode here. Instead, we focus on the hydrodynamic limit, in which both redundant and massive Goldstone modes are integrated out. The resulting low-energy effective theory is therefore governed by a single gapless excitation.

\subsubsection{Implementing the constraints}
Following the procedure outlined in Appendix~\ref{sec:inverseHiggsMechanism}, we now systematically eliminate all redundant and massive Goldstone modes by imposing a set of inverse Higgs constraints. Crucially, these constraints are compatible with all symmetries of the system and ensure that the resulting theory continues to faithfully realize the spontaneously broken symmetries in a nonlinear manner. In this sense, the inverse-Higgsed theory constitutes a minimal nonlinear realization of the symmetry breaking pattern Eq.~\eqref{eq:breakingPattern}, expressed in terms of the smallest set of independent Goldstone fields compatible with the symmetry algebra.

\paragraph{Boost Goldstone.}
The first inverse Higgs constraint follows from the nonuniform commutator between the unbroken time translation and the broken boost generator,
\be 
[\bar H, B_a] = - \bar P_a + T_a - 2\Omega \epsilon_{ab} B_b\,,
\ee 
which contains the broken translation generator $T_a$. According to the inverse Higgs criterion, this allows one to eliminate the Goldstone field associated with $B_a$ in favour of derivatives of the displacement field $u^a$.

Concretely, this is achieved by imposing the covariant constraint
\be 
D_0 u^a = \partial_0 u^a + v^a + v^b \partial_b u^a = 0\,,
\ee
which algebraically expresses the boost Goldstone $v^a$ in terms of $u^a$ and its derivatives.

To resolve this constraint explicitly, we define
$v^\mu = (1, v^a)$, in terms of which the constraint can be written as
\be
v^\mu \partial_\mu \phi^I = 0 \,.
\ee 
Since at fixed time the map $\phi_t : x^a \mapsto \phi^I(x,t)$ is a diffeomorphism, the Jacobian $\partial_a \phi^I$ is invertible and the above equation uniquely determines $v^a$.

The most general solution is given by the velocity field of the lattice points,\footnote{This velocity admits a clear geometrical interpretation as the conserved current associated with the vortex-space volume form,
$v = \star \mathrm{vol}(\mathcal M_\phi) = \star \frac{1}{2} \epsilon_{IJ} d\phi^I \wedge d\phi^J$.}
\be 
v^\mu = \frac{1}{|\partial \phi|}\,
\epsilon^{\mu \nu \rho}\,\epsilon_{IJ}\,
\partial_\nu \phi^I \partial_\rho \phi^J\,,
\ee 
where
\be
|\partial \phi| = \epsilon^{ij} \epsilon_{IJ} \partial_i \phi^I \partial_j \phi^J\,.
\ee
In particular, this yields
\be 
v^a = - (\partial \phi^{-1})^a{}_b \, \partial_0 \phi^b \,. 
\ee 
For $\phi^a = x^a + u^a$ one has
\be 
(\partial \phi^{-1})^a{}_b = \delta^a_b - \partial_b u^a + \dots \,,
\ee 
and consequently
\be \label{eq:boostIHC}\boxed{
v^a
= -\partial_0 u^a
+ \partial_0 u^b \partial_b u^a
+ \dots \,.}
\ee

\paragraph{Rotational Goldstone.} The inverse Higgs constraint associated with the rotational Goldstone follows from the nonuniform commutator
\be 
[L,\bar P_a] = \epsilon_{ab} (\bar P_b - T_b)\,,
\ee 
which implies that the rotational Goldstone can be consistently removed from the theory by imposing the inverse Higgs constraint 
\be 
\epsilon_{ab} D_a u^b =0\,.
\ee 
%%%%%%%
This constraint yields an algebraic relation between the matrix $R_{ab}(\gamma)$ and the displacement Goldstone 
\be 
\epsilon_{ab} D_a u^b = \epsilon_{ab} \big(R_{ca}(\gamma) \partial_c \phi^b - \delta^b_a \big) = 0\,. 
\ee 
Expanding $\phi^a = x^a + u^a$, this condition reduces to
\be 
\epsilon_{ab} \partial_a u_b \cos{\gamma} + (2 + \partial_a u^a ) \sin{\gamma} = 0 \,, 
\ee 
which can be solved iteratively to express the rotational Goldstone $\gamma$ in terms of the displacement field and its derivatives
\be \label{eq:gammaIHC} \boxed{
\gamma = -\frac{1}{2} \epsilon_{ab}\partial_a u^b + \frac{1}{4} \epsilon_{ab}\partial_a u^b \partial_c u^c + \dots}
\ee 

\paragraph{Kohn mode.} 
Finally, we consider the nonuiform commutation relation for the vortex translations
 \be
 [\bar P_a, T_b] = -2m\Omega \epsilon_{ab} N\,,
 \ee 
in order to eliminate the displacement field $u^a$ in favour of the $\mathrm U(1)$ Goldstone $\varphi$.

To achieve this, we fix 
 \be \label{eq:constarintKohn}
 D_a \varphi  = \big(\partial_b \varphi +2m\Omega \epsilon_{bc} u^c  - m  v^b   + m\Omega \epsilon_{cd} \partial_b u^c u^d \big) R_{ba}(\gamma)  =0\,.
 \ee 
 By doing so, we are effectively integrating out the massive Kohn mode, which corresponds to a physical propagating excitation with mass $2\Omega$ \cite{Glodkowski:2025krf}. 
 
Solving Eq.~\eqref{eq:constarintKohn} iteratively yields
\be \label{eq:IHC}\boxed{
u^a = \frac{1}{2m\Omega} \epsilon_{ab} \partial_b \varphi  + \frac{1}{8m^2\Omega^2} \epsilon_{ab} \epsilon_{cd} \partial_b \partial_c \varphi \partial_d \varphi + \dots }
\ee 
This constraint eliminates the longitudinal fluctuations of the displacement field,
implying that the low-energy dynamics of the vortex lattice is incompressible,
\be
\partial_a u^a \simeq 0\,,
\ee
where the symbol $\simeq$ indicates that this relation holds only in the low-energy limit.

\subsubsection{Effective field theory}
In this section, we construct an effective field theory description for the long-wavelength oscillations in superfluid vortex crystals. We focus on the low-energy regime where the dynamic is governed by a single massless NG mode with a quadratic dispersion relation $\omega \sim k^2$. We construct both the leading-order and next-to-leading-order effective theories, including nonlinear cubic interactions, and determine the decay rate of the low-energy mode.

\paragraph{Derivative expansion.} \label{sec:derivativeExpansion}
In the limit of slow processes, spacetime derivatives are naturally suppressed by a small expansion parameter.

To organize the effective theory systematically, we introduce a derivative counting scheme and assign
\be 
\mathcal{O}(\partial_t) \sim O(\epsilon^{2})\,, \quad \mathcal{O}(\partial_i) \sim O(\epsilon)\,.
\ee 
Furthermore, based on our analysis of the inverse Higgs constraints in Section~\ref{sec:inverseHiggs}, we infer the following scaling of the Goldstone fields,
\be
\varphi \sim \mathcal O(\epsilon^{-1})\,, \quad u^a \sim \mathcal O(\epsilon^0)\,, \quad \gamma \sim \mathcal O(\epsilon)\,, \quad v^a \sim \mathcal O(\epsilon^2)\,.
\ee
%%%%%%%%%%
%%%%%%%%
Applying this counting, the covariant derivatives Eq. \eqref{eq:covariantDerivatives} can be systematically organized in a perturbative expansion. In particular, we find
\be  \begin{split}\label{eq:covariantLowEnergy}
 D_a u^b &= \partial_a u^b + \gamma \epsilon_{ab}  + \gamma  \epsilon_{ac} \partial_c u^b - \frac{\gamma^2}{2} \delta_{ab} + \mathcal{O}(\epsilon^3) \,, \\
 D_0 \varphi &= \partial_0 \varphi + v^a \left( \partial_a \varphi + m\Omega \epsilon_{ab} u^b \right) + \mathcal{O}(\epsilon^4)\,.
 \end{split}
\ee 

We now substitute the solutions of the inverse Higgs constraints,
Eqs.~\eqref{eq:boostIHC}, \eqref{eq:gammaIHC}, and \eqref{eq:IHC}, thereby expressing all Goldstone fields in terms of the $\mathrm U(1)$ Goldstone $\varphi$. The resulting expressions can be organized consistently within the derivative expansion as
\be \begin{split}
u^a &= \underbrace{ \Theta  \tilde \partial_a \varphi}_{\mathcal{O}(\epsilon^0)}  \, + \, \underbrace{\frac{\Theta^2}{2}   \tilde \partial_a \partial_b \varphi \tilde \partial_b \varphi }_{\mathcal{O}(\epsilon)}  \, + \,\, \mathcal{O}(\epsilon^2)\,, \\
v^a &= - \underbrace{\Theta  \partial_0 \tilde \partial_a \varphi}_{\mathcal{O}(\epsilon^2)} \, + \, \underbrace{ \frac{\Theta^2}{2} \left( \partial_0 \tilde \partial_b \varphi \partial_b \tilde \partial_a \varphi - \tilde \partial_b \varphi  \partial_0 \partial_b \tilde \partial_a \varphi \right)}_{\mathcal{O}(\epsilon^3)}  \,  + \, \mathcal{O}(\epsilon^4)\,, \\
 \gamma &= \underbrace{\frac{\Theta}{2} \partial^2 \varphi}_{\mathcal{O}(\epsilon)} + \underbrace{\frac{\Theta^2}{4} \partial^2 \partial_a \varphi \tilde \partial_a \varphi}_{\mathcal{O}(\epsilon^2)} + \mathcal{O}(\epsilon^3)\,, 
\end{split}
\ee 
where $\Theta = \frac{1}{2m\Omega}$ and $\tilde \partial_a = \epsilon_{ab} \partial_b$. 

After substituting these expressions into the covariant derivatives \eqref{eq:covariantLowEnergy}, we identify the following low-energy building blocks, 
 \be
\begin{split}\label{eq:defects}
       D_a u^b &= \underbrace{\Theta \left(  \partial_a \tilde \partial_b \varphi  + \frac{\epsilon_{ab}}{2} \partial^2 \varphi  \right) }_{\mathcal O(\epsilon)}  \\ 
       &+    \underbrace{\frac{\Theta^2}{2} \left[ \partial_a \left( \tilde \partial_b \partial_c \varphi \tilde \partial_c \varphi \right) + \frac{\epsilon_{ab}}{2} \partial^2 \partial_c \varphi \tilde \partial_c \varphi  + \partial^2 \varphi \left(\tilde \partial_a \tilde \partial_b \varphi - \frac{\delta_{ab}}{4} \partial^2 \varphi \right) \right] }_{\mathcal O(\epsilon^2)} \, + \, \mathcal{O}(\epsilon^3) \,, \\
         D_0 \varphi 
         &= \underbrace{\partial_0 \varphi}_{\mathcal O(\epsilon)}  \, - \, \underbrace{\frac{\Theta}{2} \partial_0 \tilde \partial_a  \varphi \partial_a \varphi}_{\mathcal{O}(\epsilon^2)} \, + \, \underbrace{\frac{\Theta^2}{4} \partial_0 \partial_a \partial_b \varphi \tilde \partial_a   \varphi \tilde \partial_b \varphi }_{\mathcal{O}(\epsilon^3)}\, + \, \mathcal{O}(\epsilon^4)\,.
\end{split}
\ee 
The second term in $D_0 \varphi$ is a total derivative that accounts for the Berry phase acquired by the vortices. For regular field configurations, the expression for $D_0 \varphi$ can, up to total derivatives, be rewritten as
\be
D_0 \varphi \simeq  \underbrace{ \frac{\Theta^2}{4} \partial_0 \varphi \left[ (\partial_a \partial_b \varphi)^2 - (\partial^2 \varphi)^2 \right]}_{\mathcal{O}(\epsilon^3)}  + \, \mathcal{O}(\epsilon^4)\,,
\ee
where the symbol $\simeq$ denotes equivalence up to total derivative terms. Similarly, we find
\be 
(D_0 \varphi)^2 \simeq \underbrace{(\partial_0 \varphi)^2}_{\mathcal{O}(\epsilon^2)}  + \, \mathcal{O}(\epsilon^4) \,,  \quad (D_0 \varphi)^3 \simeq \underbrace{(\partial_0 \varphi)^3}_{\mathcal{O}(\epsilon^3)}  + \, \mathcal{O}(\epsilon^5) \,.
\ee 
Having identified the appropriate low-energy structures, it is now straightforward to construct the effective action up to leading $\sim \mathcal{O}(\epsilon^2)$ and next-to-leading $\sim \mathcal{O}(\epsilon^3)$ order in the perturbative expansion. 

\paragraph{Leading-order theory.}
The leading-order effective theory is described by a quadratic Lagrangian,
\be \label{eq:actionTkachenko}
\mathcal L_{(2)} =  n_0  D_0 \varphi + \frac{\chi}{2} (D_0 \varphi)^2   - E^{(2)}(\partial u) + \mathcal{O}(\epsilon^3) \,,
\ee 
where the elastic energy of the vortex crystal is given by
\be 
E^{(2)}(\partial u) = \lambda^{abcd} D_a u^b D_c u^d \,,
\ee 
with the elastic modulus tensor
\be 
\lambda^{abcd}= \lambda_1 \delta_{ab} \delta_{cd} + \lambda_2 \delta_{a\langle c} \delta_{d \rangle b} \,. 
\ee 

Using \eqref{eq:defects}, we find that it admits the following expansion in terms of the phase field 
\be\begin{split}
E^{(2)}(\partial u) &= \frac{\Theta^2 \lambda_2}{2}\big( 2(\partial_a \partial_b \varphi)^2-(\partial^2 \varphi)^2 \big) + \frac{\Theta^3 \lambda_2 }{4} \Big[ \left( \partial_a \tilde \partial_b \varphi + \tilde \partial_a \partial_b \varphi \right) \partial_a \left( \tilde \partial_b \partial_c \varphi \tilde \partial_c \varphi \right)  \Big]  \, + \, \mathcal{O}(\epsilon^4)\,, \\ 
&\simeq \frac{\Theta^2 \lambda_2}{2}(\partial^2 \varphi)^2  + \mathcal{O}(\epsilon^4)\,.
\end{split}
\ee
Therefore, up to a total derivative, the Lagrangian in Eq.~\eqref{eq:actionTkachenko} can be cast into the following form
\be \label{eq:lifshitz}
\mathcal L_{(2)} \simeq  \frac{\chi}{2} (\partial_0 \varphi)^2   - \frac{\Theta^2 \lambda_2}{2} (\partial^2 \varphi)^2 + \mathcal{O}(\epsilon^3) \,.
\ee 
This effective theory describes a single gapless excitation--the Tkachenko mode--propagating with a quadratic dispersion relation,
\be 
\omega^2 = \frac{\Theta^2 \lambda_2}{\chi} k^4 \,.
\ee 
%%%%%%%
%break%
%%%%%%%%
\paragraph{Next-to-leading order theory.}
At next-to-leading order the effective theory becomes cubic. It can be expressed as follows
\be 
\mathcal L_{(3)} =  \mathcal L_{(2)} + g_1 (D_0 \varphi)^3 + g_2 D_0 \varphi E^{(2)}(\partial u) + E^{(3)}(\partial u) + \mathcal{O}(\epsilon^4) \,,
\ee 
where 
\be 
E^{(3)}(\partial u) = g_3 \rm{Im}(\partial_z \partial_z \varphi)^3 
\ee
with $z=x+iy$ is a $C_6$-invariant cubic structure. 

More explicitly, the theory can be written as
\be \label{eq:actionTkachenkoInteracting}
\begin{split}
\mathcal L_{(3)} = \mathcal L_{(2)} +  g_1 (\partial_0 \varphi)^3 + \tilde g_2 \partial_0 \varphi  (\partial_a \partial_b \varphi)^2 - \bar g_2 \partial_0 \varphi (\partial^2 \varphi)^2  + g_3 \text{Im}(\partial_z \partial_z \varphi)^3 + \mathcal{O}(\epsilon^4) \,,
\end{split} \ee 
where 
\be 
\tilde g_2 =  \Theta^2 \left( g_2 \lambda_2  + \frac{n_0}{4} \right)\,, \quad  \bar g_2 = \Theta^2 \left(\frac{g_2 \lambda_2}{2}  + \frac{n_0}{4} \right) \,.
\ee
%%%%%%%%%%
Following Ref.~\cite{PhysRevResearch.6.L012040}, we employ a simple
dimensional analysis to determine the low-energy scaling of the decay
width of the Tkachenko mode into two Tkachenko modes. 

In our scaling scheme, where $E \sim k^2$, the Goldstone field $\varphi$
is dimensionless, while all cubic interaction terms in
Eq.~\eqref{eq:actionTkachenkoInteracting} contain six spatial derivatives
and therefore have couplings with scaling dimension
\be 
[g] \sim k^{-2} \sim E^{-1} \,.
\ee 
The decay width $\Gamma$ has the dimension of energy. Since the leading
contribution to the decay rate is quadratic in the cubic couplings,
dimensional analysis alone fixes
\be 
\Gamma(E) \sim g^2 E^3 \,.
\ee 
We therefore conclude that, at zero temperature, the decay rate of the
Tkachenko mode exhibits a cubic dependence on the energy.

%%%%%%%%%
%%%%%%%%%%%
%%%%%%%%%%%
%%%%%%%%%
%%%%%%%%%%%
%%%%%%%%%%%
\part{Many-body dynamics with nonuniform symmetries}\label{part:2}

\section{Hydrodynamics with nonuniform symmetries}\label{sec:nonuniformHydrodynamics}
In this section, we begin by reviewing the hydrodynamic paradigm, in which the long-wavelength dynamics of thermal many-body systems is controlled by a small set of slow variables associated with conserved densities. We begin with diffusion as the simplest hydrodynamic process, formulating its hydrodynamic theory and classifying diffusive transport by symmetry. We extend this framework to nonuniform symmetries and demonstrate that their presence generically modifies diffusion, leading to qualitatively new relaxation dynamics, including the possibility of anomalously slow transport.

We then present a detailed hydrodynamic construction of \textit{Aristotelian fluids}, namely homogeneous and isotropic systems without boost symmetry. They provide a minimal setting in which a nonuniform symmetry, such as spatial rotations generated by angular momentum, is realized in the fluid phase. We analyze their global symmetries and the associated slow degrees of freedom, with particular emphasis on nonuniform currents, and systematically derive the corresponding hydrodynamic modes. We also consider Aristotelian fluids flowing in nontrivial external electromagnetic backgrounds, and show how these backgrounds deform the symmetry structure and qualitatively modify the hydrodynamic mode spectrum.

While many of the results presented in this section can be found in the existing literature \cite{deBoer:2017ing,deBoer:2020xlc,10.21468/SciPostPhys.5.2.014,Hongo:2021ona}, our presentation differs from conventional treatments by systematically incorporating the full set of global symmetries, including in the presence of external electromagnetic backgrounds, and carefully accounting for nonuniform symmetries, which are often treated only implicitly. This framework sets the stage for the subsequent sections, where we apply the hydrodynamic paradigm to analyze the collective dynamics of fracton fluids.

\subsection{The hydrodynamic paradigm}
In this subsection we review the basic logic underlying the hydrodynamic description of thermal many-body systems, with particular emphasis on the role of conserved quantities in identifying the relevant collective variables.

Low-energy dynamics in a generic many-body system is governed by gapless excitations, as these can be excited with arbitrarily small energy. At finite temperature, however, a thermal background of such excitations is already populated. As a result, generic gapless modes that are not protected by symmetry scatter efficiently off this bath and acquire a finite relaxation time $\tau$, rendering them irrelevant at long timescales $t \gg \tau$. In other words, they decay rapidly back to a state of local equilibrium. Importantly, since this relaxation takes place locally, the timescale $\tau$ is independent of the wavelength of the perturbation.

\begin{figure}[h!]
    \centering
\includegraphics[width=0.35\linewidth]{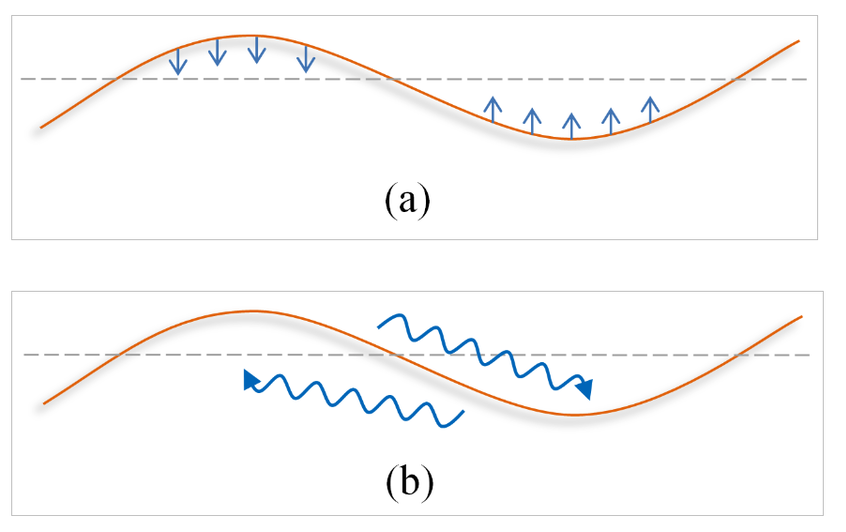}
    \caption{Relaxation of perturbations in a generic many-body system. The dashed line represents the state of thermodynamic equilibrium, while the solid line denotes a perturbed quantity. (a) Perturbations of non-conserved quantities relax locally on a timescale $\tau$, independently of the wavelength. (b) Perturbations of conserved quantities can relax only via transport, which leads to parametrically long relaxation times at large wavelengths. Figure adapted from \cite{Liu:2018kfw}.}
    \label{fig:relaxation}
\end{figure}

In contrast, excitations associated with conserved quantities cannot simply decay, as this would violate the underlying conservation law. Instead, they correspond to local excesses or deficits of the conserved charge, which can relax only through transport, see Fig.~\ref{fig:relaxation}. In addition, Goldstone modes arising from spontaneously broken continuous symmetries are protected by symmetry and therefore remain long-lived excitations, as guaranteed by the Goldstone theorem. As shown in Sec.~\ref{sec:nonuniformGoldstone}, this correspondence is modified in the presence of nonuniform symmetries, where the number and nature of gapless modes is determined by the nonuniform counting formula. Therefore, in the \textit{hydrodynamic limit}, $t \gg \tau$ and $l \gg l_{mf}$, conserved densities and Goldstone bosons remain the only relevant degrees of freedom governing the infrared dynamics.\footnote{Near a continuous phase transition, the order parameter itself can become parametrically slow as the correlation length diverges.}

These \textit{hydrodynamic variables} encode the macroscopic state of a given many-body system. Their dynamics is governed by \textit{hydrodynamic equations}, whose structure is constrained by symmetries and organized in a \textit{derivative expansion}. In what follows, we illustrate these general principles, beginning with the hydrodynamic theory of diffusion and proceeding to more general fluid systems, while keeping track of the role of nonuniform symmetries.

\subsection{Hydrodynamic theory of diffusion}
We now apply the hydrodynamic paradigm to charge transport. For uniform conserved charges, we show that hydrodynamics generically leads to Fick’s law of diffusion. We then extend the analysis to nonuniform symmetries. In this case, the correspondence between conserved charges and universal diffusive relaxation breaks down, and Fick’s law is no longer enforced. Instead, the dispersion relations of the hydrodynamic modes depend on the structure of the nonuniform symmetry algebra and may include anomalously slow subdiffusive modes.

\subsubsection{Fick’s law of diffusion}
We begin with a many-body system carrying a set of conserved uniform abelian charges, labeled by an index $A$,
\be \label{eq:chargesHydro}
Q^A = \int  \star j^A\,, 
\ee 
together with the associated local conservation laws
\be 
\quad  d \star j^A = 0\,.
\ee 
They constitute the hydrodynamic equations governing the relaxation of small, long-wavelength charge density modulations around the state of thermodynamic equilibrium.

It is useful to explicitly separate the charge densities from the spatial currents, as the densities play a distinguished role, serving as the fundamental hydrodynamic variables. In components, the conservation laws take the form
\be \label{eq:hydroEQInternal}
 \partial_t j^{0A} + \partial_i j^{iA} = 0\,.
\ee 
In hydrodynamics, one aims to solve the hydrodynamic equations in order to determine the time evolution of the hydrodynamic variables. At first sight, however, the number of unknowns exceeds the number of available equations, since the spatial currents are not known \textit{a priori} and constitute additional unknowns.

\paragraph{Constitutive relations.} To close the system, one must express the currents in terms of the hydrodynamic densities,
\be 
j^{iA} \equiv j^{iA}[j^{0A}]\,.
\ee 
These are known as the \textit{constitutive relations}. They are organized in a \textit{derivative expansion}, valid at long wavelengths where gradients are small. The hydrodynamic equations can then be solved order by order in this expansion.

The structure of the constitutive relations is further constrained by the second law of thermodynamics, which requires the existence of an entropy current with non-negative divergence. This condition places powerful constraints on the allowed transport coefficients and will be used repeatedly in the rest of this thesis to classify admissible hydrodynamic transport. We refer to this approach as the \textit{entropy current formalism}. We now apply this framework to analyze hydrodynamic diffusion.

As a simple toy setup, we consider a system in which energy and momentum are not conserved and therefore do not give rise to hydrodynamic modes. Accordingly, the entropy density depends only on the conserved charge densities,
\be \label{eq:entropUniform}
s \equiv s[j^{0A}] \,.
\ee
The time evolution of the entropy density is therefore given by
\be
\partial_t s = - \mu_A\,\partial_t j^{0A}\,,
\ee
where the chemical potentials are defined by
\be \label{eq:chemicalPotential}
\mu_A = -\frac{\delta s}{\delta j^{0A}}\,.
\ee
Using the conservation laws Eq.~\eqref{eq:hydroEQInternal}, we find
\be
\partial_t s + \partial_i \left( -\mu_A j^{iA} \right)
= - j^{iA} \partial_i \mu_A\,.
\ee
This identifies the entropy current as
\be
s^i = -\mu_A j^{iA}\,,
\ee
and the local entropy production as
\be
\Delta \equiv \partial_t s + \partial_i s^i
= - j^{iA} \partial_i \mu_A\,.
\ee
The second law of thermodynamics requires $\Delta \ge 0$ for all configurations of hydrodynamic densities. At leading order in a gradient expansion, this fixes the constitutive relations to take the form
\be \label{eq:constitutiveRelation}
j^{iA} = - \Sigma^{ijAB} \partial_j \mu_B + \mathcal O(\nabla^2)\,,
\ee
where the symmetric transport tensor $\Sigma^{ijAB}$ is positive semi-definite, so that
\be
\Delta
= \Sigma^{ijAB}
\partial_i \mu_A
\partial_j \mu_B
\ge 0\,.
\ee
If the system admits no background spatial vectors or tensors, rotational covariance implies that the dissipative tensor reduces to the isotropic form
\be \label{eq:isotropicForm}
\Sigma^{ijAB}=\delta^{ij} \Lambda^{AB}\,,
\ee
where $\Lambda^{AB}$ is a symmetric positive semi-definite matrix in charge space. 

\paragraph{Hydrodynamic modes.} To identify the hydrodynamic modes, we work within linear response around a homogeneous equilibrium,
\be \label{eq:equilibriumLinearized}
j^{0A} = j^{0A}_{\rm eq} + \delta j^{0A}\,.
\ee
We expand the entropy density around equilibrium up to quadratic order in the fluctuations,
\be \label{eq:entropyDensityFluctuation}
\delta s
= -\mu^{\rm eq}_A \delta j^{0A}
-\frac{1}{2} \chi_{AB} \delta j^{0A} \delta j^{0B}
+\ldots\,,
\ee
Here $\chi_{AB}$ is the susceptibility matrix, defined by
\be \label{eq:defChi}
\chi_{AB} = - \frac{\partial^2 s}{\partial j^{0A} \partial j^{0B}} \,,
\ee
which is symmetric and positive definite as a consequence of thermodynamic stability.

The definitions in Eqs.~\eqref{eq:chemicalPotential}, \eqref{eq:defChi}, and the expansion \eqref{eq:entropyDensityFluctuation}, imply the linearized relation between chemical potentials and densities,
\be \label{eq:chemicalPotentialLinearized}
\delta \mu_A = \chi_{AB} \delta j^{0B}\,.
\ee
%%%%%%%%%%%%%%%%%%%%
Using charge conservation Eq.~\eqref{eq:hydroEQInternal} together with the constitutive relation Eq.~\eqref{eq:constitutiveRelation}, and assuming isotropic form Eq.~\eqref{eq:isotropicForm}, we obtain the linearized hydrodynamic equations
\be \label{eq:hydroEQInternal2}
\partial_t \delta j^{0A} - D^A{}_B \partial^2 \delta j^{0B} = 0\,,
\ee
where we have defined the diffusion matrix
\be 
D^A{}_B \equiv \Lambda^{AC}\chi_{CB}\,.
\ee 
%%%%%%%%
After going to Fourier space, the spectrum of the hydrodynamic excitations follows from solving the characteristic equation
\be \label{eq:diffusion}
\det \left(-i\omega \delta^{A}{}_{B} + k^2 D^{A}{}_{B} \right)=0\,.
\ee
The resulting dispersion relations are generically diffusive,
\be
\omega_A(k)=-i \lambda_A k^2 \,,
\ee
where $\lambda_A$ denote the eigenvalues of the diffusion matrix $D^{A}{}_{B}$. 
The corresponding eigenvectors determine the linear combinations of charge densities that evolve independently at long wavelengths. 

Thus, a system with $N$ conserved uniform charges generically exhibits $N$ diffusive hydrodynamic modes.\footnote{
For non-Abelian symmetries, the hydrodynamic modes can mix and acquire additional nondissipative contributions, leading to dispersion relations with both real and imaginary parts at order $k^2$. The diffusive scaling and the number of hydrodynamic modes, however, remain unchanged. See \cite{10.21468/SciPostPhys.10.1.015}.} 
In this sense, each conserved quantity produces a diffusive excitation describing the relaxation of long-wavelength charge fluctuations. 

This conclusion, however, relies crucially on the assumption that the conserved currents transform homogeneously under spacetime translations. When this assumption is relaxed, the structure of the hydrodynamic equations is modified, and diffusion is no longer guaranteed.
%%%%%%%%%%%%%%%

%%%%%%%%%
\subsubsection{Diffusion with nonuniform symmetries}\label{sec:diffusionNonuniform}
We now consider symmetries that act nonuniformly in space, which is encoded in their algebra with spacetime translations,
\be \label{eq:nonuniformAlgebraHydro}
[P_\mu,Q^A]=(\lambda_\mu)^A{}_B Q^B\,,
\ee
where $(\lambda_\mu)^A{}_B$ are constant coefficients characterizing the nonuniformity of the symmetry action. Here $P_\mu$ denotes the generators of spacetime translations. Importantly, these need not correspond to conserved charges. In particular, we will assume that momentum and energy are not conserved in the present setup. For simplicity, we restrict to abelian charges.

\paragraph{Nonuniform currents and hydrodynamic equations.}
Expressing the charges in terms of their currents,
\be
Q^A=\int \star j^A\,,
\ee
the nonuniform algebra Eq.~\eqref{eq:nonuniformAlgebraHydro} implies that these currents transform inhomogeneously under the action of translations,
\be
[P_\mu, j^A]
= -\partial_\mu j^A
+ (\lambda_\mu)^A{}_B j^B\,.
\ee
It is then possible to factor out the explicit coordinate dependence by defining 
\be \label{eq:nonunifromCurrent}
j^A = (e^{x^\mu \lambda_\mu})^A{}_B \tilde j^B\,,
\ee 
where we have introduced the \emph{uniform currents} $\tilde j^A$ that transform homogeneously under spacetime translations
\be
[P_\mu, \tilde j^A]
= -\partial_\mu \tilde j^A\,.
\ee
%%%%%%%
The uniform currents are not conserved in general. Rather, they satisfy a covariant conservation law,
\be \label{eq:covariantLaw}
D \star \tilde j = 0 \,,
\ee
with the covariant derivative given by
\be
D \equiv d + \lambda \,,
\ee
where $\lambda \equiv \lambda_\mu dx^\mu$ is a matrix-valued one-form acting in charge space. Here we have assembled the components $\tilde j^A$ into a vector $\tilde j$ in charge space. 
Eq.~\eqref{eq:covariantLaw} reflects the mixing of the individual components $\tilde j^A$ under spacetime evolution. As a result, only particular linear combinations correspond to genuine conservation laws.

To make this precise, we follow Sec.~\ref{sec:nonuniformGoldstone} and denote the charge space by $V^\star$, with dual $V$ and canonical pairing $\langle \cdot , \cdot \rangle$. The matrices $\lambda$ act on $V^\star$, while $\lambda^T$ act on $V$. We now consider a linear combination 
\be 
\tilde j_v \equiv \langle \tilde j, v \rangle
\ee
for some $v \in V$. Using the covariant conservation law Eq.~\eqref{eq:covariantLaw}, we find
\be 
d \star \tilde j_v
= \langle d \star \tilde j, v \rangle
= \langle D \star \tilde j, v \rangle - \langle \lambda \wedge \star \tilde j, v \rangle
= - \langle \star \tilde j, \lambda^T v \rangle\,.
\ee 
Thus, $\tilde j_v$ is conserved if and only if $\lambda^T v = 0$.

It follows that conserved currents are in one-to-one correspondence with the common kernel of the matrices $\lambda_\mu^T$,
\be \label{eq:kernel2} 
K \equiv \bigcap_\mu \ker \lambda_\mu^T \subset V \,.
\ee
These currents carry independent conserved densities and therefore give rise to gapless hydrodynamic modes. The number of conserved densities is therefore
\be
N_{\rm CD} = \dim K \,.
\ee
Furthermore, since $\dim K = \dim K^\star$, where $K^\star$ is the common kernel of the matrices $\lambda$, the counting can equivalently be performed in the charge representation $\lambda$, in direct analogy with the counting rule for nonuniform NG modes in Sec.~\ref{sec:nonuniformGoldstone}.

%%%%%%%%%%%
%%%%%%%%%%
%%%%%%%%%%
%
%
\paragraph{Constitutive relations.} To identify the constitutive relations, it is helpful to express the covariant conservation law Eq.~\eqref{eq:covariantLaw} in components
\be \label{eq:hydroEQInternalInhomogenous}
\partial_t \tilde j^{0 A} + \partial_i \tilde j^{i A}
= -  (\lambda_0)^A{}_B \tilde j^{0 B}
  -  (\lambda_i)^A{}_B \tilde j^{i B}\,.
\ee 
As in Eq.~\eqref{eq:entropUniform}, we take the entropy density to be a local function of the conserved densities.
Using Eq.~\eqref{eq:nonunifromCurrent} we can express the time evolution of entropy in terms of the uniform densities 
\be
\partial_t s = - \tilde \mu_A \partial_t \tilde j^{0A} - \tilde \mu_A (\lambda_0)^A{}_B \tilde j^{0B}\,,
\ee
where 
\be 
 \tilde \mu_A = - \frac{\partial s}{\partial \tilde j^{0A}} 
= - \frac{\partial s}{\partial j^{0B}}  \frac{\partial j^{0B}}{\partial \tilde j^{0A}} 
=  \mu_B (e^{x^\mu \lambda_\mu})^B{}_A  \,.
\ee
Notice that the chemical potentials $\tilde \mu^A$ transform homogeneously under spacetime translations, in contrast to $\mu^A$, which transform inhomogeneously.
Using the continuity equation Eq.~\eqref{eq:hydroEQInternalInhomogenous}, we find
\be
\partial_t s + \partial_i \left( -\tilde \mu_A \tilde j^{iA} \right)
= \tilde \mu_A (\lambda_i)^A{}_B \tilde j^{iB}
- \tilde j^{iA} \partial_i \tilde \mu_A\,.
\ee
It is therefore natural to define the entropy current as
\be
s^i = -\tilde \mu_A \tilde j^{iA}\,,
\ee
so that the local entropy production takes the form
\be
\Delta \equiv \partial_t s + \partial_i s^i
= - \tilde j^{iA} \left( \partial_i \tilde \mu_A
- \tilde \mu_B (\lambda_i)^B{}_A \right)\,.
\ee
Imposing the second law of thermodynamics locally,
\be
\Delta \ge 0\,,
\ee
then constrains the constitutive relation for the uniform currents. In an isotropic system, at leading order in derivatives, one obtains
\be
\tilde j^{iA}
=
- \Lambda^{AB}
\left(
\partial_i \tilde \mu_B
- \tilde \mu_C (\lambda_i)^C{}_B
\right)\,,
\ee
where $\Lambda^{AB}$ is a positive semi-definite symmetric matrix.
%%%%%%%%%%%%%

%%%%%%%%%%%%
\paragraph{Hydrodynamic fluctuations.} Expanding the entropy density around homogeneous equilibrium to quadratic order, as in Eq.~\eqref{eq:entropyDensityFluctuation}, the linearized relation between the uniform chemical potentials and the charge densities takes the form 
\be 
\delta \tilde \mu_A = \chi_{AB} \delta \tilde j^{0B}\,,
\ee 
where the susceptibility matrix is positive definite and defined by
\be
\chi_{AB} \equiv - \frac{\partial^2 s}{\partial \tilde j^{0A}\,\partial \tilde j^{0B}}  \,.
\ee
Then, the hydrodynamic equations~\eqref{eq:hydroEQInternalInhomogenous},
linearized around the homogeneous equilibrium configuration Eq.~\eqref{eq:equilibriumLinearized},
take the form
\be \label{eq:hydroEQInternal3}
\partial_t  \delta \tilde j^{0A}  + (V_i)^A{}_B  \partial_i  \delta \tilde j^{0B} - D^A{}_B \partial^2 \delta \tilde j^{0B} = -  \Gamma^A{}_B \delta  \tilde j^{0 B} \,.
\ee
In writing Eq.~\eqref{eq:hydroEQInternal3} we have defined the diffusion matrix
\be 
D^A{}_B \equiv \Lambda^{AC}\chi_{CB}\,,
\ee 
the drift matrix
\be 
(V_i)^A{}_B = \Lambda^{AC} (\lambda_i)^D{}_C  \chi_{DB} - (\lambda_i)^A{}_C \Lambda^{CD} \chi_{DB}\,,
\ee 
and the relaxation matrix 
\be \label{eq:relaxationMatrix}
\Gamma^A{}_B = (\lambda_0)^A{}_B + (\lambda_i)^A{}_C \Lambda^{CD} (\lambda_i)^E{}_D \chi_{EB}\,.
\ee 
%%%%%%%%
%%%%%%%%
%
%
We now choose a basis of $V$ adapted to the kernel decomposition introduced in Sec.~\ref{sec:kernelDecomposition}, writing
\be \label{eq:kernelDecomposition}
V = K \oplus K^\perp = \mathrm{span}\{e^a\} \oplus \mathrm{span}\{e^\alpha\}\,.
\ee
The corresponding dual basis $\{E_A\}$ of $V^\star$ is defined by the canonical pairing
\be
\langle E_A, e^B \rangle = \delta_A{}^B\,.
\ee
Then, the relaxation matrix vanishes along the kernel directions. 
To see this, we express the relaxation matrix Eq.~\eqref{eq:relaxationMatrix} in operator form,
\be
\Gamma = \lambda_0 + \lambda_i \Lambda \lambda_i^T \chi \,.
\ee
Projecting onto the kernel directions, we obtain
\be
\Gamma^a{}_A
=
\langle \Gamma E_A, e^a \rangle
=
\langle \lambda_0 E_A, e^a \rangle
+
\langle \lambda_i \Lambda \lambda_i^T \chi\, E_A, e^a \rangle \,.
\ee
Both terms vanish, since $\lambda_\mu^T e^a = 0$ by definition of the kernel. Therefore,
\be\label{eq:gamma}
\Gamma^a{}_A = 0 \,,
\ee
reflecting the fact that the hydrodynamic sector does not relax due to damping, as this is forbidden by the underlying conservation laws.

%%
%%%%%
\paragraph{Mode spectrum.} To determine the spectrum of hydrodynamic excitations, we Fourier transform the hydrodynamic equations \eqref{eq:hydroEQInternal3}, obtaining the eigenvalue problem
\be \label{eq:eigenvalueModes}
 L^A{}_B  \tilde j^{0B} = 0\,,
\ee 
%%%%%%%%
where we have defined the operator
\be
L^A{}_B \equiv -i\omega \delta^{A}{}_{B} + i k_i (V_i)^A{}_B + k^2 D^{A}{}_{B} + \Gamma^A{}_B\,.
\ee
To isolate the hydrodynamic sector, we express the charge fluctuation in a basis adapted to the kernel decomposition Eq.~\eqref{eq:kernelDecomposition},
\be
\delta \tilde j^0 = E_A \delta \tilde j^{0A} = E_a \delta \tilde j^{0a} + E_\alpha \delta \tilde j^{0\alpha}\,.
\ee
In this basis, the eigenvalue equation \eqref{eq:eigenvalueModes} decomposes into the block form
\bs \label{eq:hydroEqs}
L^a{}_b \delta \tilde j^{0b} + L^a{}_\alpha \delta \tilde j^{0\alpha} = 0\,, \\
L^\alpha{}_a \delta \tilde j^{0a} + L^\alpha{}_\beta \delta \tilde j^{0\beta} = 0\,.
\es
%%%%%%%%
%%%%%%%
%%%%%
We now analyze the low-energy, long-wavelength limit. In this regime, $\omega \to 0$ and $k \to 0$, the block
\be
L^\alpha{}_\beta = \Gamma^\alpha{}_\beta + \mathcal{O}(\omega, k)
\ee
is dominated by the relaxation matrix $\Gamma^\alpha{}_\beta$. On the other hand, the zero modes of the relaxation matrix $\Gamma^A{}_B$ correspond to conserved currents that lie in the kernel $K$. Consequently, $\Gamma^A{}_B$ has no zero eigenvalues on the complement subspace $K^\perp$, and $\Gamma^\alpha{}_\beta$ is generically invertible. In other words, the modes in $K^\perp$ relax on microscopic timescales and can be integrated out. This is achieved by solving the second equation in \eqref{eq:hydroEqs} for $\delta \tilde j^{0\alpha}$, which gives
\be \label{eq:fastMode}
\delta \tilde j^{0\alpha}
= - (L_\perp^{-1})^\alpha{}_\beta L^\beta{}_a  \delta \tilde j^{0a}\,,
\ee
where $(L_\perp^{-1})^\alpha{}_\beta$ denotes the inverse of the $K_\perp$ block $L^\alpha{}_\beta$.

Substituting \eqref{eq:fastMode} into the first equation in \eqref{eq:hydroEqs} yields the effective equation for the hydrodynamic sector,
\be \label{eq:hydroEQSSSS}
\left(L^a{}_b - L^a{}_\alpha (L_\perp^{-1})^\alpha{}_\beta L^\beta{}_b \right)\delta \tilde j^{0b} = 0\,,
\ee
where the operator in parentheses is the Schur complement of the $K^\perp$ block $L^\alpha{}_\beta$ in $L^A{}_B$.

The hydrodynamic modes therefore correspond to the solutions of the eigenvalue problem
\be
\det \left( L^a{}_b - L^a{}_\alpha (L_\perp^{-1})^\alpha{}_\beta L^\beta{}_b \right) = 0\,.
\ee
In general, this equation depends on the matrices $\lambda_\mu$ in a highly nontrivial way, and a universal classification of solutions is not available. Nevertheless, some general conclusions can be drawn. 

Since the relaxation matrix satisfies Eq.~\eqref{eq:gamma}, the effective operator in Eq.~\eqref{eq:hydroEQSSSS} contains no momentum-independent term on the kernel sector. Therefore, all modes associated with $K$ are necessarily gapless in the long-wavelength limit. Moreover, in contrast to the case of uniform symmetries described by Eq.~\eqref{eq:diffusion}, the structure of the effective operator does not enforce a universal diffusive scaling $\omega \simeq - i k^2$. Instead, the leading behavior of the dispersion relations is determined by the detailed structure of the effective operator Eq.~\eqref{eq:hydroEQSSSS} and must be analyzed case by case. This opens up the possibility of subdiffusive transport, with dispersion relations controlled by higher-order gradients. An explicit realization of this mechanism will be discussed in Sec.~\ref{sec:dipoleSubdiffusion2}, where nonuniform dipole symmetry gives rise to modes with dispersion $\omega \simeq - i k^4$.
%%%%%%%%%
%%%%%%%%%

So far, we have restricted our discussion to diffusive systems in which energy and momentum are not treated as dynamical fields. 
The description of genuine fluids, however, requires their inclusion as hydrodynamic variables. 
We therefore now extend the framework to fluid systems with dynamical momentum and energy densities. 
A natural setting for this is provided by \textit{Aristotelian fluids}, which define the minimal hydrodynamic theory compatible with homogeneity and isotropy, and therefore exhibit nonuniform rotational symmetry.

%%%%%%%
%%%%%%%%
%Aristotelian fluids%
%%%%%%%%%
%%%%%%%%%% 
\subsection{Aristotelian fluids}\label{sec:aristotelian}

In this section we construct the hydrodynamic theory of such fluids directly from symmetry principles and thermodynamics, without assuming invariance under Galilean or Lorentz boosts. 
This boost-agnostic formulation provides a unified framework from which both relativistic and nonrelativistic fluids emerge upon imposing the corresponding boost symmetries, while naturally accommodating systems in which boost symmetry is absent.%%%%%

For the remainder of this section, we specialize to $2+1$--dimensions, which simplifies the tensor structures, without affecting any essential physical features.

\subsubsection{Aristotelian spacetime and symmetries}

An Aristotelian spacetime is a flat manifold 
$\mathcal M \simeq \mathbb{R}_t \times \mathbb{R}^2$
equipped with a spatial metric $g =  \delta_{ij} dx^i \otimes dx^j$,\footnote{Since spatial indices are raised and lowered with the Euclidean metric, we will not carefully distinguish between upper and lower spatial indices.} a clock one-form 
$\tau = dt$, and a vector field $v = \partial_t$ defining an absolute frame. This structure represents the minimal geometric framework compatible with absolute time and spatial covariance. In Galilean (Newton–Cartan) spacetime, the notion of a preferred observer is lacking, leading to a redundancy in the description encoded in (Milne) boost transformations relating different local frames \cite{Jensen:2014aia}. A Lorentzian spacetime, on the other hand, arises when the clock one-form and spatial metric combine into a single Minkowski metric. 
More generally, one may consider curved Aristotelian backgrounds \cite{deBoer:2020xlc,Armas:2020mpr}, but for our purposes it suffices to work on a flat background.

In addition, we consider charged fluids invariant under an abstract uniform $\mathrm U(1)$ symmetry generated by $Q$. 
This symmetry is incorporated by coupling the system to a background $\mathrm U(1)$ gauge field $a = a_\mu dx^\mu$ sourcing the electromagnetic current, so that the background data becomes
$\left( g\,, \tau\,, v\,, a \right)$.

\paragraph{Background variations.} A general variation of the background geometric data is parametrized by a vector field
\be
\boldsymbol{\xi} = \xi^\mu(\boldsymbol{x}, t) \partial_\mu \,,
\ee
generating diffeomorphisms of the manifold $\mathcal M$, together with a function
\be
\lambda \equiv \lambda(\boldsymbol{x}, t) \,,
\ee
representing $\mathrm U(1)$ gauge transformations. We denote collectively the parameters
\be 
\chi = \{ \xi, \lambda\}\,.
\ee 
The infinitesimal action on the background geometry is 
\bs
\delta_\chi g &= \mathcal L_\xi g \,, \\
\delta_\chi \tau &= \mathcal L_\xi \tau \,, \\
\delta_\chi v &= \mathcal L_\xi v \,, \\
\delta_\chi a &= \mathcal L_\xi a + d\lambda\,.
\es 
The global symmetries of the system are those transformations $\chi = \{\xi, \lambda\}$ that leave the background geometry invariant,
\be 
\delta_\chi \left( g\,, \tau \,, v\,, a \right) = \left( 0\,, 0 \,, 0\,, 0 \right)\,.
\ee 
They form the stabilizer subgroup of the chosen background configuration.
The commutator of two transformations closes onto a transformation of the same form,
\be 
\delta_{\chi_2} \chi_1 = [\chi_1, \chi_2] = \{ \mathcal L_{\xi_1} \xi_2,  \mathcal L_{\xi_1} \lambda_2 - \mathcal L_{\xi_2} \lambda_1 \} \,,
\ee 
with the induced bracket given by 
\be \label{eq:lieAlgebroid}
[\delta_{\chi_1}, \delta_{\chi_2}] = \delta_{[\chi_1, \chi_2]}\,.
\ee 
The resulting structure forms an infinite-dimensional Lie algebroid, reflecting the semidirect product of spacetime diffeomorphisms with $\mathrm U(1)$ gauge transformations.

\paragraph{Isometries.} We consider fluids propagating on the flat Aristotelian background in the absence of external electromagnetic fields. The background data are 
\be \label{eq:flatAristotelian}
\left( g\,, \tau \,, v\,, a \right) = \left( \delta_{ij} dx^i \otimes dx^j \,, dt \,, \partial_t\,, 0\right)\,.
\ee 
We now determine explicitly the stabilizer subgroup of the flat background Eq.~\eqref{eq:flatAristotelian}. The isometry variations $\chi = \{ \xi, \lambda\}$ satisfy
\bs
\mathcal L_\xi g = 0 
&\quad \Longrightarrow \quad 
\partial_i \xi_j + \partial_j \xi_i = 0 \,, \\
\mathcal L_\xi \tau = 0 
&\quad \Longrightarrow \quad 
\partial_\mu \xi^t = 0 \,, \\
\mathcal L_\xi v = 0 
&\quad \Longrightarrow \quad 
\partial_t \xi^\mu = 0 \,, \\
\mathcal L_\xi a + d\lambda = 0 
&\quad \Longrightarrow \quad 
\partial_\mu \lambda = 0 \,.
\es 
Thus, the most general Killing transformations of the flat Aristotelian spacetime can be expressed as 
\be \label{eq:flatIsometry}
\delta_\chi
=
 c_0\{\partial_t,\,0\}
+ c_i\{\partial_i, 0 \} + \theta  \{\epsilon_{ij}x_i\partial_j,\,0\}
+ \lambda\,\{0,1\}\,,
\ee
where $c_0, \theta,\lambda \in \mathbb R$ and $c_i \in \mathbb R^2$.
These transformations
generate time translations, spatial translations, rotations, and global $\mathrm U(1)$ shifts, respectively.

It is convenient to introduce the corresponding generators
\bs
H &= \{\partial_t, 0 \}\,, \\
P_i &= \{\partial_i, 0 \}\\
L &= \{\epsilon_{ij}x_i\partial_j, 0\}\,,\\
Q &= \{0,1\}\,.
\es
Using the Lie bracket defined in Eq.~\eqref{eq:lieAlgebroid}, one finds the non-vanishing
commutation relations
\be \label{eq:nonuniformAlgebraAristotelian}
[L,P_i]=\epsilon_{ij} P_j\,.
\ee
Hence, rotations are realized as a nonuniform symmetry.
%%%%%%%

%%%%%
\paragraph{The global symmetry group.} The associated finite transformations correspond to translations in time and space as well as rotations
\begin{align}
\text{Time translation:}\quad &(t,x_i)\ \mapsto\ (t+c_0,\ x_i)\,, \\
\text{Spatial translation:}\quad &(t, x_i)\ \mapsto\ (t,\ x_i + c_i)\,, \\
\text{Rotation:}\quad &(t, x_i)\ \mapsto\ (t,\ R_{ij}(\theta)\, x_j)\,,
\end{align}
where $c_0 \in\mathbb R$, $c_i\in\mathbb R^2$, and $R(\theta)\in \mathrm{SO}(2)$.

In total, the spacetime symmetry group is the $(2+1)$-dimensional Aristotelian group
\begin{equation}
G_{\mathrm{sp}} \simeq \mathbb{R}_t \times \big(\mathrm{SO}(2)\ltimes \mathbb{R}^2_x \big)\,,
\end{equation}
where $\mathrm{SO}(2)$ acts on $\mathbb{R}^2_x$ in the standard way.
The group $G_{\mathrm{sp}}$ acts transitively on $\mathcal M\simeq \mathbb R_t \times \mathbb R^2$, 
so that $\mathcal M$ is a homogeneous space.
Choosing a reference point $(t_0,\boldsymbol x_0)$, the stabilizer subgroup is
\begin{equation}
H \simeq \mathrm{SO}(2)\,,
\end{equation}
and therefore the manifold $\mathcal M$ can be identified with the coset space 
\begin{equation}
\mathcal M \simeq G_{\mathrm{sp}}/H\,.
\end{equation}
Including the internal symmetry generated by $Q$, the full global symmetry group factorizes as
\begin{equation}
G_{\mathrm{tot}} \simeq G_{\mathrm{sp}} \times \mathrm{U}(1)\,.
\end{equation}
The action of the spacetime symmetry group on the manifold $\mathcal M$ admits a faithful affine representation, which we summarize in Appendix~\ref{app:aristotelian}.

%%%%%%%%%%
%%%%%%%%%
%%%%%%%%%%%

\subsubsection{Hydrodynamics of Aristotelian fluids}
Having established the spacetime structure on which Aristotelian fluids propagate,
we now formulate the corresponding hydrodynamic theory. We begin with the conservation laws associated with global symmetries.
By Noether’s theorem, each continuous global symmetry gives rise to a conserved current, whose integral defines the corresponding charge,
\be
Q = \int \star j \,, \quad
P_i = \int \star \tau_i \,, \quad
H = \int \star \varepsilon \,, \quad
L = \int  \star l \,.
\ee
The conservation of the respective charges then follows from the local form of the continuity equations  
\be 
d \star j = 0\,, \quad d \star \tau_i =0 \,, \quad d \star \varepsilon = 0\,, \quad d \star l = 0\,.
\ee 
The conserved charges obey the algebraic relations Eq.~\eqref{eq:nonuniformAlgebraAristotelian}, which implies that the angular momentum current transforms inhomogeneously under spatial translations 
\be 
[P_i, l] = -\partial_i l - \epsilon_{ij} \tau_j\,,
\ee 
reflecting the nonuniform character of the rotational symmetry.

We can therefore isolate the uniform part of the angular momentum, which in the literature is known as the intrinsic spin density and the orbital part 
\be 
l = \sigma - \epsilon_{ij} x_i \tau_j\,.
\ee 
The uniform spin density current satisfies the local equation 
\be 
d \star \sigma = \epsilon_{ij} dx_i \wedge \star \tau_j \,.
\ee 
%%%%%%
We find it helpful to express the hydrodynamic equations in component form 
\be \begin{split} \label{eq:eomsAristotelian}
\partial_t j^0 + \partial_i j_i &= 0\,, \\
\partial_t \varepsilon^0 + \partial_i \varepsilon_i &= 0\,, \\
\partial_t \tau^0_i + \partial_j \tau_{ji} &=0\,, \\
\partial_t \sigma^0 + \partial_i \sigma_i &= \hat \tau \,,
\end{split}
\ee  
where 
\be
\hat \tau \equiv \epsilon_{ij} \tau_{ij}\,,
\ee 
is the pseudoscalar contraction of the antisymmetric part of the stress tensor, which acts as a source for the spin current.

\paragraph{Entropy production.}
The entropy density must be invariant under the global symmetries. 
In particular, invariance under spatial translations implies that it can depend on the angular momentum density $l^0$ only via its uniform component $\sigma^0$. 
We therefore take the entropy density to be a function\footnote{The inclusion of spin density in the entropy has been proposed in the context of spin hydrodynamics \cite{Hattori:2019lfp,Hongo:2021ona}.}
\begin{equation}
s\equiv s\big(j^0,\varepsilon^0,\tau_i^0,\sigma^0\big)\,.
\end{equation}
Its differential can be written as
\begin{equation}
ds
=
\frac{1}{T}\, d\varepsilon^0
-
\frac{\mu}{T}\, dj^0
-
\frac{v_i}{T}\, d\tau_i^0
-
\frac{\omega}{T}\, d\sigma^0 \,,
\end{equation}
where the thermodynamic conjugate variables are defined by
\be \label{eq:thermodynamicConjugate}
\frac{1}{T} \equiv \frac{\partial s}{\partial \varepsilon^0},
\qquad
\frac{\mu}{T} \equiv -\,\frac{\partial s}{\partial j^0},
\qquad
\frac{v_i}{T} \equiv -\,\frac{\partial s}{\partial \tau_i^0},
\qquad
\frac{\omega}{T} \equiv -\,\frac{\partial s}{\partial \sigma^0}.
\ee
Taking a time derivative of the entropy density and using the local conservation laws~\eqref{eq:eomsAristotelian}, we find
\begin{align}
\partial_t s
&=
\frac{1}{T} \partial_t \varepsilon^0
-\frac{\mu}{T} \partial_t j^0
-\frac{v_i}{T} \partial_t \tau_i^0
-\frac{\omega}{T} \partial_t \sigma^0 \nonumber\\
&=
-\frac{1}{T} \partial_i \varepsilon_i
+\frac{\mu}{T} \partial_i j_i
+\frac{v_i}{T} \partial_j \tau_{ji}
+\frac{\omega}{T} \partial_i \sigma_i
-\frac{\omega}{T} \hat \tau \,.
\end{align}
Rearranging total derivatives, this can be written as an entropy balance equation
\begin{equation}\label{eq:entropyBalance}
\partial_t s + \partial_i s_i = \Delta \,,
\end{equation}
with the canonical entropy current
\begin{equation}\label{eq:entropyCurrentCanonical}
s_i \equiv \frac{1}{T} \left( p v_i + \varepsilon_i
-\mu j_i
-v_j \tau_{ij}
-\omega \sigma_i \right) \,,
\end{equation}
and entropy production rate
\bs\label{eq:entropyProduction}
\Delta
&=
\left( \varepsilon_i -(p+\varepsilon^0) v_i \right) \partial_i\!\left(\frac{1}{T}\right)
- \left( j_i - j^0 v_i \right) \partial_i\!\left(\frac{\mu}{T}\right)
 - \left( \sigma_i - \sigma^0 v_i \right) \partial_i\!\left(\frac{\omega}{T}\right) \\
 &- \left( \tau_{ij} - p \delta_{ij} - v_i \tau^0_j \right)    \partial_i\! \left( \frac{v_j}{T} \right) -  \hat \tau \frac{\omega}{T}
 \,.
\es
In the above, we have introduced the pressure through the local Euler relation
\be
p + \varepsilon^0 = T s + \mu j^0 + v_i \tau_i^0 + \omega \sigma^0 \,.
\ee
%%%%%%
To determine the ideal constitutive relations we set $\Delta = 0$, which yields
\bs \label{eq:idealConstitutive}
\varepsilon_i &= (\varepsilon^0 + p)\, v_i + \tilde{\varepsilon}_i \,, \\
j_i &= j^0 \, v_i + \tilde{j}_i \,, \\
\sigma_i &= \sigma^0 \, v_i + \tilde{\sigma}_i \,, \\
\tau_{ij} &= p\,\delta_{ij} + v_i \tau^0_j + \tilde{\tau}_{ij} \,,
\es
where $\tilde{\varepsilon}_i$, $\tilde{j}_i$, $\tilde{\sigma}_i$, and $\tilde{\tau}_{ij}$ denote dissipative corrections.

%%%%%%%%%%%%%%
%%%%%%%%%%%%%%%
%%%%%%%%%%%%%%%%
It is useful to decompose the dissipative stress tensor into irreducible representations of $SO(2)$. 
In two spatial dimensions a rank-two tensor decomposes into a symmetric traceless part $\bar{\tau}_{ij}$, a scalar trace $\tilde{\tau}$, and a pseudoscalar antisymmetric component $\hat{\tau}$, as follows
\be
\tilde{\tau}_{ij}
=
\bar{\tau}_{ij}
+
\tilde{\tau} \delta_{ij}
+
\frac12 \epsilon_{ij}\hat{\tau}\,,
\ee
where
\be
\bar{\tau}_{ij}
\equiv
\frac12(\tilde{\tau}_{ij}+\tilde{\tau}_{ji})
-
\tilde{\tau}\,\delta_{ij},
\qquad
\tilde{\tau}
\equiv
\frac12 \tilde{\tau}_{ii}\,.
\ee
In addition, in two spatial dimensions it is important to distinguish between vectors and pseudovectors. In particular,
\be
\partial_i\!\left(\frac{1}{T}\right)\,, \quad
\partial_i\!\left(\frac{\mu}{T}\right)
\ee
transform as vectors, whereas
\be
\partial_i\!\left(\frac{\omega}{T}\right)
\ee
transforms as a pseudovector since $\omega$ is a pseudoscalar. 

The entropy production can therefore be decomposed into irreducible representations of $SO(2)\times \mathbb{Z}_2^{\rm parity}$ as
\be\label{eq:entropyProductionDecomp}
\Delta
= \Delta_{\rm vec} + \Delta_{\rm ps\text{-}vec} + \Delta_{\rm ten} + \Delta_{\rm sca} + \Delta_{\rm ps\text{-}sca}\,,
\ee
where
\bs
\Delta_{\rm vec} &\equiv 
\tilde{\varepsilon}_i \partial_i\!\left(\frac{1}{T}\right)
- \tilde{j}_i \partial_i\!\left(\frac{\mu}{T}\right)\,, \\
\Delta_{\rm ps\text{-}vec} &\equiv 
- \tilde{\sigma}_i \partial_i\!\left(\frac{\omega}{T}\right)\,, \\
\Delta_{\rm ten} &\equiv 
- \bar{\tau}_{ij}\,\partial_i\!\left(\frac{v_j}{T}\right)\,, \\
\Delta_{\rm sca} &\equiv 
- \tilde{\tau}\,\partial_i\!\left(\frac{v_i}{T}\right)\,, \\
\Delta_{\rm ps\text{-}sca} &\equiv 
- \hat{\tau}\!\left(\Omega+\frac{\omega}{T}\right),
\es
and we have defined the fluid vorticity
\be
\Omega=\frac12\,\epsilon^{ij}\partial_i\!\left(\frac{v_j}{T}\right)\,.
\ee
Since the entropy production is a scalar under spatial rotations and parity, quadratic terms built from objects transforming in inequivalent irreducible representations of $SO(2)\times \mathbb{Z}_2^{\rm parity}$ cannot mix. 
Positivity of the entropy production therefore requires that each sector define a non-negative quadratic form independently,
\be \label{eq:positivityConstraints}
\Delta_{\rm vec} \ge 0\,, \quad
\Delta_{\rm ps\text{-}vec} \ge 0\,, \quad
\Delta_{\rm ten} \ge 0\,, \quad
\Delta_{\rm sca} \ge 0\,, \quad
\Delta_{\rm ps} \ge 0\,,
\ee
for arbitrary fluid flows.

This determines the allowed form of the constitutive relations.
The vector sector takes the form\footnote{In writing the constitutive relations in the vector sector we have imposed the Onsager reciprocity relations \cite{PhysRev.37.405,PhysRev.38.2265}.}
\bs \label{eq:constitutiveRelsVec}
\begin{pmatrix}
\tilde \varepsilon_i \\
\tilde j_i
\end{pmatrix}
&=
\begin{pmatrix}
\kappa & \alpha \\
\alpha & \lambda
\end{pmatrix}
\begin{pmatrix}
\partial_i\!\left(\frac{1}{T}\right) \\
-\partial_i\!\left(\frac{\mu}{T}\right)
\end{pmatrix}\,,
\es
while the pseudovector sector gives
\be \label{eq:constitutiveRelsSpinCurrent}
\tilde \sigma_i
=
-\theta \partial_i\!\left(\frac{\omega}{T}\right)\,.
\ee
The dissipative stress components are
\bs \label{eq:constitutiveRelsStress}
\bar \tau_{ij} &=
-  \frac{\eta}{2}
\left(
\partial_i \frac{v_j}{T}
+
\partial_j \frac{v_i}{T}
-
\delta_{ij}\partial_k \frac{v_k}{T}
\right)\,, \\
\tilde \tau &= - \zeta\,\partial_i \frac{v_i}{T}\,, \\
\hat \tau &= - \beta \left( \Omega + \frac{\omega}{T} \right)\,.
\es
Demanding Eq.~\eqref{eq:positivityConstraints} for arbitrary fluid flows implies
\be
\kappa \ge 0\,, \quad
\lambda \ge 0\,,\quad
\theta \ge 0\,,\quad
\eta \ge 0\,,\quad
\zeta \ge 0\,,\quad
\beta \ge 0\,,
\ee
together with
\be
\kappa\lambda-\alpha^2 \ge 0\,.
\ee

%%%%%%%%%%%%
%%%%%%%%%%%%%%%%%%%
%%%%%%%%%%

% --- Box ---
%%
%
\begin{scipostbox}[A remark on spin density in equilibrium] 
Notice that the combination
\be 
\Omega + \frac{\omega}{T}
\ee
enters as a thermodynamic force in the entropy production. 
In hydrostatic equilibrium, there is no entropy production, and therefore all such forces vanish. It follows that 
\be 
\Omega + \frac{\omega}{T} = 0 \,.
\ee
Thus, the spin chemical potential dynamically adjusts to compensate the local fluid vorticity.  
In particular, for a non-rotating fluid with $\Omega = 0$, one finds 
\be
\omega = 0 \,,
\ee
so that the equilibrium spin density vanishes in the absence of rotation.
\end{scipostbox}
% --- Box ---
%%
%
%
%%%%%%%
%%%%%%

\paragraph{Spectrum of linearized excitations.}
Having established the hydrodynamic equations and constitutive relations, we now analyze the spectrum of small fluctuations around equilibrium. In Aristotelian fluids, states with different background velocity fields are not related by symmetry and therefore correspond to genuinely distinct equilibrium configurations. 
As a result, they may exhibit different spectra of linear excitations and qualitatively different long-wavelength dynamics \cite{deBoer:2017ing}. 
In the following, we focus on the simplest homogeneous equilibrium state with vanishing velocity field, for which the equilibrium spin chemical potential also vanishes.

For the remainder of this section, we denote the conserved densities by
\be
n \equiv j^0 , \qquad
\varepsilon \equiv \varepsilon^0 , \qquad
\tau_i \equiv \tau_i^0 , \qquad
\sigma \equiv \sigma^0 \,,
\ee
and consider small fluctuations around a homogeneous and stationary equilibrium of the form
\be
n = n_0 + \delta n , \qquad
\varepsilon = \varepsilon_0 + \delta \varepsilon , \qquad
\tau_i = \delta \tau_i , \qquad
\sigma = \delta \sigma \,.
\ee
Expanding the entropy density around equilibrium to quadratic order gives
\be
s =
s_0
+
\frac{1}{T_0} \delta \varepsilon
-
\frac{\mu_0}{T_0} \delta n
-
\frac12
\begin{pmatrix}
\delta \varepsilon & \delta n
\end{pmatrix}
\begin{pmatrix}
\chi_{\varepsilon\varepsilon} & \chi_{\varepsilon n} \\
\chi_{\varepsilon n} & \chi_{nn}
\end{pmatrix}
\begin{pmatrix}
\delta \varepsilon \\
\delta n
\end{pmatrix}
-
\frac{\chi_\tau}{2} \delta \tau_i^2
-
\frac{\chi_\sigma}{2} \delta \sigma^2 .
\ee
Here $T_0$ and $\mu_0$ denote the equilibrium temperature and chemical potential, while the coefficients 
\be 
\chi_{\varepsilon \varepsilon},\chi_{nn},\chi_\tau,\chi_\sigma \ge 0\,, \quad 
\chi_{\varepsilon \varepsilon} \chi_{nn} - \chi_{\varepsilon n}^2 \geq 0 
\ee 
are thermodynamic susceptibilities.

Using Eq.~\eqref{eq:thermodynamicConjugate}, the corresponding thermodynamic potentials admit the linear expansion
\bs
\frac{1}{T}
&=
\frac{1}{T_0}
-
\chi_{\varepsilon\varepsilon} \delta \varepsilon
-
\chi_{\varepsilon n} \delta n \,, \\
\frac{\mu}{T}
&=
\frac{\mu_0}{T_0}
+
\chi_{\varepsilon n} \delta \varepsilon
+
\chi_{nn} \delta n \,, \\
\frac{v_i}{T} &= \chi_\tau \delta \tau_i \,, 
\qquad
\frac{\omega}{T} = \chi_\sigma\,\delta \sigma \,.
\es
Substituting these expressions into the constitutive relations \eqref{eq:idealConstitutive} with the dissipative terms \eqref{eq:constitutiveRelsVec}, we obtain the linearized energy and charge currents
\be
\begin{pmatrix}
\delta \varepsilon_i \\
\delta j_i
\end{pmatrix}
=
T_0 \chi_\tau
\begin{pmatrix}
\varepsilon_0 + p_0 \\
n_0
\end{pmatrix}
\delta \tau_i
-
\begin{pmatrix}
D_{\varepsilon\varepsilon} & D_{\varepsilon n} \\
D_{n\varepsilon} & D_{nn}
\end{pmatrix}
\begin{pmatrix}
\partial_i \delta \varepsilon \\
\partial_i \delta n
\end{pmatrix}\,,
\ee
where we introduced the diffusion matrix
\be
\begin{pmatrix}
D_{\varepsilon\varepsilon} & D_{\varepsilon n} \\
D_{n\varepsilon} & D_{nn}
\end{pmatrix}
\equiv
\begin{pmatrix}
\kappa & \alpha \\
\alpha & \lambda
\end{pmatrix}
\begin{pmatrix}
\chi_{\varepsilon\varepsilon} & \chi_{\varepsilon n} \\
\chi_{\varepsilon n} & \chi_{nn}
\end{pmatrix} \,.
\ee
%%%%%%%%%%%
From Eq.~\eqref{eq:constitutiveRelsSpinCurrent}, we also obtain the linearized spin current 
\be
\delta \sigma_i
=
-\theta \chi_\sigma \partial_i \delta\sigma\,.
\ee
Finally, using the constitutive relation for the stress tensor \eqref{eq:idealConstitutive} with the dissipative contributions \eqref{eq:constitutiveRelsStress}, we obtain
\be
\delta \tau_{ij}
=
\left( \delta p + \delta \tilde \tau  \right) \delta_{ij}
+
\delta \bar\tau_{ij}
+
\frac12 \delta \hat \tau \epsilon_{ij}\,,
\ee
where the pressure fluctuation is
\be
\delta p
=
T_0
\begin{pmatrix}
\varepsilon_0 + p_0 & n_0
\end{pmatrix}
\begin{pmatrix}
\chi_{\varepsilon\varepsilon} & \chi_{\varepsilon n} \\
\chi_{\varepsilon n} & \chi_{nn}
\end{pmatrix}
\begin{pmatrix}
\delta \varepsilon \\
\delta n
\end{pmatrix}
=
\begin{pmatrix}
    c_\varepsilon & c_n 
\end{pmatrix} \begin{pmatrix}
\delta \varepsilon \\
\delta n
\end{pmatrix}\,,
\ee
with 
\be
c_\varepsilon \equiv \left( \frac{\partial p}{\partial \varepsilon} \right)_n\,, \quad 
c_n \equiv \left( \frac{\partial p}{\partial n} \right)_\varepsilon\,,
\ee
and the dissipative contributions are 
\bs
\delta \tilde \tau
&=
-\zeta\chi_\tau \partial_k\delta\tau_k \,, \\
\delta \bar\tau_{ij}
&=
-\frac{\eta\chi_\tau}{2}
\left(
\partial_i\delta\tau_j
+
\partial_j\delta\tau_i
-
\delta_{ij}\partial_k\delta\tau_k
\right)\,, \\
\delta \hat \tau
&=
-\beta
\left(
\frac{\chi_\tau}{2}\epsilon^{ij}\partial_i\delta\tau_j
+
\chi_\sigma \delta\sigma
\right) \,.
\es
After substituting the linearized constitutive relations into the hydrodynamic equations~\eqref{eq:eomsAristotelian} and passing to Fourier space, $\partial_t \mapsto -i\omega$ and $\partial_i \mapsto i k_i$, it is convenient to decompose the momentum density fluctuation into longitudinal and transverse components,
\be
\delta\tau_i = \hat k_i\,\delta\tau_L + \hat k_i^\perp\,\delta\tau_T\,,
\qquad
\hat k_i \equiv \frac{k_i}{k},
\qquad
\hat k_i^\perp \equiv \epsilon_{ij}\hat k_j \,.
\ee
The linearized equations then split into a longitudinal sector, involving $(\delta\varepsilon,\delta n,\delta\tau_L)$, and a transverse sector, involving $(\delta\sigma,\delta\tau_T)$.

The longitudinal fluctuations obey
\be
M_L(\omega,k)
\begin{pmatrix}
\delta\varepsilon\\
\delta n\\
\delta\tau_L
\end{pmatrix}
=0\,,
\ee
with
\be
M_L(\omega,k)=
\begin{pmatrix}
-i\omega+k^2 D_{\varepsilon\varepsilon} &
k^2 D_{\varepsilon n} &
ik\,T_0\chi_\tau(\varepsilon_0+p_0)
\\[0.5em]
k^2 D_{n\varepsilon} &
-i\omega+k^2 D_{nn} &
ik\,T_0\chi_\tau n_0
\\[0.5em]
ik\,c_\varepsilon &
ik\,c_n &
-i\omega+k^2\chi_\tau\left(\zeta+\frac{\eta}{2}\right)
\end{pmatrix}\,.
\ee

While the transverse and spin fluctuations obey
\be
M_T(\omega,k)
\begin{pmatrix}
\delta\sigma\\
\delta\tau_T
\end{pmatrix}
=0\,,
\ee
where
\be
M_T(\omega,k)=
\begin{pmatrix}
-i\omega+ \beta\chi_\sigma +\theta\chi_\sigma k^2 &
\frac{i}{2}\beta\chi_\tau k
\\[0.5em]
    \frac{i}{2}\beta\chi_\sigma k &
-i\omega+\frac{\chi_\tau}{2}(\eta-\frac{\beta}{2})k^2
\end{pmatrix} \,.
\ee
The hydrodynamic modes are determined by the conditions
\be
\det M_L(\omega,k) = 0\,, \quad 
\det M_T(\omega,k) = 0\,.
\ee
\paragraph{Transverse sector.} The transverse sector contains a shear diffusive mode
\be
\omega_\perp = - i D_\perp k^2 \,, \quad D_\perp = \frac{\eta \chi_\tau}{2}\,,
\ee
and a gapped spin-relaxation mode
\be 
\omega_{\rm spin} = - i \beta \chi_\sigma \,. 
\ee
At $k=0$, the two modes decouple and correspond to pure transverse momentum and spin fluctuations.

Notice that the gapped spin mode describes exponential relaxation of the spin density with a finite relaxation time 
\be
\tau_{\rm spin} = \frac{1}{\beta \chi_\sigma} \,.
\ee
At late times $t \gg \tau_{\rm spin}$, one can approximate 
\be
\partial_t + \frac{1}{\tau_{\rm spin}} \simeq \frac{1}{\tau_{\rm spin}} \,,
\ee 
in the spin equation.
In this regime, the spin sector ceases to be dynamical and instead imposes the algebraic constraint
\be
\hat \tau = \partial_i \sigma_i \,,
\ee
which enforces the symmetry of the stress tensor $\tau_{ij} = \tau_{ji}$.\footnote{Strictly speaking, this holds up to higher-derivative terms, which can be absorbed by a redefinition of the currents. The stress tensor can therefore be chosen symmetric at all orders in the derivative expansion.} Therefore, in the hydrodynamic limit of long times and long wavelengths, rotational invariance is implemented as a constraint on the stress tensor rather than as an independent dynamical equation. This illustrates a general feature of nonuniform symmetries in shaping hydrodynamic transport, namely that they do not introduce additional gapless modes but instead constrain the conserved currents associated with uniform charges.

\paragraph{Longitudinal sector.} In the longitudinal sector, we find two propagating sound modes
\be 
\omega_\pm = \pm v_s k - i \Gamma k^2 \,, 
\ee 
where 
\bs 
v_s^2 &= T_0 \chi_\tau \begin{pmatrix}
    c_\varepsilon & c_n 
\end{pmatrix} \begin{pmatrix}
    \varepsilon_0 + p_0 \\
    n_0
\end{pmatrix} \,, \\ 
\Gamma &= \frac{T_0^2 \chi_\tau^2}{2v_s^2} \begin{pmatrix}
    c_\varepsilon & c_n 
\end{pmatrix} \begin{pmatrix}
D_{\varepsilon\varepsilon} & D_{\varepsilon n} \\
D_{n\varepsilon} & D_{nn}
\end{pmatrix} \begin{pmatrix}
    \varepsilon_0 + p_0 \\
    n_0
\end{pmatrix} + \frac{1}{4} \left( 2 \zeta + \eta \right) \chi_\tau \,.
\es
In addition to the two propagating sound modes, the longitudinal sector contains also a purely diffusive mode
\be
\omega_{\rm diff} = - i D_{\rm diff} k^2 \,.
\ee
Its diffusion constant is given by
\be
D_{\rm diff}
=
\frac{T_0 \chi_\tau}{v_s^2} \begin{pmatrix}
    c_\varepsilon & c_n 
\end{pmatrix} \begin{pmatrix}
D_{nn} & -D_{\varepsilon n} \\
-D_{n\varepsilon} & D_{\varepsilon\varepsilon}
\end{pmatrix} \begin{pmatrix}
    \varepsilon_0 + p_0 \\
    n_0
\end{pmatrix} \,.
\ee
This mode corresponds to a diffusive linear combination of energy and charge densities orthogonal to pressure fluctuations, and therefore decouples from sound at leading order in the gradient expansion.

%%%%%%%%%
%%%%%%%%%%%
%%%%%%%%%%
\subsubsection{Aristotelian fluids in an electromagnetic background}
In this section we consider Aristotelian fluids propagating on nontrivial electromagnetic backgrounds, focusing on constant magnetic and electric fields. These backgrounds modify the global symmetry structure, giving rise to additional nonuniform symmetries and altering the hydrodynamic equations.

Rather than pursuing a detailed classification of transport, we focus on the symmetry structure and its hydrodynamic implications. In particular, we identify the symmetry algebra, derive the corresponding hydrodynamic equations, and extract general features of the resulting dynamics. A complete hydrodynamic analysis is conceptually straightforward, albeit technically involved.

\paragraph{Constant magnetic field.}
We begin by considering Aristotelian fluids in a homogeneous background magnetic field,
\be
b = \epsilon_{ij}\partial_i a_j \,.
\ee
We fix the gauge by choosing the rotationally invariant vector potential
\be
a_i = -\frac{b}{2}\,\epsilon_{ij}x_j \,,
\ee
so that the background geometric data takes the form
\be \label{eq:magneticBackground}
\left(g_{ij},\,\tau,\,v,\,a\right)
=
\left(\delta_{ij},\,dt,\,\partial_t,\,-\frac{b}{2}\epsilon_{ij}x_j\right)\,.
\ee
The most general infinitesimal transformation $\chi=\{\xi,\lambda\}$ preserving the geometric background Eq.~\eqref{eq:magneticBackground} is
\be \label{eq:magneticIsometry}
\delta_\chi
=
 c_0\,\{\partial_t,\,0\}
+ c_i\,\{\partial_i,\,\tfrac{b}{2}\epsilon_{ij}x_j\} + \theta\,\{\epsilon_{ij}x_i\partial_j,\,0\}
+ \lambda\,\{0,1\}\,,
\ee
where $c_0, \theta,\lambda \in \mathbb R$ and $c_i \in \mathbb R^2$ are constant parameters. The individual terms in Eq.~\eqref{eq:magneticIsometry} correspond to time translations, magnetic translations, rotations, and global $\mathrm U(1)$ transformations, respectively. Compared to the flat background Eq.~\eqref{eq:flatIsometry}, spatial translations are replaced by \textit{magnetic translations}, reflecting the presence of a nontrivial gauge field.

Introducing the generators 
\bs
H &= \{\partial_t, 0 \}\,, \\
P_i &= \{\partial_i,\,\tfrac{b}{2}\epsilon_{ij}x_j\}\,,\\
L &= \{\epsilon_{ij}x_i\partial_j,\,0\}\,,\\
Q &= \{0,1\}\,,
\es
and using the Lie bracket defined in Eq.~\eqref{eq:lieAlgebroid}, one finds the non-vanishing commutation relations
\be \label{eq:magneticAlgebra}
[L,P_i]=\epsilon_{ij}P_j\,,\qquad
[P_i, P_j]=b \epsilon_{ij} Q\,.
\ee
The central extension proportional to $Q$ reflects the noncommutativity of spatial translations in the presence of a magnetic field. It provides an interesting example of a nonuniform symmetry in which spatial translations themselves become nonuniform.

To each one of the conserved charges we again associate a hydrodynamic current,
\be 
 Q = \int \star j \,, \quad P_i = \int \star \tau_i \,, \quad  
 H = \int \star \varepsilon \,, \quad
 L = \int \star  l \,,
\ee  
satisfying their respective local conservation equations
\bs
d \star j &= 0 \,, \quad d \star \tau_i = 0 \,, \quad d \star \varepsilon = 0 \,, \quad d \star l = 0\,.
\es
The nonuniform structure of the algebra, Eq.~\eqref{eq:magneticAlgebra}, implies the existence of inhomogeneous currents that contain explicit dependence on coordinates. In particular, the inhomogeneous currents transform as follows
\bs 
[P_i, l] &= -\partial_i l - \epsilon_{ij} \tau_j \,, \\
[ P_i, \tau_j ] &= - \partial_i \tau_j + \epsilon_{ij} b j\,.
\es 
Therefore, we have the following  algebraic relations between the hydrodynamic currents
\bs
\tau_i &= \pi_i - b  \epsilon_{ij} x_j j\,, \\
l &= \sigma - \epsilon_{ij} x_i \tau_j + \tfrac{b}{2} x^2 j \,,\\
&= \sigma - \epsilon_{ij} x_i \pi_j - \tfrac{b}{2} x^2 j\,,
\es 
where currents $\sigma$ and $\pi_i$ are homogeneous but not conserved. They satisfy the following local equations
\bs 
d \star \pi_i &= b \epsilon_{ij} dx_j \wedge \star j\,, \\
d \star \sigma &= \epsilon_{ij} dx_i \wedge \star \pi_j + b x_i dx_i \wedge \star j\,.
\es 
The right-hand sides encode the Lorentz force exerted by the magnetic field on the charged fluid and the associated torque, which act as sources for the homogeneous momentum and spin currents. In particular, the first equation shows that the homogeneous momentum current is no longer conserved in the magnetic background, reflecting the nonuniform nature of spatial translations. Therefore, the momentum sector does not give rise to an independent gapless hydrodynamic mode, but instead participates in magnetically gapped collective excitations.

 \begin{scipostbox}[Emergent conservation of dipole moment in a strong magnetic field]
Notice that we can express the conserved magnetic momentum as 
\be 
P_i = \Pi_i - b \epsilon_{ij} D_j\,,
\ee 
where 
\be 
\Pi_i = \int \star \pi_i\,, \quad D_i  = \int  x_i \star j\,,
\ee 
represent the invariant momentum and dipole moment, respectively. A priori, neither $\Pi_i$ nor $D_i$ are conserved. In the limit of vanishing magnetic field, $b \rightarrow 0$, the conserved momentum becomes homogeneous and the dipole contribution drops out, $P_i \rightarrow \Pi_i$. Conversely, in the limit of large magnetic field, $b \rightarrow \infty$, the contribution from the dipole sector dominates, $P_i \rightarrow - b \epsilon_{ij} D_j$, leading to an emergent conservation of dipole moment.
\end{scipostbox}
% --- Box ---
%%
%
%
Finally, in the presence of a magnetic field the equilibrium state need not have vanishing spin density. Since both $b$ and $\sigma$ are pseudoscalars, a term proportional to $b \sigma$ is allowed in the thermodynamic potential. As a result, even a homogeneous and static equilibrium state generically carries a finite spin density, corresponding to a magnetic polarization of the fluid.

\paragraph{Constant electric field.} We now consider a homogeneous background with a constant electric field $e_i$. We choose a gauge with $a_0 = - e_i x_i$, so that the background is
\be
\left( g_{ij}\,, \tau \,, v\,, a \right) = \left( \delta_{ij}\,, dt \,, \partial_t\,,  -e_i x_i dt \right)\,.
\ee 
The vector space of isometries consists of transformations of the form
\be
\delta_\chi
=
 c_0 \{\partial_t,\,0\}
+ c_i \{\partial_i, t e_i \} 
+ \lambda \{0,1\}\,,
\ee
generated by
\bs
H &= \{\partial_t, 0 \}\,, \\
P_i &= \{\partial_i, t e_i \} \,,\\
Q &= \{0,1\}\,,
\es
corresponding to time translations, modified ``electric'' spatial translations and global $\mathrm U(1)$ shifts. 
These generators satisfy the nonuniform algebra
\be \label{eq:electricAlgebra}
[H, P_i]= e_i  Q\,.
\ee
In the presence of an electric field, rotational symmetry is explicitly broken by the external vector field, while spatial translations are realized nontrivially, generating time-dependent phase shifts as encoded in the algebra Eq.~\eqref{eq:electricAlgebra}.

The nonuniform algebra implies the existence of inhomogeneous momentum currents. In particular, the action of time translations on the momentum current takes the form
\be
[H, \tau_i] = -\partial_t \tau_i + e_i j\,,
\ee
which leads to the relation between the conserved and homogeneous momentum currents
\be
\tau_i = \pi_i - t e_i j\,,
\ee
where $\pi_i$ is homogeneous but not conserved. The corresponding local equation for $\pi_i$ follows from the algebra and reads
\be
d \star \pi_i = e_i dt \wedge \star j\,,
\ee
which encodes the force exerted by the electric field on the charged fluid. Thus, the homogeneous momentum current is no longer conserved in the presence of an electric field, reflecting the nonuniform realization of spatial translations. 

However, this effect is anisotropic. Only the component of $\pi_i$ parallel to $e_i$ is sourced by the electric field, while the transverse component remains unsourced at this level and can still participate in the gapless hydrodynamic dynamics. This describes an anisotropic lifting of the momentum sector induced by the external electric field.

\section{Fracton phases of matter}\label{sec:fracton}

In this section we study a class of phases of matter hosting \emph{fractons}, 
quasiparticle excitations with restricted mobility. 
Unlike conventional particles, isolated fractons are either completely immobile 
or restricted to move along lower-dimensional submanifolds, while suitable 
bound states may propagate collectively. In line with the perspective developed in this thesis, we focus on realizations of fracton phenomenology in systems with nonuniform symmetries that constrain the dynamics of elementary excitations.

We begin with a brief introduction to fracton physics and its relation to 
nonuniform symmetries, including both subsystem and multipole symmetries. Then, we construct the dipole sigma model, which describes a zero-temperature 
fluid phase with spontaneously broken monopole and dipole symmetries. This model can be viewed as a natural generalization of the conventional zero-temperature $\mathrm U(1)$ superfluid to a system with nonuniform symmetry. We develop an analogue of the charge--vortex duality for this phase, 
showing that the zero-temperature theory can be mapped onto an emergent 
higher-rank gauge theory, in which superfluid vortices play the role of 
gauge charges.

Finally, we consider the finite-temperature dynamics of systems with 
both monopole and dipole symmetry, providing an explicit example of 
transport in the presence of nonuniform symmetries. We show that dipole 
conservation qualitatively modifies the hydrodynamic universality class, 
leading to subdiffusive relaxation of charge density modulations. We also discuss a physical realization 
of this phenomenon in tilted Bose--Hubbard chains, where the dipole 
conservation emerges dynamically at long times.

\subsection{Introduction to fractons}
Fracton phases were originally discovered in exactly solvable lattice models, 
such as the X-cube model and Haah's code \cite{PhysRevA.83.042330,PhysRevB.94.235157} (see also Refs.~\cite{Nandkishore:2018sel,Pretko:2020cko} for reviews). These models exhibit unconventional forms of order and restricted mobility of excitations. In many important cases, including the X-cube model, these features can be understood as a consequence of \emph{subsystem symmetries}. Unlike ordinary global symmetries, which act on the entire system, subsystem symmetries act on lower-dimensional submanifolds, such as lines or planes \cite{Williamson:2016jiq,You:2018oai,Devakul:2018fhz,Stephen:2019zas,10.21468/SciPostPhys.6.4.041}. In this sense, they share certain features of both 
global and gauge symmetries.\footnote{
This analogy is only heuristic as gauge ``symmetries'' are merely redundancies of the 
description rather than physical symmetries, whereas subsystem symmetries are formally still global symmetries that act nontrivially on the physical Hilbert space.
}

\begin{scipostbox}[Subsystem symmetry in a continuum field theory]
A simple continuum realization of subsystem symmetry is provided by a scalar 
field $\varphi$ in $2+1$ dimensions with Lagrangian
\be 
\mathcal{L} = \frac{1}{2} (\partial_t \varphi)^2 
- \frac{\kappa}{2} (\partial_x \partial_y \varphi)^2 \,,
\ee 
which arises upon coarse-graining the $2+1$d $XY$-plaquette model \cite{Gorantla:2021bda}. This theory is invariant under transformations of the form
\be
\varphi(x,y,t) \;\rightarrow\; \varphi(x,y,t) + f(x) + g(y)\,,
\ee 
where $f$ and $g$ are arbitrary functions. These transformations act 
independently on lines of constant $x$ or $y$, and therefore constitute 
subsystem symmetries. Importantly, these symmetries do not commute with spatial translations. 
For instance, under a translation generated by $P_x$, one finds
\be 
[P_x, \delta_f] \varphi = -\delta_{f'} \varphi \,, \quad f'(x) = \partial_x f(x)\,,
\ee
demonstrating that the symmetry transformation depends explicitly on 
spatial coordinates. Therefore, subsystem symmetries are nonuniform symmetries.
\end{scipostbox}
%%%%%%%%%%%
When present, subsystem symmetries give rise to a number of unusual 
phenomena. In particular, many fracton models exhibit an extensive 
ground-state degeneracy, making them promising candidates for robust 
quantum information storage, with the number of encoded degrees of 
freedom scaling with system size \cite{PhysRevB.94.235157}. Another characteristic feature of systems with subsystem symmetries is the presence of UV/IR mixing, whereby the IR dynamics depends sensitively on the UV, challenging conventional local effective field theory descriptions of such lattice models~\cite{Gorantla:2021bda,Seiberg:2020bhn,Casasola:2023tot}.

While subsystem symmetries provide a promising microscopic realization of fracton phases, we will not pursue them further in this thesis. Instead, we focus on a simpler class of nonuniform symmetries, namely multipole symmetries \cite{PhysRevX.9.031035}, corresponding to the conservation of higher spatial moments of charge.

Multipole symmetries impose strong kinematic constraints on the dynamics, which can be treated analytically within effective field theory. They capture the essential kinematic features of subsystem symmetries, including constrained dynamics and fractonic behavior \cite{Pretko:2016kxt,Pretkoo_2017}, and can give rise, in appropriate microscopic realizations, to phenomena such as Hilbert space fragmentation and anomalously slow thermalization \cite{Sala:2019zru,PhysRevB.101.174204,Kohlert:2023rdh,Lydzba:2024zml}.
%%%%%%%%%%%
%%%%%%%%%%%%
%%%%%%%%%%%

%%%%%%%%%%%%%%
%%%%%%%%%%%%%%
%%%%%%%%%%%%%%
\subsection{Fractons from dipole conservation}

A simple route to realize fractons is to impose, in addition 
to charge conservation, the conservation of total dipole moment. 
This idea, originally emphasized by Pretko \cite{Pretko:2016kxt}, provides a minimal mechanism 
for generating restricted mobility. It can be understood already at the 
level of classical particle systems.

Consider an ensemble of classical particles with phase space variables
\be
\{ x^i_{(n)}, p^j_{(m)} \} = \delta_{nm} \delta^{ij}\,,
\ee
where $n,m$ label the particles. We define the total charge, dipole moment,
and momentum as
\be
Q = \sum_{n=1}^{N} q_{(n)}\,, \quad
D^i = \sum_{n=1}^{N} q_{(n)} x_{(n)}^i\,, \quad
P^i = \sum_{n=1}^{N} p_{(n)}^i\,.
\ee
A straightforward computation yields the algebra
\be
\{ D^i, P^j \} = \delta^{ij} Q\,,
\ee
which reflects the nonuniform nature of dipole symmetry.

Assuming that the particle charges $q_{(n)}$ are time-independent, conservation of dipole moment implies
\be
\dot D^i = \sum_{n=1}^{N} q_{(n)} \dot x_{(n)}^i = 0\,,
\ee
which imposes strong constraints on the allowed motion of the particles.

In particular, consider attempting to move a single particle while keeping all others fixed. In that case,
\be
\dot D^i = q \dot x^i\,.
\ee
For a particle carrying nonzero charge $q \neq 0$, dipole conservation requires
\be
q \dot x^i = 0 \quad \Longrightarrow  \quad \dot x^i = 0\,.
\ee
Therefore, a charged particle cannot move independently.

By contrast, neutral bound states can propagate without violating dipole conservation.
For example, consider a dipole state consisting of two particles with charges
\be
q_{(1)} = +q\,, \quad q_{(2)} = -q\,.
\ee
Then
\be
D^i = q \bigl(x_{(1)}^i - x_{(2)}^i\bigr)\,,
\ee
so that
\be
\dot D^i = q \bigl(\dot x_{(1)}^i - \dot x_{(2)}^i\bigr) = 0\,.
\ee
This condition is satisfied provided the particles move collectively,
\be
\dot x_{(1)}^i = \dot x_{(2)}^i\,,
\ee
in which case the relative separation remains fixed, so that the internal structure of the dipole is preserved, while the bound state can propagate as a whole.

This provides a simple realization of fractonic behavior where charged excitations are immobile, but neutral composites can propagate freely.

\subsection{Dipole sigma model}\label{sec:dipole-sigma-model}

\textit{This section is partially based on Ref.~\cite{Glodkowski:2024ova}.} \newline 

\noindent In this section, we construct the effective zero-temperature theory for systems with conserved charge and dipole moment in the superfluid phase where both monopole and dipole symmetries are spontaneously broken \cite{Chen_fractonic_2020,Stahl:2023prt,Glodkowski:2024ova}. The derivation is based on the nonlinear realization of the underlying nonuniform symmetry algebra established in Sec.~\ref{sec:nonuniformGoldstone}. We identify the invariant building blocks and formulate the most general low-energy effective field theory consistent with these symmetries. We then provide a hydrodynamic interpretation of the resulting theory. Finally, we study the spectrum of linearized excitations and present a dual formulation in terms of emergent higher-form gauge fields, leading to a dipole--charge--vortex duality.

\subsubsection{Goldstone theory and symmetry breaking}

We now construct the zero-temperature effective theory in the superfluid phase, where both the monopole charge $Q$ and the dipole generators $D_i$ are spontaneously broken. The relevant nonuniform symmetry algebra is
\be \label{eq:nonuniformDipoleAlgebra}
[ D_i, P_j ] = \delta_{ij} Q\,,
\ee
supplemented by the standard commutation relations with spatial rotations, which we leave implicit.

To derive the effective theory, we follow the general prescription for nonlinear realizations of nonuniform symmetries developed in Sec.~\ref{sec:nonuniformGoldstone}. To this aim we introduce the Goldstone fields $\theta$ and $\psi_i$ associated to spontaneously broken charges $Q$ and $D_i$, respectively, which obey
\be 
[\theta, Q] = 1\,, \quad [\psi_i, D_j] = \delta_{ij}\,.
\ee 
We work in the homogeneous basis where the corresponding gapless Goldstone field $\theta$ transforms homogeneously under spatial translations,
\be \label{eq:unifromMOmentnumTranss}
[\theta, P_i] = - \partial_i \theta\,, \quad [\psi_i, P_j] = - \partial_j \psi_i\,.
\ee
The nonuniform dipole algebra Eq.~\eqref{eq:nonuniformDipoleAlgebra} then implies that the homogeneous Goldstone $\theta$ transforms nontrivially under dipole symmetry,
\be
[\theta, D_i] = x_i\,.
\ee
It is then straightforward to identify the covariant derivatives as
\be
D_t \theta \equiv \partial_t \theta \,, \quad
D_i \theta \equiv \partial_i \theta - \psi_i\,, \quad D_t \psi_i \equiv \partial_t \psi_i \,, \quad D_i \psi_j \equiv \partial_i \psi_j \,.
\ee
These operators provide the invariant building blocks of the low-energy theory.

The most general zero-temperature effective Lagrangian therefore takes the form
\be \label{eq:effectiveDipoleSigma}
\mathcal L_{\mathrm{eff}} = \mathcal L_{\mathrm{eff}}
\bigl(
\partial_t \theta\,,\,
\partial_t \psi_i\,,\,
\partial_i \theta - \psi_i\,,\,
\partial_i \psi_j
\bigr)\,.
\ee
The Euler-Lagrange equations are
\begin{align}\label{eq:euler-lagrange}
\partial_t n + \partial_i J^i &= 0 \,, \\
\partial_t \pi_i + \partial_j K^{ji} &= - J^i
\end{align}
where we have introduced the following notation
\be
n = \frac{\partial \mathcal{L}_{\text{eff}}}{\partial (\partial_t \theta)}\,, \quad  J^i = \frac{\partial \mathcal{L}_{\text{eff}}}{\partial (\partial_i \theta)}\,, \quad \pi_i =  \frac{\partial \mathcal{L}_{\text{eff}}}{\partial (\partial_t \psi_i)}\,, \quad K^{ij} =  \frac{\partial \mathcal{L}_{\text{eff}}}{\partial (\partial_i \psi_j)}\,. 
\ee 
These equations encode charge conservation and local continuity equation for internal dipole density $\pi_i$ sourced by the charge current. 

\subsubsection{A hydrodynamic perspective}
It is convenient to switch to a Hamiltonian description via a Legendre transform, which will be useful for the hydrodynamic description at finite temperature developed later. We obtain
\begin{equation}
    h(n,\pi_i,v_i^s,\xi_{ij}) = n\partial_t\theta +\pi_i\partial_t\psi_i -\mathcal L_{\text{eff}}\,, 
\end{equation}
where we have identified the canonically conjugate momenta with the charge and dipole densities. 

It is then possible to interpret the Hamiltonian density as a microcanonical equation of state given as 
\be \label{eq:microcanonical}
\epsilon \equiv h(n\,, \pi_i\,, v^s_i\,, \xi_{ij})
\ee 
with an analogue of the local first law
\be 
d\epsilon = \mu dn + \mu_i d\pi_i + \lambda_i dv_i^s+F_{ij}d\xi_{ij} 
\ee 
where we have introduced the conjugate variables 
\be \label{eq:thermDefs}
\mu = \frac{\partial h}{\partial n}\,, \quad \mu_i = \frac{\partial h}{\partial \pi_i}\,, \quad \lambda_i = \frac{\partial h}{\partial v^s_i}\,, \quad F_{ij} = \frac{\partial h}{\partial \xi_{ij}}\,.
\ee 
In addition, the canonical Poisson bracket for this field theory reads 
\be \label{eq:canonical}
\{F, G\}_C  = \int d^d x  \Big[ \Big(\frac{\delta F}{\delta \theta} \frac{\delta G}{\delta n} - \frac{\delta F}{\delta n} \frac{\delta G}{\delta \theta} \Big)
+ \Big( \frac{\delta F}{\delta \psi_i} \frac{\delta G}{\delta \pi_i} - \frac{\delta F}{\delta \pi_i} \frac{\delta G}{\delta \psi_i} \Big)
\Big]\,.
\ee
Equations of motion, as well as conservation laws, can be obtained by computing time evolution using the canonical Poisson bracket \eqref{eq:canonical} with the Hamiltonian
\be 
\partial_t \mathcal F = \{\mathcal F, \mathcal{H}\}_C \,,
\ee 
where
\be
\mathcal{H}=\int d^d x \, h \,.
\ee 
In particular, this yields the time evolution of the invariant superfluid velocities 
\bs\label{eq:superfluidVelocities}
\partial_t  v^s_i  &=  \partial_i  \mu- \mu_i \,, \\
\partial_t \xi_{ij} &= \partial_i  \mu_j \,.
\es
Equivalently, using $\partial_t \theta = \mu$, the Josephson relations take the form
\be\begin{split}
    \label{eq:josephson}
\partial_t \theta &= \mu\,, \\
\partial_t v^s_i &= \partial_i  \mu- \mu_i \,.
\end{split}
\ee
On the other hand, the evolution of the densities $n$ and $\pi_i$ is governed by the same local equations as computed in Lagrangian formalism \eqref{eq:euler-lagrange} with the identification
\be 
J^i = - \lambda_i\,, \quad K^{ij} = - F_{ij}\,.
\ee 
Therefore, the full set of hydrodynamic equations governing the long-time dynamics of the zero-temperature Goldstone theory are given as 
\be \begin{split}
\partial_t \delta n - \partial_i \lambda_i &= 0\,, \\
\partial_t \pi_i - \partial_j F_{ji} -\lambda_i  &= 0\,,\\
\partial_t  v^s_i  +\mu_i    - \partial_i \mu    &= 0 \,, \\
\partial_t \xi_{ij} - \partial_i \mu_j &= 0\,.
\end{split}
\ee 
We can also define the momentum density as the conserved density associated with spatial translations,
\be 
p_i = - n \partial_i \theta - \pi_j \partial_i \psi_j 
\ee 
obtained via Noether’s procedure. The Poisson brackets follow from Eq.~\eqref{eq:canonical} and take the form
\be \begin{split}\label{eq:poisson}
\{ p_i(\textbf x)  \,, \theta(\textbf y) \}_C &=  \partial_i \theta \, \delta(\textbf x-\textbf y)\,, \\
\{ p_i(\textbf x) \,, \psi_j (\textbf y) \}_C &= \partial_i \psi_j \, \delta(\textbf x- \textbf y)\,, \\
\{ p_i(\textbf x) \,, n(\textbf y) \}_C &=  -n \, \partial_i \delta(\textbf x-\textbf y)\,, \\
\{ p_i(\textbf x) \,, \pi_j (\textbf y) \}_C &= - \pi_j   \, \partial_i \delta(\textbf x- \textbf y)\,.
\end{split}
\ee 
Notice that above relations are compatible with the identification of the total momentum of a system as the generator of space translations Eq.~\eqref{eq:unifromMOmentnumTranss}.

%%%%%%%%%%%%%%%%%%%%%%%%%%%%

%%%%%%%%%%%%%%%%%%%
\subsubsection{Linear fluctuations}
In order to study the hydrodynamic modes of the theory, we consider small perturbations around a homogeneous fluid configuration with $\lambda_i=0$, $F_{ij}=0$ and $\mu_i =0$. This regime corresponds to superfluids at rest and zero dipole chemical potential.

Expanding the Hamiltonian density around a homogeneous equilibrium up to the second order in perturbations 
\be \label{eq:hamiltonian}
h = h_0 + \mu_0 \delta n + \frac{1}{2} \chi_n \delta n^2 + \frac{1}{2} \chi_\pi \delta \pi_i^2 + \frac{1}{2} \chi_v ( v^s_i)^2  + \frac{1}{2} \lambda_1 \xi^2 + \frac{1}{2} \lambda_2  \xi_{\langle i j \rangle}^2 +\frac{1}{2} \lambda_3 \xi_{[ i j ]}^2 
\ee 
with $\delta n\,, \delta \pi_i\,, v^s_i$ and $\xi_{ij}$ regarded as small. The condition for the stability of a stationary solution implies that all susceptibilities are non-negative 
\be \label{eq:stability}
\chi_n\,, \chi_\pi\,, \chi_v\,, \lambda_1\,, \lambda_2\,, \lambda_3 \geq 0\,.
\ee
Using definitions \eqref{eq:thermDefs} we find the following identities
\be \begin{gathered}
\mu = \mu_0 + \chi_n \delta n\,, \quad \mu_i = \chi_\pi \delta \pi_i\,, \\
\lambda_i = \chi_v v^s_i\,, \quad F_{ij} =   \lambda_1 \xi \delta_{ij} +  \lambda_2  \xi_{\langle i j \rangle} + \lambda_3 \xi_{[ i j ]}\,.
\end{gathered}
\ee 
Substituting these into the equations of motion \eqref{eq:euler-lagrange} and \eqref{eq:superfluidVelocities} we obtain a set of linear differential equations
\be \begin{split}
\partial_t \delta n - \chi_v \partial_i v^s_i &= 0\,, \\
\partial_t \pi_i - \lambda_{1} \partial_i \xi - \lambda_{2} \partial_j \xi_{\langle j i \rangle} - \lambda_{3} \partial_j \xi_{[ j i ]} - \chi_v v^s_i &= 0\,,\\
\partial_t  v^s_i  - \chi_n \partial_i \delta n + \chi_\pi \delta  \pi_i   &= 0 \,, \\
\partial_t \xi_{ij} - \chi_\pi \partial_i \delta \pi_j &= 0\,.
\end{split}
\ee 
After performing a Fourier transformation and decomposing the variables into their parallel and perpendicular components: $\tilde \psi_i = \tilde \psi_{||} \hat k_i + \boldsymbol{\tilde \psi_{\perp}}$, $\tilde \pi_i = \tilde \pi_{||} \hat k_i + \boldsymbol{\tilde \pi_{\perp}}$ and $\tilde  v^s_i = \tilde v^s_{||} \hat k_i + \boldsymbol{\tilde v^s_{\perp}}$, the linearized equations of motion can be elegantly expressed in a block diagonal matrix form
\be
\begin{pmatrix}
\mathcal M^{4 \times 4}_{||}  & \mbox{\Large 0} \\ 
\mbox{\Large 0}  & \mathcal M^{3 \times 3}_{\perp} \\
\end{pmatrix} \begin{pmatrix}
\textbf{v}_{||}  \\ 
 \textbf{v}_{\perp}   \\
\end{pmatrix} = 0
\ee 
where
\be\begin{split}
\mathcal M^{3 \times 3}_{\perp} &= \begin{pmatrix}
    - i \omega & -\chi_v & k^2 \lambda_{\perp} \\
\chi_\pi & -i \omega & 0 \\
-\chi_\pi & 0 & -i \omega
\end{pmatrix}\,, \quad \textbf{v}_{\perp} = \begin{pmatrix}
\boldsymbol{\tilde \pi_{\perp}}\, &
\boldsymbol{\tilde v^s_{\perp}}\, &
\boldsymbol{\tilde \psi_{\perp}}
\end{pmatrix}^{\intercal} \,, \\
\mathcal M^{4 \times 4}_{||} &= \begin{pmatrix}
- i \omega & 0 & -i \chi_v k & 0 \\
0 & -i \omega & -\chi_v & k^2 \lambda \\
- i k \chi_n & \chi_\pi & -i \omega & 0 \\
0 & -\chi_\pi & 0 & -i \omega
\end{pmatrix}\,, \quad \textbf{v}_{||} = \begin{pmatrix}
\tilde n\, &
\tilde \pi_{||}\, &
\tilde v^s_{||}\, &
\tilde \psi_{||}
\end{pmatrix}^{\intercal}
\end{split}
\ee
correspond to transverse and longitudinal sectors respectively and we have defined $\lambda = \lambda_1 + \lambda_2 \frac{d-1}{d}$ and $\lambda_{\perp} =  \frac{\lambda_2 + \lambda_3}{2}$. In the transverse sector we find a solution at zero $\omega_{\perp}(k)=0$, as well as a pair of gapped propagating modes 
\be\label{eq:transDisp}
\omega_{\perp}(k) = \pm \sqrt{\chi_{\pi}\chi_v + \lambda_{\perp} \chi_{\pi} k^2}\,.
\ee 
Interestingly, we observe that the system possesses an intrinsic lengthscale, which we quantify by introducing the \textit{transverse dipole wavevector}
\be 
k^{\perp}_0 = \sqrt{\frac{\chi_v}{\lambda_{\perp} }}   \,.
\ee 
Depending on the ratio $\kappa = \frac{k}{k^{\perp}_0}$, the transverse modes exhibit the following asymptotic dependence on the momentum wave vector: 
\be
\omega_{\perp}(\kappa) \approx 
     \begin{cases}
        \pm m_0 (1 + \frac{\kappa^2}{2}) &\quad\text{for } \kappa\ll 1\\
       \pm m_0 \kappa  &\quad\text{for } \kappa \gg 1 \\ 
     \end{cases}
\ee
Here, we have introduced a mass term $m_0= \sqrt{\chi_{\pi}\chi_v}$. Therefore, transverse perturbations for which $\kappa \ll 1$ correspond to massive modes with a quadratic dispersion whereas in the regime when $\kappa \gg 1$ the perturbations travel with a soundlike dispersion.

For the longitudinal sector, we find four modes propagating with the following dispersion relations
\be \label{eq:disLong}
\omega^{\pm}_{||} = \pm \frac{\sqrt{ m^2_0 + \Lambda k^2 \pm \sqrt{ \left(m^2_0 + \Lambda k^2\right)^2 -4 m^2_0 \lambda \chi_n k^4  }}}{\sqrt{2}}
\ee 
where the superscript in $\omega^{\pm}_{||}$ refers to the second $(\pm)$ sign and $\Lambda = \lambda \chi_{\pi} + \chi_n \chi_v$. 
Here, as well, we identify a characteristic lengthscale that we dub \textit{longitudinal dipole wave vector}
\be 
k^{||}_{0} = \frac{m_0}{\sqrt{\Lambda}}\,,
\ee 
and examine the behavior of the dispersion relations as a function of the ratio $\kappa = \frac{k} {k^{||}_{0}}$. In terms of $\kappa$  \eqref{eq:disLong} takes the simple form
\be 
\omega^{\pm}_{||} = \pm m_0 \frac{\sqrt{ 1 + \kappa^2 \pm \sqrt{ \left(1 + \kappa^2 \right)^2 -4 a  \kappa^4  }}}{\sqrt{2}}
\ee 
where $a=\frac{\lambda \chi_n m^2_0}{\Lambda^2} \leq \frac{1}{4}$ is a free parameter. In the small $\kappa$ expansion, the pair of modes associated to $\omega^{-}_{||}$ displays a magnonlike dispersion relation, while the second pair corresponds to massive excitations that also propagate with quadratic momentum dependence
\be \begin{split}
\omega^{-}_{||} &= \pm m_0 \sqrt{a} \kappa^2 + \mathcal{O}(\kappa^4)\,, \quad \text{for } \kappa \ll 1, \\
\omega^{+}_{||} &= \pm m_0  \Big( 1 + \frac{\kappa^2}{2}\Big) + \mathcal{O}(\kappa^4) \,, \quad \text{for } \kappa \ll 1 \,.
\end{split}
\ee   
In contrast, in the large $\kappa$ regime we find two pairs of soundlike modes propagating with different velocities 
\be \
\omega^{\pm}_{||} = \pm m_0 \frac{\sqrt{1\pm\sqrt{1-4a}}}{2} \kappa\,, \quad \text{for } \kappa \gg 1\,.
\ee 
We thus see that in the regime where the perturbations have a wavelength smaller than the macroscopic dipole lengthscale $k_0^{-1}=\frac{\sqrt{\Lambda}}{m_0}$, there are two distinct linearly propagating modes propagating with different velocities.
Finally, let us point out that all modes are purely real, which follows from the condition for thermodynamic stability \eqref{eq:stability}. 
%%%%%%%%%%%%%%%%%%%%%

%%%%%%%%%%%%%%%%%%%
\subsubsection{Dipole--charge--vortex duality}\label{sec:chargeDipoleVortexDuality}
In this section we develop an analogue of charge--vortex duality in the presence of nonuniform dipole symmetry, whereby a zero-temperature dipole superfluid is dualized to an emergent higher-rank gauge theory. The topological defects of the original theory are mapped to gauge charges, which inherit the mobility constraints of the underlying dynamics and therefore exhibit fractonic behavior. For this section, we specialize to $2+1$--dimensions and adopt a different notation better suited to the duality construction.

We start from the effective theory Eq.~\eqref{eq:effectiveDipoleSigma} and consider the quadratic action 
\be \label{eq:actioNDualDipoel}
S[\theta, \psi_i] =\frac{1}{2} \int d^2 x dt \, \left[ \frac{1}{ v^2} (\partial_t \theta)^2
- {(\partial_i \theta - \psi_i)}^2
+ \tau^2  \partial_t \psi_i \partial_t \psi_i
- l^2 \lambda_{ijkl} \partial_i \psi_j \partial_k \psi_l \right] \,,
\ee
The normalization is chosen such that $[\theta]=T^{-1/2}$, which implies
\bs
[\psi_i] &= L^{-1}T^{-1/2}\,, \\
[v] &= LT^{-1}\,, \\
[\tau] &= T\,, \\
[l] &= L \,, \\
[\lambda_{ijkl}] &= 1 \,.
\es
We then find it useful to introduce the covariant derivatives 
\bs \label{eq:covariantDipoleDerivatives}
D_\mu \psi_i &\equiv \left(  \tau \partial_t \psi_i, l \partial_j \psi_i \right) \,, \\
D_\mu \theta & \equiv \left( \frac{1}{v}\partial_t \theta, \partial_i \theta - \psi_i \right) \,,
\es 
such that the Lagrangian can be written as
\begin{equation}
\mathcal L
=
-\frac{1}{2}\eta^{\mu\nu} D_\mu \theta D_\nu \theta
-\frac{1}{2}\Sigma^{\mu\nu ij} D_\mu \psi_i D_\nu \psi_j ,
\end{equation}
where the derivative-space indices $\mu,\nu=0,1,2$ are raised and lowered with the emergent Minkowski metric
\begin{equation}
\eta^{\mu\nu}
=
-\delta^\mu_0\delta^\nu_0+\delta^\mu_k\delta^\nu_k 
\end{equation}
and the tensor $\Sigma^{\mu\nu ij}$ is given by
\begin{equation}
\Sigma^{\mu\nu 
ij}
=
-\delta^\mu_0\delta^\nu_0 \delta^{ij}
+
\delta^\mu_k\delta^\nu_l \lambda^{klij}\,.
\end{equation}
%%%%%%%%%%%%%%%%%%%
%%%%%%%%%%%%%%
For the action of the form 
\bs
S[\theta, \psi_i] = \int d^2 x dt \, \mathcal{L}(D_\mu \psi_i, D_\mu \theta)\,.
\es
the general structure of the equations of motion is
\be \label{eq:eomsDipoleDuality}
\partial_\mu \tau^{\mu i} = - j^i\,,  \quad \partial_\mu j^\mu = 0\,,
\ee 
where 
\be 
\tau^{\mu i} = \frac{\partial \mathcal L}{\partial \left(D_\mu \psi_i \right) }\,, \quad j^\mu = \frac{\partial \mathcal L}{\partial \left(D_\mu \theta \right) }\,.
\ee 
For the quadratic action in particular we have
\be 
j^\mu = - \eta^{\mu \nu} D_\nu \theta \,, \quad \tau^{\mu i} = - \Sigma^{\mu\nu i j}  D_\nu \psi_j \,.
\ee 
To expose the dual description, it is convenient to treat the covariant derivatives as independent variables and introduce their conjugate currents via a Legendre transform. In this way, the Lagrangian can be expressed as
\be \label{eq:legendreLagrangian}
\mathcal L = j^\mu D_\mu \theta + \frac{1}{2} \eta_{\mu \nu} j^\mu j^\nu + \tau^{\mu i} D_\mu \psi_i + \frac{1}{2} \Sigma^{-1}_{\mu \nu ij} \tau^{\mu i} \tau^{\nu j} \,,
\ee 
where $\Sigma^{-1}_{\mu \nu ij}$ denotes the inverse of $\Sigma^{\mu \nu ij}$.

We now decompose the Goldstone fields into smooth and singular parts, with the latter encoding topological defects. Integrating out the smooth part imposes the equations of motion \eqref{eq:eomsDipoleDuality}, which can be resolved with the following gauge fields 
\bs \label{eq:conjugateCurrents}
j^\mu &= \epsilon^{\mu \nu \rho} \partial_\nu a_\rho \,, \\
\tau^{\mu i} &= \epsilon^{\mu \nu \rho} \partial_\nu b_\rho^i - \epsilon^{i \mu \rho} a_\rho \,,
\es 
where $a_\mu$ and $b^i_\mu$ are emergent monopole and dipole gauge fields, respectively. These fields exhibit the following gauge redundancies
\bs \label{eq:gaugeFreedom}
a_\mu &\mapsto a_\mu + \partial_\mu \alpha \,, \\ 
b^i_\mu &\mapsto b^i_\mu + \partial_\mu \beta^i + \delta^i_\mu \alpha \,.
\es 
%%%%%%%%%%
%%%%%%%%%%%%
%%%%%%%%%%%%
Substituting the expressions Eq.~\eqref{eq:conjugateCurrents} back into the Lagrangian Eq.~\eqref{eq:legendreLagrangian} yields the dual gauge theory
\be \label{eq:dualLagrangian}
\mathcal L_{\rm dual}
=
-\frac{1}{4} f_{\mu \nu} f^{\mu \nu}
+
\frac{1}{2} \Lambda^{\alpha\beta\gamma\delta}_{ij} F^i_{\alpha\beta} F^j_{\gamma\delta}
-
a_\mu J^\mu
-
b^i_\mu K^\mu_i \,,
\ee
where the tensor
\be
\Lambda^{\alpha\beta\gamma\delta}_{ij}
\equiv
\frac{1}{8}\Sigma^{-1}_{\mu \nu ij}\epsilon^{\mu \alpha \beta}\epsilon^{\nu \gamma \delta}
\ee
encodes the constitutive structure inherited from the original theory, the field strengths of the emergent gauge fields are defined as
\bs 
f_{\mu \nu} &\equiv \partial_\mu a_\nu - \partial_\nu a_\mu \,, \\
F^i_{\mu \nu} &\equiv \partial_\mu b^i_\nu - \partial_\nu b^i_\mu + \delta^i_\mu a_\nu - \delta^i_\nu a_\mu \,,
\es 
and the topological defect currents take the form
\bs 
J^\mu &\equiv \epsilon^{\mu \nu \rho} \left( \partial_\nu D_\rho \theta - \delta^i_\rho D_\nu \psi_i \right) \,, \\
K^\mu_i &\equiv \epsilon^{\mu \nu \rho} \partial_\rho D_\nu \psi_i \,.
\es 
%%%%%%%%%%%%%%%%%%%%%
%%%%%%%%%%%
The dual theory Eq.~\eqref{eq:dualLagrangian} describes an emergent higher-rank gauge theory, where the gauge charges are identified with topological defects of the original theory Eq.~\eqref{eq:actioNDualDipoel}.

%%%%%%%%%%%%
As a result of gauge freedom Eq.~\eqref{eq:gaugeFreedom} the defect currents obey the following equations of motion 
\be 
\partial_\mu K^\mu_i = 0 \,, \quad \partial_\mu J^\mu = K^i_i\,.
\ee 
The first equation expresses the conservation of the dipole defect current, whereas the second shows that the monopole defect current is sourced by the trace of the dipole defect current. In particular, any process involving a nonzero trace component $K^i_i$ necessarily requires the creation or annihilation of monopole defects. Such processes are typically energetically suppressed, which implies that the corresponding component of the dipole current cannot contribute to transport, reflecting the effective fractonic constraint on the dynamics.\footnote{This constraint may be lifted if monopole defects proliferate.}

Moreover, notice that the field strengths satisfy the following Bianchi identities 
\be 
\epsilon^{\mu \nu \rho} \partial_\mu f_{\nu \rho} = 0 \,, \quad \epsilon^{\mu \nu \rho} \partial_\mu F^i_{\nu \rho} = -  \epsilon^{i \mu \nu} f_{\mu \nu} \,.
\ee 
The second identity reflects the nontrivial coupling between the monopole and dipole gauge sectors, whereby the dipole field strength is not closed but sourced by the monopole field strength.

This completes the dualization of the dipole superfluid theory Eq.~\eqref{eq:actioNDualDipoel} into an emergent higher-rank gauge theory Eq.~\eqref{eq:dualLagrangian}, thereby generalizing the charge--vortex duality to the case of a $\mathrm U(1)$ superfluid with dipole symmetry. A closely related duality arises in the theory of elasticity, where a similar procedure maps elastic defects onto charges of an emergent higher-rank gauge theory, giving rise to the so-called fracton--elasticity duality \cite{duality,gromov_duality_2020,Glodkowski:2024qsf}.%%%%%%%%%%%%%
%%%%%%%%%%%%%%%%%%%%%%

\subsection{Subdiffusion from dipole symmetry} \label{sec:dipoleSubdiffusion2}
In this section we present an explicit example of diffusion in the presence of nonuniform symmetries discussed in Sec.~\ref{sec:diffusionNonuniform}. We show that dipole conservation qualitatively modifies charge transport, resulting in subdiffusive relaxation of charge density modulations \cite{Guardado_Sanchez_2020,Gromov_hydro_2020,PhysRevLett.125.245303,PhysRevB.105.205127,Nandy:2023bop}. We also provide a physical realization of this phenomenon in the tilted Bose--Hubbard model, which has been experimentally realized in cold-atom systems \cite{Guardado_Sanchez_2020}.

\subsubsection{Hydrodynamic theory}\label{sec:dipoleSubdiffusion}

\textit{Parts of the discussion in this section are based on Sec.~2 of Ref.~\cite{Glodkowski:2024ova}.} \newline 

\noindent We now construct the hydrodynamic theory of charge diffusion in a system with dipole symmetry, providing an explicit construction illustrating the diffusion of nonuniform charges discussed in Sec.~\ref{sec:diffusionNonuniform}.

We consider a many-body system whose microscopic dynamics conserves both monopole and dipole moments of charge, satisfying the nonuniform dipole algebra
\be \label{eq:dipoleAlgebra}
[P_i,D_j]=-\delta_{ij}Q \,.
\ee
In the long-wavelength limit, the corresponding conserved charges can be expressed in terms of their respective currents 
\be 
Q = \int \star j\,, \quad D_i = \int \star J_i\,,
\ee 
that obey local conservation laws
\be 
d \star j = 0\,, \quad d \star J_i = 0\,.
\ee 
The nonuniform dipole algebra Eq.~\eqref{eq:dipoleAlgebra} implies that the dipole current transforms inhomogeneously under translations 
\be 
[P_i, J_j] = - \partial_i J_j - \delta_{ij} j\,.  
\ee 
Therefore, the dipole current carries explicit coordinate dependence, which can be isolated as
\be 
J_i = \tilde J_i - x_i  j\,.
\ee 
The uniform current $\tilde J_i$ transforms homogeneously,
\be  
[P_i, \tilde J_j] = - \partial_i \tilde J_j \,,  
\ee 
but is not conserved 
\be \label{eq:internalDipoleEq}
d \star \tilde J_i = dx_i \wedge \star j\,.
\ee 
We write the currents in component form as
\be 
j = j^0 dt + j_i dx^i\,, \quad J_i = J^0_i dt + J_{ij} dx^j \,.
\ee 
Define the entropy density as a function of the conserved densities,
\be
s \equiv s(j^0, J_i^0) \,,
\ee
with differential
\be
ds = -\frac{\mu}{T_0} d j^0 -\frac{\mu_i}{T_0} d J_i^0 \,.
\ee
When written directly in terms of the conserved densities, translations are realized nonlinearly on the entropy density function,
\be \label{eq:entropTransfom} 
[P_i, s] = - \partial_i s + \frac{\mu_i}{T_0} j^0\,. 
\ee 
As a consequence, the associated chemical potential,
\be \label{eq:defMu}
\mu=-T_0\frac{\partial s}{\partial j^0} \,,
\ee
transforms inhomogeneously under spatial translations. This follows by using Eq.~\eqref{eq:entropTransfom} together with Eq.~\eqref{eq:defMu}, which implies 
\be 
[P_i, \mu] = - \partial_i \mu + \frac{\mu_i}{T_0}\,.
\ee
It is therefore convenient to introduce the uniform chemical potential
\be
\tilde\mu=\mu-\mu_i x_i\,,
\ee
which transforms homogeneously under translations. The first law then takes the form
\be \label{eq:1stLawUniform}
ds = - \frac{\tilde\mu}{T_0} dj^0 - \frac{\mu_i}{T_0} d\tilde J_i^0 \,.
\ee
Passing from $s(j^0,J^0)$ to $s(j^0,\tilde J^0)$ corresponds merely to a change of variables. When viewed as a function of the conserved dipole density, the entropy density exhibits an inhomogeneous transformation under translations, reflecting the nonlinear realization of the symmetry in this parametrization. The same applies to the associated chemical potentials. In contrast, the uniform variables furnish a uniform representation, and the entropy density written as $s(j^0,\tilde J^0)$ transforms manifestly as a scalar under translations. It is therefore natural to formulate thermodynamics in terms of the uniform variables, especially when translations are imposed as a symmetry, as this makes translational invariance manifest.

\paragraph{Relaxation and hydrodynamic limit.} We now determine the hydrodynamic consequences of the dipole symmetry. Following the general procedure outlined in Sec.~\ref{sec:diffusionNonuniform}, after using the conservation laws together with the first law Eq.~\eqref{eq:1stLawUniform}, we can express the entropy production in the form
\be \label{eq:secondLawInternal}
\Delta \equiv \partial_t s + \partial_i s_i =  - j_i  \Big(\partial_i \frac{\tilde \mu}{T_0} + \frac{\mu_i}{T_0} \Big) - \tilde J_{ij} \partial_j  \frac{\mu_i}{T_0} \,,
\ee
where we have defined the entropy current 
\be 
s_i \equiv  -\frac{\tilde \mu}{T_0}  j_i - \frac{\mu_j}{T_0}  \tilde J_{ji}\,.
\ee 
%%%%%%%%%%%
Imposing the local second law, i.e.\ demanding $\Delta \geq 0$ for
arbitrary configurations of thermodynamic forces, we conclude that the
leading-order constitutive relation for the charge current takes the
form
\bs
j_i &=
- \beta
\left(
\partial_i \frac{\tilde \mu}{T_0}
+ \frac{\mu_i}{T_0}
\right)\,,  \\
\tilde J_{ij} &= -\sigma_{ijkl} \partial_k \mu_l \,, 
\es
where 
\be 
\sigma_{ijkl} = \sigma_1 \delta_{ij}\delta_{kl} + \sigma_2 \delta_{i\langle k}\delta_{l \rangle j} + \sigma_3\delta_{i[k}\delta_{l]j}
\ee 
 and $\beta$ are dissipative transport coefficients satisfying $\beta, \sigma_1, \sigma_2, \sigma_3 \geq 0$.

We define the dipole susceptibility $\xi$ through the
thermodynamic relation
\be \label{eq:thermoIdentities}
\xi
\equiv
\frac{\partial  \mu_i}{\partial \tilde J_i^0}
=
- T_0
\frac{\partial^2 s}{\partial (\tilde J_i^0)^2} \,,
\ee
which follows from the definition of the chemical potentials and the
quadratic expansion of the entropy density in small fluctuations.
Thermodynamic stability requires the entropy density to be a concave
function of the conserved densities, implying
\be
\xi \ge 0 \,.
\ee
%%%%%
%%%%%%%%
%%%%%%%
Using \eqref{eq:thermoIdentities} we can express the local equations as follows
\bs \label{eq:someHydroEqs}
    \partial_t \tilde j^0 - \beta \partial_i \left(  \mu_i + \partial_i \tilde \mu \right) &= 0 \,, \\
    \partial_t  \mu_i - \xi \sigma_{jikl} \partial_j \partial_k \mu_l &= - \frac{1}{\tau} \left(  \mu_i + \partial_i \tilde \mu \right)  \,,
\es  
where we have defined the relaxation time 
\be 
\tau \equiv \frac{T_0 }{\beta \xi}\,.
\ee 
%%%%%%%%%%
It is helpful to reorganize the dipole equation as
\be 
\left( \partial_t + \frac{1}{\tau} \right) \mu_i = - \frac{1}{\tau} \partial_i \tilde \mu -  \xi \sigma_{jikl} \partial_j \partial_k \mu_l \,.
\ee 
%%%%%%%%%
At long times $t\gg\tau$, the relaxation term dominates over temporal derivatives and one can drop the temporal derivative
\be 
\left( \partial_t + \frac{1}{\tau} \right) \mu_i \simeq \frac{1}{\tau} \mu_i \,.
\ee 
In this regime, $\mu_i$ can be systematically eliminated in terms of $\tilde \mu$, as 
\be \label{eq:hydrostaticDipoleInternal}
\mu_i = - \partial_i \tilde \mu - \tau \xi  \sigma_{jikl} \partial_j \partial_k \partial_l \tilde \mu + \cdots   
\ee 
so that $\mu_i$ does
not constitute an independent hydrodynamic degree of freedom.

After substituting \eqref{eq:hydrostaticDipoleInternal} into the charge equation in \eqref{eq:someHydroEqs}, we end up with 
\be 
    \partial_t \tilde j^0 + T_0 \sigma_{jikl} \partial_i \partial_j \partial_k \partial_l \tilde \mu  = 0 \,.
\ee 
This predicts subdiffusive relaxation of charge fluctuations
\be 
\omega = - i D k^4 \,,
\ee 
where 
\be 
D = T_0 \chi \big( \sigma_1 +\frac{d-1}{d} \sigma_2\big)\,, 
\ee  
and we have defined the charge susceptibility 
\be
\chi 
\equiv
\frac{\partial  \tilde \mu}{\partial \tilde j^0}
=
- T_0
\frac{\partial^2 s}{\partial (\tilde j^0)^2} \geq 0 \,.
\ee
The quartic dispersion reflects the suppression of ordinary diffusive transport due to the presence of nonuniform dipole symmetry, which forces relaxation to proceed through higher-order gradient processes. This anomalous subdiffusive scaling of charge transport in systems with dipole conservation was derived in \cite{Gromov_hydro_2020}, where dipole conservation was implemented by coupling to a symmetric tensor gauge field.

This example illustrates the mechanism discussed in Sec.~\ref{sec:diffusionNonuniform} where integrating out a non-hydrodynamic mode associated with a nonuniform symmetry can qualitatively modify the relaxation of conserved uniform charges. In the dipole-conserving case, this leads to subdiffusion.
%%%%%%%

%%%%%%%%%%%
\subsubsection{Physical realization in the tilted Bose--Hubbard chain}

\textit{This section builds in part on Ref.~\cite{Nandy:2023bop}.} \newline 

\paragraph{The tilted Bose--Hubbard model.} Consider the Bose-Hubbard model in the presence of a tilt 
\bs \label{eq:boseHubbar}
H &=  H_0 + E D \,, \\
H_0 &= \sum_i \left( \hat c_i^\dag  \hat c_{i+1}  +  \hat c_{i+1}^\dag \hat c_i  + g \hat n_i \hat n_{i+1} \right) \,, \\
D &= \sum_i \, i \hat n_i\,,  
\es  
where $\hat c_i$ creates a particle at lattice site $i$, $E$ denotes the strength of the applied electric field, and $\hat n_i=\hat c_i^\dagger \hat c_i$ is the particle number operator.

Let us introduce the translation operator $T$, which acts on the lattice operators via
\be 
T \hat c_i T^{-1} = \hat c_{i+1}\,.
\ee 
The Bose--Hubbard Hamiltonian $H_0$ is invariant under translations,
\be 
T H_0 T^{-1} = H_0 \,.
\ee 
On the other hand, the dipole operator shifts
\be 
T D T^{-1} = D - Q \,, 
\ee 
where 
\be 
Q = \sum_i \hat n_i\,,
\ee 
is the total charge. Therefore, the tilted Bose--Hubbard Hamiltonian Eq.~\eqref{eq:boseHubbar} transforms as
\be 
T H T^{-1} = H - E Q\,.
\ee 
Introducing the anti-Hermitian generator of translations $ P$ via
\be
T = e^{ P},
\ee
the adjoint action of translations takes the infinitesimal form
\be \label{eq:boseHubbardAlgebra}
[ P, H] = E Q \,.
\ee
We have thus established that the tilted Bose--Hubbard chain Eq.~\eqref{eq:boseHubbar} realizes a nonuniform algebra Eq.~\eqref{eq:boseHubbardAlgebra}.
\paragraph{Coarse-grained picture.} Upon coarse-graining over blocks of lattice sites, 
we introduce smoothly varying density fields 
$j^0(x,t)$ and $\varepsilon^0(x,t)$ obtained by spatial averaging 
the microscopic operators $\hat n_i$ and $\hat h_i$, where
\be
H = \sum_i  \hat h_i \,.
\ee
In the hydrodynamic regime these coarse-grained densities are treated 
as classical fields describing the expectation values of the conserved 
quantities.

We introduce the particle and energy currents
\be
j = j^0 dt + j^x dx \,, \quad \varepsilon = \varepsilon^0 dt + \varepsilon^x dx \,,
\ee
which describe the local densities and fluxes of the conserved charge $Q$ and energy $H$.

At long wavelengths, their dynamics is governed by local conservation laws,
\be
\begin{split}
d \star j &= 0 \,, \\
d \star \varepsilon &= 0 \,,
\end{split}
\ee
which constitute the hydrodynamic equations of motion.

The nonuniform algebra Eq.~\eqref{eq:boseHubbardAlgebra} implies that the energy density current transforms inhomogeneously under translations 
\be 
[P, \varepsilon] =  - \partial_x \varepsilon + E j\,.
\ee 
It follows that we can decompose the energy current into the uniform piece and explicit coordinate dependence 
\be 
 \varepsilon = \tilde  \varepsilon - E x  j\,.
\ee 
In this decomposition, the uniform component $\tilde  \varepsilon$ represents the energy current associated with $H_0$, while the full energy current includes the explicit coordinate-dependent contribution arising from the linear potential.

%%%%%%%%%%%%
The uniform energy current satisfies the dynamical equation
\be  \label{eq:nonuniformEnergy}
d \star \tilde \varepsilon = E dx \wedge j\,,
\ee 
where the source term on the right-hand side represents the power density supplied by the electric field and describes Joule heating.

After rescaling the uniform energy current
\be 
\tilde \varepsilon \mapsto E^{-1} \tilde \varepsilon \,,
\ee 
Eq.~\eqref{eq:nonuniformEnergy} coincides with the dipole equation Eq.~\eqref{eq:internalDipoleEq}. The tilted Bose--Hubbard chain therefore belongs to the dipole-conserving universality class discussed in Sec.~\ref{sec:dipoleSubdiffusion}. Consequently, its long-wavelength dynamics exhibits subdiffusive relaxation of charge fluctuations,
\be 
\omega \sim -i k^4\,.
\ee 
 This behavior was indeed confirmed numerically in Ref.~\cite{Nandy:2023bop} and observed in experiments on cold atoms \cite{Guardado_Sanchez_2020}. In our framework, it arises from the presence of a nonuniform dipole symmetry, which, upon integrating out the associated gapped mode, qualitatively modifies the hydrodynamic relaxation of the monopole charge.
%%%%%%%%%%%%%%%%%%%%%%%%%%

\section{Fracton fluids}\label{sec:fractonFluids}
\textit{This chapter is based on the results obtained in Refs.~\cite{Glodkowski:2022xje,Glodkowski:2024ova}.} \newline 

\noindent In this section we introduce fracton fluids \cite{PhysRevResearch.3.043186,GloriosoLucas22,Jain:2023nbf,Armas:2023ouk}, defined as homogeneous and isotropic many-body systems which, in addition to the ordinary monopole charge, conserve its first spatial moment, i.e. the dipole moment. 

From the perspective developed in this thesis, dipole symmetry is nonuniform and therefore does not give rise to additional hydrodynamic modes. Instead, it imposes kinematic constraints on the independent hydrodynamic degrees of freedom determined by the kernel of the translation action. As discussed in Sec.~\ref{sec:dipoleSubdiffusion}, dipole symmetry qualitatively modifies charge transport, leading to anomalously slow diffusion. In the present context, we analyze how these constraints apply to fluid systems, which also conserve energy and momentum. 

We begin by analyzing the constraints imposed by dipole symmetry on homogeneous equilibrium states. In particular, we show that the simultaneous presence of nonzero charge density and finite fluid velocity is incompatible with an unbroken realization of dipole symmetry. This leads to a natural classification of fracton fluids into distinct phases, depending on how the symmetry is realized at low energies. We then construct the corresponding hydrodynamic theories for these phases, elucidating their macroscopic dynamics and highlighting the interplay with the kinematic constraints imposed at the microscopic level.

\subsection{The no-flow theorem}\label{sec:dipoleSSB}
We begin by recalling the no-flow theorem for systems with conserved dipole moment~\cite{Armas:2023ouk}. It states that, at nonzero charge density, a homogeneous equilibrium state cannot support a finite fluid velocity unless dipole symmetry is spontaneously broken. 

To see this, consider a homogeneous equilibrium state at finite temperature and chemical potential for both the monopole charge and momentum. The equilibrium state is described by the Gibbs density matrix
\be
\rho = e^{-\beta \left( H - \mu Q - v^i P_i \right)} \,,
\ee
where $\mu$ is the chemical potential associated with the conserved charge $Q$, and $v^i$ denotes the fluid velocity.

We now examine the action of the dipole symmetry generated by $D_i$. Using the dipole algebra
\begin{equation}
[P_i, D_j] = -\delta_{ij} Q \,,
\end{equation}
one finds that the density matrix transforms as
\begin{equation}
\rho \mapsto e^{-\lambda^i D_i}\rho  e^{\lambda^i D_i}
=
e^{-\beta \left( H - (\mu + \lambda_i v^i) Q - v^i P_i \right)} \,.
\end{equation}
Thus, the density matrix is not invariant under dipole transformations unless either the fluid is stationary, $v^i=0$, or the state is charge neutral, $\langle Q \rangle = 0$. At finite charge density, this implies that homogeneous equilibrium states with nonzero velocity are incompatible with an unbroken realization of dipole symmetry. Therefore, generic flowing fracton fluids at finite density necessarily realize dipole symmetry nonlinearly.

In addition, the monopole symmetry may be spontaneously broken. In this case, dipole symmetry is necessarily broken as well, leading to a qualitatively distinct superfluid phase. We will therefore refer to phases with unbroken monopole symmetry as \textit{dipole-conserving fluids}, even though dipole symmetry may itself be spontaneously broken in generic flowing states. By contrast, when the monopole symmetry is also spontaneously broken, we will call it a \textit{fracton superfluid}. The dipole sigma model constructed in Sec.~\ref{sec:dipole-sigma-model} provides an effective description of this phase at zero temperature.

In the remainder of this section, we develop the hydrodynamic description of these phases and analyze their characteristic transport and collective excitations.

%%%%%%%%%%%%%%

\subsection{Dipole-conserving fluids}
\textit{This section closely follows Ref.~\cite{Glodkowski:2022xje}.} \newline 

\noindent In this section, we construct a hydrodynamic theory for dipole-conserving fluids, i.e. phases in which the monopole symmetry remains unbroken. In doing so, we examine how the constraints imposed by dipole conservation at the microscopic level manifest themselves in macroscopic transport. This provides the minimal hydrodynamic realization of a fracton fluid, capturing the essential kinematic restrictions associated with dipole symmetry while allowing for a consistent description of long-wavelength dynamics.

\subsubsection{Symmetries and hydrodynamic equations}\label{sec:symmetries}
We consider a many-body system exhibiting simultaneous conservation of energy $H$, momentum $P_i$, monopole charge $Q$, dipole moment $ D_i$ and angular momentum $J_{ij}$. 

These generators satisfy the algebra
\bs  \label{eq:dipoleAlgebraFluids}
     \{   J_{ij},    J_{kl}\} &= 2\delta_{i[k}    J_{l]j} + 2\delta_{j[l}    J_{k]i}\,, \\
     \{   J_{ij},   P_k\} &=  2\delta_{k[i}   P_{j]}\,,\\
      \{   J_{ij},   D_k\} &=  2\delta_{k[i}   D_{j]}\,, \\  
\{  D_{i},   P_j\} &= \delta_{ij}   Q\,.
\es
In particular, both the angular momentum and dipole generators are nonuniform symmetries and therefore do not give rise to independent hydrodynamic degrees of freedom in the long-wavelength limit. In the following, we work in the strict hydrodynamic regime, in which the dynamics is governed solely by the gapless hydrodynamic variables $(n, p_i, \epsilon )$, corresponding to the densities of charge, momentum, and energy associated with uniform conserved generators. 

Accordingly, in the long-wavelength regime, all the conserved charges can be expressed in terms of the hydrodynamic variables
\begin{equation}\label{eq:conserved}
\begin{split}
       H   &= \int d^d x \,\epsilon \,, \quad   P_i  = \int d^d x \,p_i \,,  \\
   Q   &= \int d^d x \,n\,, \quad   D_i  = \int d^d x \hspace{2pt}x_i n \,, \\
   J_{ij}  &=  \int d^d x \,\big( x_i p_j  - x_j p_i  \big)\,.
\end{split}
\end{equation}
%%%%
The conservation of the respective charges is then encoded in the local conservation laws
\begin{equation}
\label{eq:continuity}
\begin{split}
 \partial_t  n &+ \partial_i \partial_j J^{ij} = 0\,,\\
  \partial_t  p_i &+ \partial_j T^{ji} = 0\,,\\
\partial_t  \epsilon  &+ \partial_i J^i_\epsilon = 0\,.
\end{split}
\end{equation}
%%%%
This construction follows a bottom-up approach, in which the system is formulated directly in the strict hydrodynamic limit, with the nonuniform symmetry sector encoded through kinematic constraints rather than independent degrees of freedom. In particular, the conservation of dipole moment implies that the charge current is a gradient, while angular momentum conservation requires that the stress tensor $T^{ij}$ be symmetric.

\subsubsection{Thermodynamics and dipole conservation}\label{sec:thermodynamics}

The nonuniform algebra Eq.~\eqref{eq:dipoleAlgebraFluids} implies that the momentum density transforms inhomogeneously under dipole symmetry,
\begin{equation}
   \{D_i, p_j \} = \delta_{ij} n \,.
\end{equation}
Consequently, the ratio $p_i/n$ exhibits a shift symmetry under dipole transformations. This observation imposes strong constraints on the equilibrium thermodynamics.

As discussed in Sec.~\ref{sec:dipoleSSB}, the nonuniform action of dipole symmetry restricts homogeneous equilibrium states. In particular, at finite charge density, a phase that supports nonzero fluid velocity must realize dipole symmetry nonlinearly.

In such phases, the ratio $p_i/n$ coincides at leading order with the associated dipole Goldstone mode~\cite{Armas:2023ouk},
\be\label{eq:dipoleGoldstoneMomentum}
\frac{p_i}{n} \simeq \psi_i + \cdots ,
\ee
where the ellipsis denotes higher-derivative corrections. Thus, whenever the fluid sustains nonzero velocity, the momentum density admits an interpretation in terms of the dipole Goldstone degree of freedom.

For the purposes of the hydrodynamic theory, however, we will adopt an agnostic formulation in terms of the conserved densities $(n, p_i, \epsilon)$, without explicitly introducing Goldstone fields. In this framework, the interpretation of $p_i$ as a Goldstone mode emerges only in phases where dipole symmetry is realized nonlinearly.

\paragraph{Dipole-invariant equation of state.}

In an ordinary fluid, the internal energy density in equilibrium is a function of the conserved densities,
\be 
\epsilon = \epsilon(s,n,p_i) \,.
\ee 
However, dipole-conserving systems are not generic fluids. Since $p_i/n$ transforms inhomogeneously under dipole symmetry, it cannot enter the equation of state directly. Instead, only dipole-invariant combinations are allowed.

The simplest local invariant built from $p_i$ and $n$ is
\begin{equation} \label{eq:tensorVij}
V_{ij} \equiv \partial_i (n^{-1} p_j).
\end{equation}
Therefore, distinct constant values of $V_{ij}$ label physically inequivalent equilibrium states. This necessitates an extension of thermodynamics in which the internal energy depends on $V_{ij}$ in addition to the usual conserved densities.

We thus postulate the generalized first law
\begin{equation}
    d\epsilon = T ds + \mu dn + F_{ij} dV_{ij} \,,
\end{equation}
where $F_{ij}$ is the thermodynamic variable conjugate to $V_{ij}$. As will become clear below, $F_{ij}$ admits an interpretation as a dipole flux.

Assuming rotational invariance and microscopic parity symmetry, the antisymmetric component $V_{[ij]}$ transforms as a pseudoscalar (in $d=2$) or pseudovector (in $d=3$). Parity invariance therefore excludes dependence on $V_{[ij]}$, and the energy density depends only on the symmetric tensor $V_{(ij)}$. In what follows we denote this symmetric tensor simply by $V_{ij}$.

The pressure is defined in the usual manner,
\begin{equation}\label{eq:pressureDef}
    P = Ts + \mu n - \epsilon,
\end{equation}
leading to
\begin{equation}
    dP = n\, d\mu + s\, dT - F_{ij}\, dV_{ij}.
\end{equation}
Since in the hydrodynamic description we take $(\epsilon,n,p_i)$ as the fundamental variables, it is convenient to regard the entropy density as the thermodynamic potential. The differential of the entropy density then reads
\begin{equation} \label{eq:diffentropy}
    ds = \frac{1}{T}\, d\epsilon - \frac{\mu}{T}\, dn - \frac{F_{ij}}{T}\, dV_{ij},
\end{equation}
with thermodynamic conjugates defined by
\be\label{eq:thermodynamicsDefinitions}
\frac{1}{T} = \frac{\partial s}{\partial \epsilon}\,, \quad
\frac{\mu}{T} = - \frac{\partial s}{\partial n}\,, \quad
\frac{F_{ij}}{T} = - \frac{\partial s}{\partial V_{ij}}\,.
\ee
The entropy density
\be 
s = s(\epsilon,n,V_{ij})
\ee 
therefore plays the role of the dipole-invariant equation of state.

\paragraph{Thermodynamic relations}
In the next section, we will study linearized hydrodynamics around the global equilibrium state $(n=n_0,\epsilon=\epsilon_0,V_{ij}=0)$. Therefore, it will be useful to introduce a set of thermodynamic identities that will allow us to relate the variations of $(\epsilon,n,V_{ij})$ with their corresponding conjugate variables. 

To do so, we first expand the entropy density function around the equilibrium state up to the second order in fluctuations
\begin{align}
 \nonumber   s &=  s_0 - \frac{\mu_0}{T_0} \delta n +\frac{1}{T_0} \delta \epsilon+ \frac{1}{2}s_{nn} \delta n ^2 + \frac{1}{2}s_{\epsilon \epsilon} \delta \epsilon ^2 + s_{n\epsilon} \delta \epsilon \delta n \\
    & - \frac{1}{2T_0} f_{||} \delta V_{kk} ^2  - \frac{1}{2T_0} f_{\perp} \delta V_{\langle ij \rangle} ^2\,,
    \end{align}
where $s_0$ is the entropy evaluated at the equilibrium state, the traceless symmetrization defined as $A_{\langle i j \rangle}= \frac{1}{2}(A_{ij}+A_{ji}-\frac{2}{d}A_{kk}\delta_{ij})$ and $A_{kk} = \delta_{ik} A_{ik}$ denoting the trace. Thermodynamic stability imposes the constraints
\begin{equation}\label{eq:stability}
    f_{||},f_\perp\geq 0\,,\quad s_{\epsilon\epsilon},s_{nn} < 0 \,,\quad s_{n\epsilon}^2- s_{n n}s_{\epsilon\epsilon} \leq 0\,.
\end{equation}
Therefore, the variations of the thermodynamic quantities Eq. \eqref{eq:thermodynamicsDefinitions} can be expressed as
\begin{equation}\label{eq:therm1}
\begin{split}
     \delta  \frac{1}{T} &= s_{\epsilon \epsilon} \delta \epsilon +  s_{\epsilon n} \delta n \,, \\
     \delta  \frac{\mu}{T} &= - s_{n \epsilon} \delta \epsilon -  s_{n n} \delta n \,, \\
     \delta  F_{ij} &=  f_{||} \delta V_{kk} \delta_{ij} + f_{\perp} \delta V_{\langle ij\rangle} \,.\\
    \end{split}
\end{equation}
Finally, after using Eq. \eqref{eq:therm1} and Eq. \eqref{eq:pressureDef} we write the variation of the pressure with respect to the thermodynamic variable as
\begin{equation}\label{eq:pressure}
      \delta P = - T_0 ( P_{\epsilon} \delta \epsilon + P_n \delta n ) \,,
\end{equation}
where we have defined
\begin{align}
\nonumber    P_\epsilon &=n_0 s_{n\epsilon} + (P_0 + \epsilon_0) s_{\epsilon \epsilon}\,,\\
\label{eq:pressureNotation}    P_n &=n_0 s_{n n} + (P_0 + \epsilon_0) s_{n \epsilon}\, .
\end{align}
%%%%%%%%%%%%
%%%%%%%%%%

%%%%%%%%%%%%
%%%%%%%%%%
\subsubsection{Dipole-conserving hydrodynamics}\label{sec:hydrodynamics}

We now construct the hydrodynamic theory of dipole-conserving fluids. To do so, we formulate the most general constitutive relations consistent with the symmetries Eq.~\eqref{eq:dipoleAlgebraFluids} and the second law of thermodynamics, organized in a systematic derivative expansion.

\paragraph{Gradient expansion.}\label{sec:grad}
To render the hydrodynamic equations Eqs.~\eqref{eq:continuity} closed,
the macroscopic currents are expressed as local functionals of the
conserved densities $n$, $\epsilon$, and $p_i$, organized in a
derivative expansion. At each order, the allowed structures are fixed by
symmetry, while the associated transport coefficients are constrained
by the second law of thermodynamics and Onsager reciprocity relations. A crucial ingredient in defining this expansion is the choice of power
counting. 

In conventional hydrodynamics the conserved densities are assigned
zeroth derivative order, as they characterize the thermodynamic
equilibrium state. The presence of dipole symmetry modifies this
hierarchy. In particular, the tensor 
\be 
V_{ij} \equiv \partial_i \frac{ p_j}{n}
\ee 
admits a thermodynamic interpretation and therefore must scale as an
equilibrium quantity. Consistency of the derivative expansion then
requires assigning the momentum density the scaling
\be 
\mathcal O(p_i) \sim \mathcal O(\partial_i)^{-1}\,,
\ee 
such that
\be 
V_{ij} \sim \mathcal O(\partial_i)^0 \,.
\ee 
This assignment is consistent with the identification of the momentum
density as the dipole Goldstone mode,
Eq.~\eqref{eq:dipoleGoldstoneMomentum}, in close analogy with ordinary
superfluids where the velocity arises as a gradient of the phase
Goldstone.

The derivative expansion is defined by truncating the equations of
motion at a fixed order in spatial gradients. We refer to $n$-th order
hydrodynamics when the conservation equations are truncated as\footnote{Notice that onshell temporal derivatives will not be independent from spatial gradients, in particular we have the hierarchy $\mathcal{O}(\partial_t) \sim \mathcal{O}(\partial_i)^2$.}
\begin{equation}
\begin{split}
\partial_t \epsilon &= -\partial_i J_\epsilon^i
    + \mathcal O(\partial_i)^{2n+3}, \\
\partial_t p_i &= -\partial_j T^{ji}
    + \mathcal O(\partial_i)^{2n+2}, \\
\partial_t n &= - \partial_i \partial_j J^{ij}
    + \mathcal O(\partial_i)^{2n+3}.
\end{split}
\end{equation}

\paragraph{Zeroth order hydrodynamics.}
We now derive the leading-order constitutive relations for dipole-conserving
fluids. The starting point is the local form of the first law of
thermodynamics, Eq.~\eqref{eq:diffentropy}, which governs the evolution
of the entropy density. Using this relation, we obtain
\begin{equation}\label{eq:passing} 
\begin{split}
      T \partial_t s  &=  \partial_t \epsilon  - \mu \partial_t n -  F_{ij} \partial_t V_{ij} \,,\\
    & =  \partial_t \epsilon  -  \partial_i \left(F_{ij} \partial_t \left(\frac{p_j}{n}\right)\right)  - \tilde{\mu} \partial_t n   -  V_j \partial_t p_j \,,
\end{split}
\end{equation}
where in the last step we have defined the effective chemical potential and velocity
\begin{equation}
\label{eq:notation}
 \tilde{\mu} = \mu - \frac{V_ip_i}{n} \,, \quad V_i = - \frac{\partial_j F_{ji} }{n} \,.
\end{equation}
Using the equations of motion Eqs. \eqref{eq:continuity} it is then possible to recast Eq. \eqref{eq:passing} into a familiar looking equation 
\begin{align}
 \label{eq:familiar}
   T \partial_t s = - \partial_i \mathcal{E}^i 
    + \tilde{\mu} \partial_i \partial_j  J^{ij} + V_i \partial_j T^{ji} \,,
\end{align} 
where we have defined a \textit{shifted energy current}
\begin{equation}\label{eq:shifted}
J_\epsilon^i  = \mathcal{E}^i - F_{ij} \partial_t \left(\frac{p_j}{n}\right)\,.
\end{equation}
Using Eq. \eqref{eq:familiar} we can express the entropy production constraint $\partial_t s +\partial_i S^i \geq0$ as 
\begin{equation}
\label{eq:entprodc0}
\partial_i S^i - \frac{1}{T}\partial_i \mathcal{E}^i 
    + \frac{\tilde{\mu}}{T} \partial_i \partial_j  J^{ij} + \frac{V_i}{T} \partial_j T^{ji} \geq 0 \,.
\end{equation}
Thus, after combining the thermodynamic relation Eq. \eqref{eq:pressureDef} with Eq.  \eqref{eq:entprodc0} and a series of tedious algebraic computations that we show in Appendix \ref{sec:nextToLeading}, it is possible to rewrite the constraint as  
\begin{equation}
\begin{gathered}
     \partial_i \Big( S^i - \frac{1}{T} \mathcal{E}^i + \frac{\tilde{\mu}}{T}  \partial_j J^{ij} - \frac{V_j}{T}\Big[  P \delta_{ij}
    +   F_{ik} \partial_j \frac{p_k}{n} +  \partial_k \big(F_{ij} \frac{p_k}{n} - F_{kj} \frac{p_i}{n}\big) - T^{ij} \Big]  \Big) + \Big( \mathcal{E}^i  - (P+\epsilon) V_i \Big) \partial_i (\frac{1}{T}) \\
    + (nV_i - \partial_j J^{ij}) \partial_i (\frac{\tilde{\mu}}{T})  +\Big[P \delta_{ij} + V_i p_j + F_{ik} \partial_j \frac{p_k}{n} +\partial_k \big(F_{ij} \frac{p_k}{n} - F_{kj} \frac{p_i}{n}\big) - T^{ij} \Big]\partial_i (\frac{V_j}{T}) \geq 0 \,. 
\end{gathered}
 \label{eq:master}
\end{equation}
Therefore, we conclude that the local version of the second law of thermodynamics will be satisfied provided that the first term in Eq. \eqref{eq:master} vanishes and the remainder is semi-positive definite for arbitrary field configurations. This constraint fixes the zeroth order currents to
\begin{equation}\label{eq:constitutive}
\begin{split}
 J_\epsilon^i &=  (\epsilon + P )V_i - F_{ij} \partial_t(\frac{p_j}{n}) + \alpha \partial_i \frac{1}{T}  \,,\\
  J^{ij} &= - F_{ij}  \,,\\
 T^{ij} &=  P \delta_{ij} + V_i p_j + V_j p_i  + \partial_k F_{ij} \frac{p_k}{n} + F_{ij} V_{kk} \,, \\ 
 S^i &= \frac{1}{T} \mathcal{E}^i  + P \frac{V_i}{T}
    - \frac{\tilde{\mu}}{T}  \partial_j J^{ij} - \frac{V_j}{T} T^{ij} \\ 
    &+ \frac{V_j}{T}  F_{ik} \partial_j \frac{p_k}{n} + \frac{V_j}{T} \partial_k \Big( F_{ij} \frac{p_k}{n} - F_{kj} \frac{p_i}{n} \Big)
\end{split}
\end{equation}
with $\alpha$ a transport coefficient that can be interpreted as the thermal conductivity of the system, satisfying the inequality 
\begin{equation}\label{eq:alpha}
    \alpha \geq 0 \,.
\end{equation}
 In addition, it can be shown that the entropy current reduces to the simple form
\begin{equation}\label{eq:entropy1}
S^i = s V_i +  \frac{\alpha}{T} \partial_i \frac{1}{T} \,.
\end{equation}
It is important to emphasize that both of the terms in the entropy current enter with a single spatial derivative of a hydrodynamic variable. Thus, in dipole conserving systems the leading order hydrodynamics is already dissipative. This happens because the lowest order contributions, in our counting scheme, allow for a dissipative transport coefficient $\alpha$.

We now turn our attention to the study of the hydrodynamic modes. To this aim, we consider linearized perturbations around the equilibrium state $(\epsilon_0,n_0,\mathbf p_0=0)$. Therefore, the fluctuations read
\begin{equation}\label{eq:equilibriumH}
    n=n_0 +\delta n, \quad \epsilon=\epsilon_0 + \delta \epsilon, \quad \mathbf p =  \mathbf{\delta p} \,,
\end{equation}
and the corresponding currents \eqref{eq:constitutive} take the form
\begin{equation}\label{eq:linearizedCurrentsZeroth}
\begin{gathered}
         J_\epsilon^i = (\epsilon_0 + P_0) V_i + \alpha \partial_i \frac{1}{T}\,, \\
          J^{ij}  = -F_{ij}\,, \hspace{10px}T^{ij} = (P_0 + \delta P) \delta_{ij} \,.  
\end{gathered}
\end{equation}
In order to solve for the evolution of the hydrodynamic variables, we express all quantities that appear in the above currents in terms of the variations of the conserved densities $\delta n, \delta \epsilon$ and  $\delta \mathbf p$. This is done using the thermodynamic relations Eqs. \eqref{eq:therm1} and  \eqref{eq:pressure}, as well as the definition of the effective velocity Eq. \eqref{eq:notation}. After doing so, the zeroth order hydrodynamic equations of motion become
\begin{equation}\label{eq:linearziedEoms}
    \begin{split}
      \partial_t \delta n -  \bar{f} \nabla^2 \nabla\cdot \mathbf{\delta p} &=0  \,, \\[2.5pt]
       \partial_t \mathbf{\delta p}    - T_0 P_{\epsilon}  \nabla \delta \epsilon -T_0 P_n  \nabla \delta n  &=0 \,, \\
        \partial_t \delta \epsilon +   \alpha s_{ee} \nabla^2 \delta \epsilon + \alpha s_{ne}  \nabla^2 \delta n -\bar{f} \frac{\epsilon_0+ p_0 }{n_0} \nabla^2 \nabla\cdot \mathbf{\delta p}  &= 0
    \end{split}
\end{equation}
where $\bar{f} = n_{0}^{-1} \Big(f_{||}+f_{\perp} \frac{d-1}{d} \Big)$. 

In order to solve the system and find the dispersion relation of the modes, we must Fourier transform the equations with frequencies and momenta ($\omega,k_i$). For this set of equations, the transverse sector ($\mathbf k\cdot\mathbf{\delta p}=0$) contains the non-dispersive mode $\omega_{shear} = 0$.

On the contrary, the longitudinal sector $(\mathbf k\times \mathbf{\delta p}= 0,\delta n, \delta \epsilon)$ is determined by the characteristic polynomial
\begin{equation}\label{eq:polynomial0}
    \left(\frac{\omega}{\omega_0}\right)^3 + i a_2 \left(\frac{\omega}{\omega_0}\right)^2   - \frac{\omega}{\omega_0} - i a_1= 0
\end{equation}
with $\omega_0 = \sqrt{a_0}k^2$, and 
\begin{equation}
\begin{split}
 a_0 &= - T_0 \bar{f} n^{-1}_0 \big[ n_0 P_{n}+(\epsilon_0 + P_0)P_{\epsilon} \big] \,, \\
    a_1 &= -a_0^{-\frac{3}{2}}\alpha T_0  \bar{f} \big[ s_{n\epsilon}  P_{\epsilon} -   s_{\epsilon \epsilon} P_{n} \big]  \,,\\
     a_2 &=  -a_0^{-\frac{1}{2}}\alpha s_{\epsilon \epsilon}  \,,
    \end{split}
\end{equation}
where the conditions Eqs. \eqref{eq:stability} and \eqref{eq:alpha} imply $a_0 \geq 0$ and $0 \leq a_1 < a_2$. 

The solutions to Eq. \eqref{eq:polynomial0} are
\begin{equation}\label{eq:dispersion0}
    \begin{split}
        \frac{\omega_1}{\omega_0} &= -i\frac{x}{3} \left(\frac{x}{ y}-\frac{y}{x}\right) -\frac{1}{3} i a_2\,,\\
      \frac{\omega_2}{\omega_0} & = \frac{x}{2\sqrt{3}} \left(\frac{x}{ y} + \frac{y}{x}\right) + i\frac{x}{6} \left(\frac{x}{ y}-\frac{y}{x}\right)- i\frac{1}{3}   a_2\,,\\
       \frac{\omega_3}{\omega_0} & = -\frac{x}{2\sqrt{3}} \left(\frac{x}{ y}+\frac{y}{x}\right) + i\frac{x}{6} \left(\frac{x}{ y}-\frac{y}{x}\right)- i\frac{1}{3}   a_2 \,.
    \end{split}
\end{equation}
with
\begin{equation}\label{eq:31}
    \begin{gathered}
        x^2=  3-a_2^2\,, \\
       2y^3= -27  a_1  + 9   a_2 - 2  a_2^3 
        \\
        +3\sqrt{81 a_1^2+6 a_1 a_2 \left(2 a_2^2-9\right)-3 a_2^2+12} \,.
    \end{gathered}
\end{equation}
Notice that all modes have a $k^2$ dependence. 
Actually, it is possible to expand the solutions for small and large values of the thermal conductivity. In the former case, which corresponds to $a_2\ll 1$ and $a_1/a_2$ fixed, the modes read
\begin{equation}
\begin{split}
    \omega_1 &\approx -i \sqrt{a_0} a_1 k^2\,,\\
    \omega_2 & \approx\sqrt{a_0} \left(1 -\frac{1}{2}i(a_2-a_1)\right)k^2\\
    \omega_3 & \approx -\sqrt{a_0} \left(1 +\frac{1}{2}i(a_2-a_1)\right)k^2 \,.
\end{split}
\end{equation}
Whereas in the latter ($a_2\gg 1$ and $a_1/a_2$ fixed) the asymptotic behaviors of the dispersion relations are
\begin{equation}
\begin{split}
    \omega_1 &\approx \sqrt{a_0}\left( \sqrt{\frac{a_1}{a_2}} - i\frac{a_2-a_1}{2a_2^2} \right)k^2\,,\\
    \omega_2 & \approx -i\sqrt{a_0}\left( a_2 - \frac{a_2-a_1}{a_2^2}\right)k^2\,,\\
    \omega_3 & \approx -\sqrt{a_0}\left(  \sqrt{\frac{a_1}{a_2}} + i\frac{a_2-a_1}{2a_2^2}\right)k^2 \,.
    \end{split}
\end{equation}

\begin{center}
    \begin{figure*}[t!]
        \centering
\includegraphics[width=0.3\columnwidth]{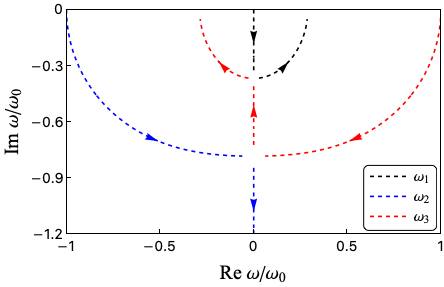}
\includegraphics[width=0.3\columnwidth]{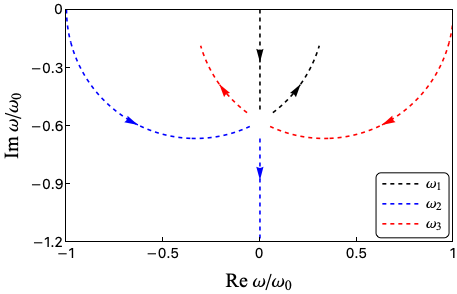}
\includegraphics[width=0.3\columnwidth]{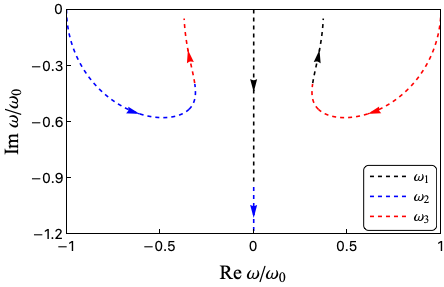}
        \caption{Trajectories of the longitudinal modes in the frequency complex plane as a function of the thermal conductivity at fixed  with arrows indicating the direction of increase in thermal conductivity. \textbf{Left:} Frequencies as a function of the thermal conductivity for $a_1/a_2=0.75(a_1/a_2)_c$ . \textbf{Middle:} Longitudinal modes in the critical regime $(a_1/a_2)_c$, the collision of the modes happens at $a_2=\sqrt{3}$. \textbf{Right:} Trajectories in Regime II for  $a_1/a_2=1.25(a_1/a_2)_c$. Reproduced from \cite{Glodkowski:2022xje}. }
        \label{fig:fig1}
    \end{figure*}
\end{center}

The full dependence of the modes as a function of the adimensional thermal conductivity $a_2$  at fixed $a_1/a_2$ is shown in Fig. \ref{fig:fig1}. We notice the existence of two qualitatively distinct regimes corresponding to $a_1/a_2<1/9$ (Regime I) and $a_1/a_2>1/9$ (Regime II). Actually, at the critical regime $(a_1/a_2)_c = 1/9$ there is a point for which the three modes are equal and purely imaginary. This situation is shown in the middle plot in Fig. \ref{fig:fig1}. The main difference between Regimes I and II is the existence of a window in the parameter space where the three modes are purely imaginary (Regime I), whereas in Regime II there will always be two modes with a non-vanishing real part. 
Finally, it is worth to point out that the thermodynamic constraints $a_1/a_2<1$ and $\alpha\geq0$ are enough to guarantee that the imaginary part of the modes is negative, and no linear instability will occur.

\paragraph{First order hydrodynamics}\label{eq:dissipativeHydrodynamics}

In the previous section, we have shown the existence of one dissipative transport coefficient $\alpha$ that controls how longitudinal fluctuations diffuse in the system. However, the shear mode $\omega_{shear}=0$ remained insensitive to the thermal conductivity. Therefore, at that level of the derivative expansion, transverse fluctuations will not diffuse. Although we may think this fact is reminiscent of the fractonic nature of the system, in this section we will prove that this is not the case. Actually, the first order transport coefficients will introduce transverse contributions to the next to leading order hydrodynamic equations of motion and predict a subdiffusive shear mode.

To proceed with the analysis, we decompose the currents into the zeroth and first order contributions according to the derivative counting scheme 
\begin{equation}
\begin{split}\label{eq:decomposition}
    & J^{ij}=J^{ij}_{0} + J^{ij}_{1}   , \quad \mathcal{E}^{i}=\mathcal{E}^i_{0} + \mathcal{E}^i_{1}\,, \\
   & T^{ij} = T^{ij}_{0} + T^{ij}_{1}  , \quad  S^{i}=S^i_{0} + S^i_{1}  \,,
    \end{split}
\end{equation}
and plug the decomposition Eq. \eqref{eq:decomposition} into Eq. \eqref{eq:passing} and cancel out the lower order terms (as these satisfy the second law provided that $\alpha \geq 0$). Then, the second law requires that 
\begin{equation}\label{eq:2ndlawD}
    \partial_i S^i_{1} - \frac{1}{T}\partial_i \mathcal{E}^i_{1}
    + \frac{\mu}{T} \partial_i \partial_j  J^{ij}_{1} + \frac{V_i}{T} \partial_j T^{ji}_{1} \geq 0.
\end{equation}
Note that for fluctuations around Eq. \eqref{eq:equilibriumH} we have that $\tilde{\mu} = \mu$ and $J^i_{\epsilon} = \mathcal E^i$ (see Eqs. \eqref{eq:notation} and \eqref{eq:shifted}). Since our goal is to identify the most general constitutive relations consistent with the above inequality, we find it convenient to rewrite Eq. \eqref{eq:2ndlawD} as follows
   \begin{equation}\label{eq:diss}
   \begin{gathered}
  \partial_i \Big(S^i_{1}  - \frac{1}{T}  \mathcal{E}^i_{1} + \frac{\mu}{T}  \partial_j J_{1}^{ij} -  \partial_j ( \frac{\mu}{T} ) J_{1}^{ij} + \frac{V_j}{T} T_{1}^{ij}  \Big) \\
 + J_{1}^{ij} \partial_i \partial_j  \frac{\mu}{T}  +  \mathcal{E}_{1}^i \partial_i \frac{1}{T} - T_{1}^{ij} \partial_i \frac{V_j}{T} \geq 0\,.
   \end{gathered}
 \end{equation}
Ignoring non-linearities, it is possible to express the energy current without loss of generality as $\mathcal{E}_1^i = \partial_j E^{ji}$. With that ansatz at hand, the entropy production constraint can be written as 
   \begin{equation}\label{eq:diss2}
   \begin{gathered}
  \partial_i \Big(S^i_{1}  - \frac{1}{T}  \partial_j E^{ji} + \partial_j \frac{1}{T} E^{ij} + \frac{\mu}{T}  \partial_j J_{1}^{ij} -  \partial_j  \frac{\mu}{T}  J_{1}^{ij} + \frac{V_j}{T} T_{1}^{ij}  \Big) \\
 + J_{1}^{ij} \partial_i \partial_j  \frac{\mu}{T}  -  E^{ij} \partial_i \partial_j \frac{1}{T} - T_{1}^{ij} \partial_i \frac{V_j}{T} \geq 0 \,.
   \end{gathered}
 \end{equation}
Then, we set the first order correction to the entropy current to be
\begin{equation}
S^i_{1}  = \frac{1}{T}  \partial_j E^{ji} - \partial_j \frac{1}{T} E^{ij} - \frac{\mu}{T}  \partial_j J_{1}^{ij} +  \partial_j  \frac{\mu}{T}  J_{1}^{ij} - \frac{V_j}{T} T_{1}^{ij} \,,
\end{equation}
such that the first term in Eq. \eqref{eq:diss2} vanishes. Therefore, the problem is reduced to finding the proper constitutive relations such that the leftover is semi-positive definite. In fact, the most general form for the currents that will allow a positive production read
\begin{equation}\label{eq:macroscopic}
\begin{split}
    J^{ij}_1 &=  \Big( \frac{\gamma_{||} }{T_0} \partial_k V_k + \sigma_{||} \nabla^2 \frac{\mu}{T} + \beta_{||} \nabla^2 \frac{1}{T} \Big) \delta_{ij} \\
    &+  \frac{\gamma_{\perp} }{T_0} \partial_{\langle i} V_{j \rangle}  + \sigma_{\perp} \partial_{\langle i} \partial_{j \rangle} \frac{\mu}{T}  + \beta_{\perp} \partial_{\langle i} \partial_{j \rangle} \frac{1}{T} \,, \\
     - T^{ij}_1 &= \Big( \frac{\zeta}{T_0} \partial_k V_k + \bar{\gamma}_{||} \nabla^2 \frac{\mu}{T} + \alpha_{||} \nabla^2 \frac{1}{T} \Big) \delta_{ij} \\
    &+ \frac{\eta}{T_0}\partial_{\langle i} V_{j \rangle}  + \bar{\gamma}_{\perp} \partial_{\langle i} \partial_{j \rangle} \frac{\mu}{T}  + \alpha_{\perp} \partial_{\langle i} \partial_{j \rangle} \frac{1}{T} \,,\\
     - E^{ij} & = \Big( \frac{\bar{\alpha}_{||}}{T_0} \partial_k V_k + \bar{\beta}_{||} \nabla^2 \frac{\mu}{T} + \kappa_{||} \nabla^2 \frac{1}{T} \Big) \delta_{ij} \\
    &+ \frac{\bar{\alpha}_{\perp} }{T_0} \partial_{\langle i} V_{j \rangle}  + \bar{\beta}_{\perp} \partial_{\langle i} \partial_{j \rangle} \frac{\mu}{T}  + \kappa_{\perp} \partial_{\langle i} \partial_{j \rangle} \frac{1}{T} \,.
    \end{split}
\end{equation}
Where we have introduced a set of 18 dissipative transport coefficients. In particular, Onsager reciprocity reduces the number of off-diagonal coefficients if time-reversal invariance is imposed 
\begin{equation}\label{eq:onsager}
    \bar{\alpha}_{||(\perp)} =  \alpha_{||(\perp)}\,, \quad \bar{\beta}_{||(\perp)} =  \beta_{||(\perp)}\,,  \quad \bar{\gamma}_{||(\perp)} =  \gamma_{||(\perp)}\,.
\end{equation}

Moreover, the entropy production constraint Eq. \eqref{eq:diss2} can be written in a compact matrix form
\begin{equation}\label{eq:matrixInequality}
    \boldmath{x}^{\intercal} \mathcal{A}_{\boldmath{||}} \boldmath{x} + \boldmath{y}^{\intercal}\mathcal{A}_{\boldmath{\perp}} \boldmath{y} \geq 0 \,,
\end{equation}
with
\begin{equation}
\begin{gathered}
    \mathcal{A}_{\boldmath{||}} = \begin{pmatrix}
\zeta & \gamma_{||} & \alpha_{||} \\
\gamma_{||} & \sigma_{||} & \beta_{||}\\
\alpha_{||} & \beta_{||} & \kappa_{||}
\end{pmatrix}, \quad    A_{\boldmath{\perp}} = \begin{pmatrix}
\eta & \gamma_{\perp} & \alpha_{\perp} \\
\gamma_{\perp} & \sigma_{\perp} & \beta_{\perp}\\
\alpha_{\perp} & \beta_{\perp} & \kappa_{\perp}
\end{pmatrix}  \,, \\
\boldmath{x} = \begin{pmatrix}
   \partial_i \frac{V_i}{T} \\ \nabla^2 \frac{\mu}{T} \\ \nabla^2 \frac{1}{T} \end{pmatrix}\,, \hspace{5px}\boldmath{y} = \begin{pmatrix}
   \partial_{\langle i} \frac{V_{i \rangle}}{T} \\ \partial_{\langle i} \partial_{j \rangle} \frac{\mu}{T} \\ \partial_{\langle i} \partial_{j \rangle} \frac{1}{T}
\end{pmatrix} \,.
\end{gathered}
\end{equation}
Since the two contributions are independent, the second law is then imposed by requiring that matrices $\mathcal{A}_{\boldmath{||}}$ and $\mathcal{A}_{\boldmath{\perp}}$ are both semi-positive definite. This poses constraints on the transport coefficients, which are summarized below. 

In total, we have found 12 independent transport coefficients that we classify in two distinct categories. The first category involves the diagonal coefficients $(\zeta,\eta,\sigma_\perp,\sigma_{||},\kappa_\perp,\kappa_{||})$ satisfying a positivity constraint
\begin{equation}\label{eq:viscosities}
    \zeta \,, \eta\,,  \sigma_{\perp} \,, \sigma_{||}\,, \kappa_{\perp}\,, \kappa_{||} \ \geq 0\,.
\end{equation}
On the other hand, the second category consists of the off-diagonal terms $(\alpha_\perp,\alpha_{||},\beta_\perp,\beta_{||},\gamma_\perp,\gamma_{||})$ obeying inequalities with the coefficients of the previous group
\begin{equation}\label{eq:coefficients}
\begin{gathered}
     \alpha^2_{\perp} \leq \sigma_{\perp} \kappa_{\perp}\,, \hspace{10px}\alpha^2_{||} \leq \sigma_{||} \kappa_{||}\,, \hspace{10px}   \beta^2_{\perp} \leq \eta \kappa_{\perp}, \\
     \beta^2_{||} \leq \zeta \kappa_{||}\,, \hspace{10px}
    \gamma^2_{\perp} \leq \eta \sigma_{\perp}\,, \hspace{10px}  \gamma^2_{||} \leq \zeta \sigma_{||}\,.
    \end{gathered}
\end{equation}
The distinction has been motivated by the fact that the value of the off-diagonal transport coefficients is bounded from above by the diagonal ones. The last constraint from semi-positivity corresponds with the positive determinant condition 
\begin{equation}
\begin{split}
       \zeta(\sigma_{||}\kappa_{||}-\beta^2_{||})-\kappa_{||}\gamma^2_{||}
        -\sigma_{||} \alpha^2_{||}\geq 0\,, \\
               \eta(\sigma_{\perp}\kappa_{\perp}-\beta^2_{\perp})-\kappa_{\perp} \gamma^2_{\perp}
               -\sigma_{\perp} \alpha^2_{\perp} \geq 0\,.
               \end{split}
\end{equation}
Having the corrections to the zeroth order hydrodynamic constitutive currents, we can plug them into the conservation equations to obtain the first order hydrodynamic equations of motion. In particular, they read
\begin{equation}\label{eq:eoms1}
    \begin{gathered}
\partial_t \delta n +  j_n \nabla^4 \delta n- (\bar{f} - j_v \nabla^2)\nabla^2(\nabla\cdot \mathbf{\delta p})  + j_e \nabla^4 \delta \epsilon  = 0\,, \\[2.5pt]
\partial_t  \mathbf{\delta p} + t_{v_{||}} \nabla^2 \nabla ( \nabla\cdot \mathbf{\delta p}) +  t_{v_{\perp}} \nabla^4 \mathbf{\delta p} \\
- (T_0 P_n  - t_n  \nabla^2) \nabla \delta n  - (T_0 P_{\epsilon}  -  t_e \nabla^2 ) \nabla \delta \epsilon 
  = 0\,, \\[2.5pt]
\partial_t \epsilon +  (  \alpha s_{ee} +  e_e \nabla^2)\nabla^2 \delta \epsilon + (\alpha s_{ne} + e_n \nabla^2)\nabla^2 \delta n\\
-\left(\bar{f} \frac{\epsilon_0+ p_0 }{n_0}  -e_v \nabla^2\right)\nabla^2( \nabla\cdot \mathbf{\delta p}   )
   = 0\,.
    \end{gathered}
    \end{equation}    
For a detailed derivation of the equations and the relation of the parameters shown in Eqs. \eqref{eq:eoms1} with the transport coefficients in Eqs. \eqref{eq:macroscopic}, we refer the reader to the Appendix \ref{sec:dissipativeCorrections}. The main output of the first order approach is the conversion of the non-dispersive shear mode into a subdiffusive one
\begin{equation}\label{eq:shear}
   \omega_{shear} = -i \eta \frac{f_\perp}{T_0n^2_0} k^4\,.
\end{equation}
On the other hand, the longitudinal modes are not strongly affected by the first order corrections, since their contribution enters at higher order in momentum. In fact, the characteristic polynomial in this case takes the same form as in Eq. \eqref{eq:polynomial0}
\begin{equation}\label{eq:polynomial1}
    \left(\frac{\omega}{\bar{\omega}_0}\right)^3 + i \bar{a}_2 \left(\frac{\omega}{\bar{\omega}_0}\right)^2   - \frac{\omega}{\bar{\omega}_0} - i \bar{a}_1= 0
\end{equation}
where $\bar{\omega}_0 = \sqrt{\bar{a}_0} k^2$ and
\begin{equation}
\begin{split}
\bar{a}_0 &= (a_0 + b_0 k^2) +\mathcal O (k^4)\,, \\
    \bar{a}_1 &= \bar{a}^{-\frac{3}{2}}_0 (a^{\frac{3}{2}}_0 a_1 + b_1 k^2) +\mathcal O (k^4)\,, \\
    \bar{a}_2 &=  \bar{a}^{-\frac{1}{2}}_0 (a^{\frac{1}{2}}_0 a_2 + b_2 k^2)+\mathcal O (k^4)\,.
    \end{split}
\end{equation}
with $b_0, b_1$ and $b_2$ derived in Appendix \ref{sec:dissipativeCorrections}, and shown in Eq. \eqref{eq:bsDef}. Actually, the solutions Eqs. \eqref{eq:dispersion0} still apply, once we substitute $a_1 \rightarrow \bar{a}_1$ and $a_2 \rightarrow \bar{a}_2$ in the Eqs. \eqref{eq:31}.

\subsection{Fracton superfluids}
\textit{This section is based on Ref.~\cite{Glodkowski:2024ova}.} \newline

\noindent In the previous section, we analyzed dipole-conserving fluids where the dipole symmetry is spontaneously broken in generic equilibrium states. We now turn to a qualitatively different fracton fluid phase in which both the monopole and dipole symmetries are spontaneously broken. As established in Sec.~\ref{sec:dipole-sigma-model}, this symmetry breaking pattern gives rise to two Goldstone fields, associated with the monopole and dipole charges, whose interplay leads to unconventional low-energy dynamics.

Our goal in this section is to construct the finite-temperature hydrodynamic theory of such a superfluid phase. This can be viewed as a natural generalization of the dipole sigma model to finite temperature, incorporating both the dynamics of the Goldstone sector and the normal component. In this sense, the resulting theory provides an analogue of the Landau--Tisza two-fluid model \cite{Landau_Theory,Tisza} adapted to systems with nonuniform multipole symmetries.

A key feature of this phase is the coexistence of multiple collective components: a monopole condensate, a dipole condensate, and a normal component carrying entropy and momentum. As we will see, the nonuniform symmetry structure imposes strong constraints on the low-energy dynamics. In particular, not all of these degrees of freedom remain independent in the hydrodynamic limit, and some become effectively gapped or enslaved variables.

We begin by formulating the ideal, nondissipative theory using the Poisson bracket approach, which allows us to systematically derive the constitutive relations from the underlying symmetry structure. We then analyze the spectrum of linearized fluctuations, identifying both gapless hydrodynamic modes and additional gapped excitations. Finally, we incorporate dissipative effects using the entropy current formalism and study the resulting long-wavelength dynamics.

\subsubsection{Ideal fracton superfluids}\label{sec:3}

We begin by constructing the ideal hydrodynamic theory of fracton superfluids and deriving the corresponding constitutive relations from the underlying Poisson bracket structure. We then analyze the spectrum of linearized fluctuations, identifying both gapless and gapped collective modes.

Throughout this section we assume that the spin density has relaxed and does not constitute an independent hydrodynamic degree of freedom. We nevertheless retain the non-hydrodynamic dipole sector at this stage, including the intrinsic dipole density and the gapped collective superfluid mode. This separation of scales assumes that spin relaxation occurs on timescales parametrically shorter than those associated with the dipole sector, $\tau_{\rm spin} \ll \tau_{\rm dipole}$. In the next section we take the long-wavelength hydrodynamic limit and integrate out all gapped excitations.

\paragraph{Three-component fluid.}
Spontaneous breaking of monopole and dipole symmetries introduces two Goldstone fields, $\theta$ and $\psi_i$, associated with the broken monopole and dipole transformations, respectively. As discussed in Sec.~\ref{sec:dipole-sigma-model}, the corresponding translation-invariant superfluid variables are
\be
v_i^s = \partial_i \theta - \psi_i\,,
\qquad
\xi_{ij} = \partial_i \psi_j\,,
\ee
which transform homogeneously under spacetime translations and therefore constitute appropriate dynamical variables.

At finite temperature, the superfluid phase contains in addition a normal component carrying energy and momentum. Accordingly, at this stage the dynamics is formulated in terms of the charge, intrinsic dipole, momentum, and energy densities, together with the superfluid variables $v_i^s$ and $\xi_{ij}$.

The equations of motion consist of the continuity equations for the conserved densities, the evolution equation for the intrinsic dipole density, and the Josephson relations governing the Goldstone sector
\be
\bg \label{eq:Idealeoms}
\partial_t n + \partial_i J^i = 0\,, \quad
\partial_t \pi_i + \partial_j K^{ji} = - J^i\,, \\
\partial_t p_i + \partial_j T^{ji} = 0\,, \quad
\partial_t \epsilon + \partial_i E^i = 0\,, \\
\partial_t v_i^s = \partial_i \mu - \mu_i\,, \quad
\partial_t \xi_{ij} = \partial_i \mu_j\,.
\eg
\ee
In principle, the Josephson relations may receive corrections. In the nondissipative regime, however, they are inherited from the canonical Poisson structure of the Goldstone fields and therefore coincide with the zero-temperature relations in Eq.~\eqref{eq:josephson}, as we verify below.

With the equations of motion at hand, we must identify the variables that parametrize the local thermodynamic state. In general, the Hamiltonian density can depend only on combinations of fields that are invariant under the global symmetries. This requirement is particularly restrictive in dipole-conserving systems, where momentum transforms nontrivially due to the nonuniform dipole algebra.

Keeping this in mind, we postulate that a finite temperature generalization of the theory Eq. \eqref{eq:microcanonical} is given by the following Hamiltonian density
\be
h \equiv h[n\,, s\,, \pi_i\,, v^s_i\,, \xi_{ij}\,, \Tilde{p}_i ]
\ee
where $v^s_i = \partial_i \theta - \psi_i$ and $\xi_{ij} = \partial_i \psi_j$ are the superfluid velocities familiar from the zero temperature theory and $\tilde{p}_i = p_i + n\partial_i\theta + \pi_j\partial_i\psi_j$ is what we dub as \textit{thermal momentum}. It is an invariant combination involving momentum that can be understood as the momentum carried by the thermal excitations. Thus, we can interpret the finite temperature fracton superfluid phase as a three-fluid model involving two superfluid and one normal components respectively that can flow independently of one another. However, as we will see, the superfluid velocity $v^s_i$ will be gapped, rendering it irrelevant in the hydrodynamic regime.

Let us also introduce a set of the conjugate variables 
\be \begin{split}\label{eq:definitions}
\mu &=  \frac{\delta h}{\delta n}\,, \quad \mu_i =  \frac{\delta h}{\delta \pi_i}\,, \\ 
T &=   \frac{\delta h}{\delta s}\,, \quad 
\Tilde{V}_i = \frac{\delta h}{\delta \Tilde{p}_i}\,, \\
\lambda_i &= \frac{\delta h}{\delta v^s_i}\,, \quad F_{ij} = \frac{\delta h}{\delta \xi_{ij}}\,,
\end{split}
\ee 
such that the differential of the Hamiltonian density is given as 
\be \label{eq:firstLaw}
dh = \mu dn + \mu_i d \pi_i + T ds + \Tilde{V}_i d \Tilde p_i + \lambda_i d v^s_i + F_{ij} d \xi_{ij}\,.
\ee 
The above relation can be understood at the first law of thermodynamics for fracton superfluids.

\paragraph{Poisson bracket method.}
A simple way to obtain the ideal form of the constitutive relations in hydrodynamics is the Poisson bracket method. Once the commutators between the hydrodynamic variables are known and the Poisson bracket structure is established, the equations of motion for any function of these variables may be determined directly via 
\be \label{eq:evolution}
\partial_t \mathcal{A} = \{\mathcal{A}\,, \mathcal{H} \}\,.
\ee 
Furthermore, the structure of the Poisson bracket is universal, determined by the generic physical considerations such as symmetries rather than the microscopic details of a particular model. In fact, Poisson brackets were successfully applied in order to derive the hydrodynamic constitutive relations for the ideal fracton fluids \cite{PhysRevResearch.3.043186}. Here we combine the developments of \cite{PhysRevResearch.3.043186} with previous studies of conventional superfluids \cite{son_hydrodynamics_2001,PhysRevD.77.025004}.

For fracton superfluids, all non-trivial commutation relations between the hydrodynamic variables may be obtained in the zero-temperature theory using the canonical Poisson bracket \eqref{eq:canonical}. Then, it is not too hard to infer that the appropriate noncanonical Poisson bracket structure capturing all commutation relations is given by
\be \begin{split}\label{eq:noncanonical}
\{F, G\}_{NC}  = &- \int d^d x  \Big[ p_i \Big(\frac{\delta F}{\delta p_j}  \partial_j \frac{\delta G}{\delta p_i} - \frac{\delta G}{\delta p_j}  \partial_j \frac{\delta F}{\delta p_i}\Big) + s \Big(\frac{\delta F}{\delta p_j}  \partial_j \frac{\delta G}{\delta s} - \frac{\delta G}{\delta p_j}  \partial_j \frac{\delta F}{\delta s}\Big) \\
&+ n \Big(\frac{\delta F}{\delta p_j}  \partial_j \frac{\delta G}{\delta n} - \frac{\delta G}{\delta p_j}  \partial_j \frac{\delta F}{\delta n}\Big) + \pi_i \Big(\frac{\delta F}{\delta p_j}  \partial_j \frac{\delta G}{\delta \pi_i} - \frac{\delta G}{\delta p_j}  \partial_j \frac{\delta F}{\delta \pi_i}\Big) \\
&-\partial_i \theta \Big(\frac{\delta F}{\delta p_i} \frac{\delta G}{\delta \theta} - \frac{\delta G}{\delta p_i} \frac{\delta F}{\delta \theta} \Big) -\partial_i \psi_j \Big(\frac{\delta F}{\delta p_i} \frac{\delta G}{\delta \psi_j} - \frac{\delta G}{\delta p_i} \frac{\delta F}{\delta \psi_j} \Big)\\
&-\Big(\frac{\delta F}{\delta \theta} \frac{\delta G}{\delta n} - \frac{\delta G}{\delta \theta} \frac{\delta F}{\delta n} \Big)- \Big( \frac{\delta F}{\delta \psi_i} \frac{\delta G}{\delta \pi_i} - \frac{\delta G}{\delta \psi_i} \frac{\delta F}{\delta \pi_i} \Big)
\Big]\,.
\end{split}
\ee 
The first four terms in the equation arise from the identification of momentum as a generator of translations, while the next two lines follow directly from \eqref{eq:poisson}. The final two lines, on the other hand, are a consequence of the fact that the charge and dipole Goldstone modes and their corresponding charges are canonically conjugated \eqref{eq:canonical}, which is a direct generalization of the ordinary $\mathrm U(1)$ case (see e.g. \cite{staruszkiewicz_quantum_1989}).

Using the noncanonical Poisson structure, as well as the definitions \eqref{eq:definitions}, it is then a straightforward exercise to determine the explicit form of the equations of motion by computing the evolution of the hydrodynamic variables via \eqref{eq:evolution}. From the equations of motion corresponding to the hydrodynamic densities we may read off the constitutive relations 
\be \begin{split}\label{eq:ideal}
J^i &= - \lambda_i \,, \\
K^{ij} &= - F_{ij}\,, \\
S^i &= s \Tilde V_i\,, \\
T^{ji}&=P \delta_{ij} + F_{jk}  \xi_{ik} + \lambda_j \partial_i \theta + \Tilde{V}_j \Tilde{p}_i\,.
\end{split}
\ee 
where the pressure has been defined as
\be \label{eq:pressure}
P  = n \mu + \pi_i \mu_i + Ts + \Tilde{V}_i \Tilde{p}_i - h \,.
\ee
Interestingly, we notice that the fractonic nature of the superfluid is manifested in the fact that the charge current receives contribution only from the superfluid velocity whereas both the superfluid velocities and normal velocity contribute to the momentum flow. In fact, if  we compute the Poisson bracket between the thermal momentum $\tilde p_i$ and the charge densities we obtain
\begin{equation}\label{eq_neutralityP}
    \{\tilde p_i (\mathbf{x}), n(\mathbf{y})\} =   \{\tilde p_i (\mathbf{x}), \pi_j(\mathbf{y})\} = 0 \,,
\end{equation}
which suggests that the thermal momentum is fracton charge ``neutral", and what in the ordinary superfluid model  would correspond to the normal component, in this case is made of thermal excitations that do not carry neither monopole nor dipole charges.

On the other hand, equations for the Goldstones lead to the same form of the Josephson's relations as derived in \eqref{eq:josephson}. In addition, it is possible to symmetrize the stress tensor by the addition of the suitable improvement terms. The improved momentum density reads 
\be 
p_i \rightarrow p_i +\frac{1}{2}\Big( \pi_j \partial_j \psi_i + \psi_i \partial_j \pi_j - \pi_i 
 \partial_j \psi_j  - \psi_j
 \partial_j \pi_i \Big)
\ee 
while the symmetric stress tensor is 
\be \begin{split}
T^{ij} &= P \delta_{ij}+\Tilde{V}_i \Tilde{p}_j  +\lambda_i v^s_j + \lambda_{(i} \psi_{j)}  - F^{(i|k} \partial_{k|} \psi_{j)}   +  F^{(ij)} \partial_k \psi_k + F^{(i |k|} \partial_{j)} \psi_k  \\
& - \partial_k F^{(i|k|} \psi_{j)}+ \partial_k F^{(ij)} \psi_k\,.
\end{split}
\ee 
Energy current cannot be directly computed in the Poisson bracket formalism, however, can be obtained using the first law of thermodynamics. Indeed, from \eqref{eq:firstLaw} after using the equations of motion \eqref{eq:Idealeoms} we find that the energy current is given as 
\be 
E^i = \Tilde{V}_i (sT + \Tilde{p}_j \Tilde{V}_j ) - \mu \lambda_i - \mu_j F_{ij}\,.
\ee 
With this we conclude our analysis of the ideal constitutive relations.
\paragraph{Hydrodynamic modes.}
Let us now turn our attention to the study of the hydrodynamic modes of ideal superfluids with dipole symmetry. For this purpose, we consider small deviations away from the stationary background: 
 \be 
 \bg
 n=n_0 + \delta n\,, \quad \Tilde{p}_i =  \delta\Tilde{p}_i \,, \quad \pi_i =  \delta \pi_i \,,  \\
 \epsilon= \epsilon_0 + \delta \epsilon\,, \quad \psi_i = \delta \psi_i\,, \quad  \theta =  \mu_0 t + \delta \theta\,.
 \eg
 \ee 
The dynamics of the perturbations is captured by the following set of linear equations 
 \be \begin{split}
\partial_t \delta n - \partial_i \delta \lambda_i &= 0\,, \\
\partial_t \pi_i -\partial_j \delta F_{ji} - \delta \lambda_i &= 0\,, \\
\partial_t \epsilon + s_0 T_0 \partial_i \delta \Tilde{V}_i - \mu_0 \partial_i \delta \lambda_i &= 0\,, \\
\partial_t \delta \tilde p_i + s_0 \partial_i \delta T   &=0 \,, \\
\partial_t  v^s_i  - \Big( \partial_i \delta \mu - \delta  \mu_i  \Big) &= 0 \,, \\
\partial_t \xi_{ij} + \partial_i \delta \mu_j &= 0\,.
\end{split}
\ee
To close the system and determine the evolution of the hydrodynamic modes, it is necessary to express the fluctuations in thermodynamics variables $\{\delta \mu\,, \delta \mu_i\,, \delta T\,, \Tilde V_i\,, \lambda_i\,, \delta F_{ij} \}$ in terms of the fundamental quantities $\{n\,, \pi_i\,, \epsilon\,, \Tilde p_i\,, v^s_i\,, \xi_{ij} \}$\footnote{These are the variables associated with quantities that are well-defined microscopically.}.

To achieve this, we expand the entropy density function up to the second order in perturbations
 \be\begin{split} \label{eq:entropyDensity}
s &= s_0+\frac{1}{T_0} \delta \epsilon  - \frac{\mu_0}{T_0} \delta n - \frac{1}{2 T_0}\chi_{nn} \delta n ^2 - \frac{1}{2 T_0}\chi_{\epsilon \epsilon} \delta \epsilon ^2 + \frac{1}{T_0} \chi_{n\epsilon} \delta \epsilon \delta n \\
     &- \frac{1}{2T_0}\lambda_{ijkl} \xi_{ij} \xi_{kl} - \frac{\chi_v}{2T_0}  (\delta v^s_i)^2 - \frac{\chi_p}{2T_0} \delta \Tilde{p}_i^2 - 
     \frac{\chi_\pi}{2T_0} \delta \pi_i^2  \,,
     \end{split}
 \ee 
where $\lambda_{ijkl} = \lambda_1 \delta_{ij} \xi + \lambda_2 \xi_{\langle i j \rangle} + \lambda_3  \xi_{[i j]}$ is the most general SO(3) invariant tensor. Using the definitions \eqref{eq:definitions} we obtain the fluctuations in the conjugate variables 
\be \begin{split} \label{eq:thermo}
 \delta T &=  T_0 \chi_{ee} \delta e - T_0 \chi_{ne} \delta n \,, \\
 \delta \mu &=  \chi_e \delta e +  \chi_n \delta n\,, \\
\delta \mu_i & =  \chi_{\pi} \delta \pi_i\,, \\
\delta \tilde V_i &=   \chi_{p} \delta \Tilde{p}_i \,,\\
\delta  \lambda_i &=    \chi_{v} \delta (\partial_i \theta - \psi_i) \,,\\
\delta F_{ij} &=   \lambda_{ijkl} \partial_k \psi_l = \lambda_1 \delta_{ij}\partial_k \psi_k + \lambda_2 \partial_{\langle i} \psi_{ j \rangle } + \lambda_3  \partial_{[ i} \psi_{ j ] } \,,
\end{split}
\ee 
where we have defined
\be
 \chi_e = \Big( \mu_0 \chi_{ee} - \chi_{ne} \Big)\,, \quad  \chi_n = \Big( \chi_{nn} - \mu_0  \chi_{ne} \Big)\,.
\ee 
Then, the set of equations can be expressed in a matrix form as 
\be
\begin{pmatrix}
\mathcal M^{6 \times 6}_{||}  & 0 \\ 
0  & \mathcal M^{3 \times 3}_{\perp} \\
\end{pmatrix} \begin{pmatrix}
\textbf{v}_{||}  \\ 
 \textbf{v}_{\perp}   \\
\end{pmatrix} = 0\,,
\ee 
where 
\be 
\textbf{v}_{||} = \begin{pmatrix}
\tilde n\, \\
\tilde \pi_{||}\, \\ \tilde \epsilon\, \\
\tilde p_{||}\, \\
\tilde \theta\, \\
\tilde \psi_{||}
\end{pmatrix}\,, \quad \mathcal M^{6 \times 6}_{||} =
\begin{pmatrix}
-i \omega & 0 & 0 & 0 & \chi_v k^2 & i \chi_v k \\
0 & -i\omega & 0 & 0 & -i\chi_v k & \chi_v + \lambda k^2 \\
0 & 0 & -i\omega & i s_0 T_0 \chi_{p} k & \mu_0 \chi_v k^2 & i\mu_0 \chi_v k \\
-i s_0 T_0\chi_{ne}  k & 0 & i s_0 T_0\chi_{ee}  k & -i\omega & 0 & 0 \\
- \chi_e & 0 & - \chi_n & 0 & -i \omega & 0 \\
0 & -\chi_{\pi} & 0 & 0 & 0 & -i\omega  \\
\end{pmatrix}
\ee 
\be
\textbf{v}_{\perp}  = \begin{pmatrix}
 \boldsymbol{\tilde\pi_{\perp}}\\
 \boldsymbol{\tilde p_{\perp}} \\
 \boldsymbol{\tilde\psi_{\perp}}
\end{pmatrix}\,, \quad \mathcal M^{3 \times 3}_{\perp} = \begin{pmatrix}
-i \omega & 0 & \chi_v + \lambda_2 k^2 \\
0 & -i \omega & 0 \\
-\chi_{\pi} & 0 & -i \omega
\end{pmatrix}
\ee 
correspond to longitudinal and transverse fluctuations respectively.

In the longitudinal sector, we find 3 pairs of modes. The first pair constitutes of the soundlike modes 

\be\label{eq:sound}
\omega = \pm v_s k + \mathcal{O}(k^2)\,, \quad v_s = 
 s_0 T_0 \sqrt{ \chi_{ee}  \chi_p }
 \,.
\ee 
For the second pair we find magnonlike dispersion 
\be\label{eq:mangon}
\omega = \pm v_m k^2 + \mathcal{O}(k^4)\,, \quad v_m = \sqrt{\frac{\lambda}{\chi_{ee}}\big(\chi_{ee}\chi_{nn}-\chi^2_{ne} \big)}\,.
\ee 
Finally, there is a pair of massive modes with a quadratic dispersion 
\be\begin{split}
\omega &= \pm m_0 \pm v_M k^2 +\mathcal{O}(k^4)\,, \\
v_M &= \frac{1}{2m_0} \Big( \lambda \chi_{\pi} + \big( \mu^2_0 \chi_{ee} - 2 \mu_0 \chi_{ne} +\chi_{nn} \big)\chi_v  \Big)\,.
\end{split}
\ee 
On the other hand, for the transverse sector we find 
\be \begin{split}
\omega_{\perp} =& \pm m_0 \sqrt{1+ \frac{\lambda_{\perp}}{\chi_v}k^2 } \\
\approx& \pm m_0 \pm \frac{\lambda_{\perp}}{2\chi_v} m_0 k^2\,.
\end{split}
\ee 

\subsubsection{Dissipative hydrodynamics}\label{sec:4}
In this section, we present a dissipative completion of the ideal theory and study (linearized) dissipative hydrodynamics of fracton superfluids. After postulating the derivative expansion, we systematically derive the most general hydrodynamic constitutive relations up to the third order in the derivative expansion by demanding consistency with the local form of the second law of thermodynamics. Then, we study the spectrum of the theory and verify the consistency of our approach. 

\paragraph{Gradient expansion.}

Before delving into the realms of dissipative hydrodynamics, we need to establish a systematic gradient counting scheme such that the hydrodynamic constitutive relations can be organized in a derivative expansion and the equations solved perturbatively. As we will shortly discuss, in the case of fracton superfluids there are certain inevitable subtleties in the power counting that arise due to the lack of the common scaling symmetry. As a result, the derivative expansion introduced here is in a sense not a standard one and it is therefore advisable to verify that it indeed represents a consistent truncation of the hydrodynamic currents and equations of motion.

In order to address this problem, let us list out the (non-exhaustive) set of consistency conditions that we demand from a feasible derivative counting scheme. In hydrodynamics, the dispersion relations $\omega_i(k)$ are organized as a power series in the wavevector
\be \label{eq:expansion}
\omega_{i}(k) = \sum^{\infty}_{n=0} a_{ni} k^n_{i}\,.
\ee 
Here, the index $i$ represents the different hydrodynamic modes. Then, we require the following conditions:
\begin{enumerate}\label{eq:conditions}
    \item \textbf{Consistent expansion.} The $n$-th order coefficients $a_{ni}$ are uniquely determined by the $(n-1)$-th order constitutive relations. For example, the $0$-th order hydrodynamics exactly fixes the coefficients corresponding to terms linear in $k$. 
    \item \textbf{Stable modes.} All modes are linearly stable $\mathrm{Im} \,\omega_i \leq0$ on account of the constraints from the second law of thermodynamics and stability of the entropy density function\footnote{In general, the Landau instability may occur at the critical superflow destroying superfluidity. However, in this work we assume expansion around the state with no superflow.}. 
    \item \textbf{Dissipation-entropy correspondence.} The only transport coefficients that contribute to the attenuation of the modes are the ones that add to the entropy production. 
\end{enumerate}

We now introduce our derivative-counting prescription. To this end, let us recall that the internal dipole density is not a conserved quantity. Likewise, the superfluid velocity $v^s_i$ corresponds to a gapped degree of freedom. Since neither is associated with a genuine hydrodynamic mode in the long-wavelength limit, both are counted as order one quantities. Therefore, the fluid variables are assigned the following orders\footnote{One could equivalently work with the invariant superfluid velocity $\xi_{ij} \sim \mathcal{O}(1)$ instead of the charge Goldstone or directly in terms of the two Goldstones $\theta \sim  \mathcal{O}(\nabla^{-2})$ and $\psi_i \sim \mathcal{O}(\nabla^{-1})$.}
\be 
\theta \sim \mathcal{O}(\nabla^{-2})\,, \quad \{\psi_i\,, p_i\} \sim \mathcal{O}(\nabla^{-1})\,, \quad \{n\,, \epsilon\,, \Tilde{p}_i\,, \xi_{ij}\} \sim \mathcal{O}(1)\,, \quad \{\pi_i\,, v^s_i\} \sim \mathcal{O}(\nabla)
\ee 
while for the conjugate quantities we have 
\be \label{eq:conjugate}
\{\mu\,, T\,, \Tilde{V}_i\,, F_{ij}\} \sim \mathcal{O}(1)\,, \quad \{\mu_i\,, \lambda_i\} \sim \mathcal{O}(\nabla)\,.
\ee 
A full set of independent hydrodynamic equations capturing the dynamics of dissipative fracton superfluids is then given by
\be \begin{split}
 \label{eq:eoms}
\partial_t n + \partial_i J^i &= 0\,, \\
\partial_t \pi_i + \partial_j K^{ji} +J^i &= 0\,, \\
\partial_t \tilde p_i + \partial_j \tilde T^{ji} -f_i&= 0\,, \\
\partial_t \epsilon + \partial_i \epsilon^i &=0\,, \\
\partial_t  \psi_i  -\Big( \mu_i+\tilde \mu_i^d \Big)  &= 0  \,, \\
\partial_t \theta-\Big( \mu+\mu^d \Big) &=0\,.
\end{split}
\ee
Notice that we have decided to parameterize our fluid in terms of the independent variables $ \{n\,, \epsilon\,,  \tilde p_i\,, \theta\,, \psi_i\,, \pi_i\}$ while also allowing for the presence of arbitrary dissipative corrections to the Josephsons relations parameterized with $\tilde \mu_i^d$ and $\mu^d$.  In particular, we find it helpful to work with the thermal momentum \(\tilde{p}_i\), this variable accounts for the momentum carried by the (non-fractonic) thermal degrees of freedom. In terms of $\tilde p_i$ the momentum conservation reads
\be 
\partial_t \tilde p_i + \partial_j \tilde T^{ji} = f_i\,,
\ee 
where
\be
\tilde T^{ji} =T^{ji}+J^j u_i + K^{jk}\xi_{ik} 
\ee
is the invariant stress tensor, while 
\be 
f_i = n  \partial_t u_j + \pi_j\partial_t\xi_{ij} + J^k \partial_i v^s_k + K^{kj} \partial_k\xi_{ij}
\ee 
can be interpreted as a force. Once the constitutive relations are expressed in terms of the fluid variables the equations may be solved to a desired order in the derivative expansion.

Let us now ascertain the appropriate truncation scheme for the equations. Firstly, the divergence of the dipole current appears in the equations akin to the charge current. Therefore, the dipole current needs to be expanded up to one order lower than the charge current, and the dipole equation is to be truncated at one order lower than that of charge conservation. Secondly, the dipole Goldstone $\psi_i$ is order minus one, hence the corresponding equation should be truncated at one order lower as compared with the equations for order zero variables. Finally, the charge Goldstone $\theta$ is actually order minus two and therefore the associated equation needs to be truncated at two orders lower. 
Therefore, the hydrodynamic equations are to be truncated as 
\be \begin{split}\label{eq:truncated}
\partial_t n + \partial_i J^i &= \mathcal{O}(\nabla^{n+2})\,, \\
\partial_t \pi_i + \partial_j K^{ji} +J^i &= \mathcal{O}(\nabla^{n+1})\,, \\
\partial_t \tilde p_i + \partial_j \tilde T^{ji} - f_i &= \mathcal O(\nabla^{n+2})\,, \\
\partial_t \epsilon + \partial_i \epsilon^i &=\mathcal{O}(\nabla^{n+2})\,, \\
\partial_t  \psi_i  -\Big( \mu_i+\tilde \mu_i^d \Big)  &= \mathcal{O}(\nabla^{n+1})  \,, \\
\partial_t \theta-\Big( \mu+\mu^d \Big) &=\mathcal{O}(\nabla^{n})\,.
\end{split}
\ee
From a practical standpoint, however, we find it convenient to parameterize our fluid in terms of the invariant superfluid velocity \(v^s_i\) rather than the dipole Goldstone \(\psi_i\). This choice is well-motivated on physical grounds, as this combination of Goldstone fields precisely corresponds to the massive degree of freedom. In terms of the new variables, the set of invariant hydrodynamic equations read
\be \label{eq:eomsM}
\begin{split}
\partial_t n + \partial_i J^i &= \mathcal O(\nabla^{n+2})\,, \\
\partial_t \pi_i + \partial_j K^{ji} +J^i &= \mathcal O(\nabla^{n+1})\,, \\
\partial_t \tilde p_i + \partial_j \tilde T^{ji} - f_i &= \mathcal O(\nabla^{n+2})\,, \\
\partial_t \epsilon + \partial_i \epsilon^i &= \mathcal O(\nabla^{n+2})\,, \\
\partial_t  v^s_i  - \Big( \partial_i  \mu -  \mu_i +\mu^d_i \Big) &=  \mathcal O(\nabla^{n+1})  \,, \\
\partial_t \theta-\Big( \mu+\mu^d \Big) &=\mathcal O(\nabla^{n})\,,
\end{split}
\ee
where we have redefined the dissipative dipole chemical potential as $\mu_i^d=\partial_i \mu^d - \tilde\mu^d
_i$, and have truncated the equations accordingly. From now on, we will work with equations \eqref{eq:eomsM} as our set of hydrodynamic equations and $\{n\,, \epsilon\,, \tilde p_i\,, \pi_i\,, \theta\,, v^s_i \}$ as the hydrodynamic variables.

Gradient expansion in hydrodynamics is typically endowed with a power counting scheme, where a certain weight $z$ is assigned to time derivatives, i.e. $\partial_t \sim \mathcal O(\nabla^z)$. For example, in ordinary hydrodynamics one has $z=1$ whereas dipole-conserving fluids exhibit $z=2$ \cite{Glodkowski:2022xje,Jain:2023nbf}. These scalings are consistent with the low-energy spectrum containing gapless modes with linear and quadratic dispersion, respectively. 

However, for fracton superfluids, the spectrum contains propagating modes with both soundlike $(\omega\sim k)$ and magnonlike $(\omega \sim k^2)$ dispersion relations (see Eq. \eqref{eq:sound} and \eqref{eq:mangon}). It is therefore not clear if one should count $\partial_t \sim \mathcal O(\nabla)$ or $\partial_t \sim \mathcal O(\nabla^2)$ as both of these counting schemes seem to be in conflict with the scaling of the magnonlike and soundlike mode, respectively. This issue was initially highlighted in \cite{Armas:2023ouk, Jain:2023nbf}. Intuitively, this tension can be attributed to the fact that the fractonic degrees of freedom vary in time more slowly than the neutral ones. Therefore, in the case of fracton superfluids it does not appear consistent to assign a common scaling to time derivatives\footnote{Unless certain coefficients are assigned anomalous scaling dimensions, though we do not explore this possibility here.}. Indeed, we have explicitly verified that assuming either $z=1$ or $z=2$ leads to the violation of at least one of the consistency conditions listed in \eqref{eq:conditions} and thus does not constitute a self-consistent power counting prescription.

In order to circumvent this issue we proceed to carry out the derivative expansion without assigning a definite scaling to the time derivatives. Our   approach to derivative expansion can be understood simply as a truncation of the hydrodynamic constitutive relations in spatial gradients according to Eq. \eqref{eq:eomsM}. Once truncated, the equations are to be solved for the time evolution of the hydrodynamic variables. Similarly, one can also truncate the entropy production equation Eq. \eqref{eq:linearEntropyProduction} in order to classify the dissipative contributions allowed by the second law of thermodynamics at a given order in the perturbative expansion.

 As we will see in the next subsections, this prescription produces dispersion relations that are organized systematically and satisfy all the consistency conditions (see Table \ref{tab:Table4}). Hence it constitutes a consistent derivative counting scheme for fracton superfluids. 

\paragraph{Entropy current analysis.}
To identify an entropy current for the system, we start with the first law of thermodynamics
\be 
d\epsilon = \mu dn + \mu_i d \pi_i + T ds + \Tilde{V}_i d \Tilde p_i + \lambda_i d v^s_i + F_{ij} d \xi_{ij}\,.
\ee 
Using $\xi_{ij}=\partial_i \psi_j = \partial_i \partial_j \theta - \partial_i v^s_j$ and $\tilde p_i = p_i + n \partial_i \theta + \pi_j \partial_i \partial_j \theta - \pi_j \partial_i v^s_j$ we arrive at
\be \label{eq_firstlaw2}
d\epsilon = \bar \mu dn + \bar \mu_i d \pi_i + T ds + \Tilde{V}_i d  p_i + \lambda_i d v^s_i + n \tilde V_i d \partial_i \theta - \bar F_{ij} d \partial_i v^s_j + \bar F_{ij} d \partial_i \partial_j \theta \,.
\ee 
where we have introduced effective variables
\be 
\bar{\mu}= \mu + \Tilde{V}_i \partial_i \theta\,, \quad \bar{\mu}_i= \mu_i + \Tilde{V}_j \partial_j \partial_i \theta - \Tilde{V}_j \partial_j v^s_i \,, \quad  \bar{F}_{ij} = F_{ij} + \tilde{V}_i \pi_j \,.
\ee 
After a series of algebraic manipulations, we arrive at the following expression that determines the constitutive relations of the currents 
\be \begin{split}
\Delta &= \partial_i \Big( s^i - P \frac{\tilde V_i}{T}-\frac{1}{T} \epsilon^i + \frac{\bar \mu}{T} J^i + \frac{\bar \mu_j}{T} K^{ij}  + \frac{\tilde V_j}{T} T^{ij} +  \frac{\mu^d_j}{T} \bar F_{ij} - \frac{\bar F_{ij}}{T} \partial_j \mu^d + \partial_j \frac{\bar F_{ji}}{T} \mu^d - \frac{ \tilde V_i}{T} n \mu^d \Big) \\
&+\Big( \epsilon^i - \big(sT+\tilde V_j \tilde p_j\big) \tilde V_i + \mu \lambda_i + \mu_j F_{ij}\Big) \partial_i\frac{1}{T}   \\&+ \Big(J^i+\lambda_i\Big) \Big( \frac{ \mu_i}{T} - \partial_i \frac{ \mu}{T} - \frac{\tilde V_j}{T} \partial_j v^s_i \Big) 
- \Big(K^{ij}+F_{ij}\Big) \partial_i \frac{\bar \mu_j}{T}   \\&-\Big( T^{ij} - P\delta_{ij} - \tilde V_i \tilde p_j - F_{ik} \partial_j \partial_k \theta + F_{ik} \partial_j  v^s_k + J^i \partial_j \theta \Big)\partial_i \frac{\Tilde{V}_j}{T}  \\
&  -  \mu^d_i \Big( \frac{\lambda_i}{T} + \partial_j \frac{\bar F_{ij}}{T} \Big) +\mu^d \Big(\partial_i \frac{ n \tilde V_i}{T}      - \partial_i \partial_j \frac{\bar F_{ij}}{T}\Big)  \geq 0\,.
\end{split}
\ee 
For the ideal sector we impose $\Delta = 0$ finding 
\be \begin{split}\label{eq:0order}
J^i &= - \lambda_i\,, \\
K^{ij} &= - F_{ij}\,, \\ 
T^{ij} & = P \delta_{ij} + \Tilde{V}_i \Tilde{p}_j + \lambda_i \partial_j \theta + F_{ik} \xi_{jk} \,, \\
 E^i & =   \Tilde{V}_i (sT + \Tilde{p}_j \Tilde{V}_j ) - \mu \lambda_i - \mu_j F_{ij}\,,\\ 
S^i &= s \Tilde{V}_i \,.
\end{split}
\ee 
These constitutive relations are in full agreement with the Poisson bracket method \eqref{eq:ideal}. On the other hand, the dissipative currents are specified by 
\be \begin{split}
\Delta &=\mathcal E^i \partial_i\frac{1}{T} + \mathcal J^i \Big( \frac{ \mu_i}{T} - \partial_i \frac{ \mu}{T} - \frac{\tilde V_j}{T} \partial_j v^s_i \Big) -\mathcal T^{ij} \partial_i \frac{\Tilde{V}_j}{T} 
  \\& - \mathcal K^{ij} \partial_i \frac{\bar \mu_j}{T}  -  \mu^d_i \Big( \frac{\lambda_i}{T} + \partial_j \frac{\bar F_{ij}}{T} \Big) +\mu^d \Big(\partial_i \frac{ n \tilde V_i}{T}      - \partial_i \partial_j \frac{\bar F_{ij}}{T}\Big)  \geq 0\,.
\end{split}
\ee 
In the remainder of this work, we focus on small deviations from the stationary configuration \eqref{eq:entropyDensity} and investigate linearized hydrodynamics by expanding the constitutive relations up to the first order in perturbations. In this special case thermal momentum $\tilde p_i$ satisfies the conservation equation $\partial_t \tilde p_i + \partial_j \tilde T^{ji} = 0$ with $\tilde T^{ij} = T^{ij} - n_0 \partial_t \theta \delta_{ij}$.

Entropy production truncated accordingly, up to the second order in deviations, can be written as 
\be \begin{split}\label{eq:linearEntropyProduction}
\Delta =  \mathcal E^i \partial_i \frac{1}{T} + \mathcal J^i \Big(  \frac{\mu_i}{T} - \partial_i  \frac{\mu}{T} \Big) -\mathcal{\tilde{T}}^{ij} \partial_i \frac{\Tilde{V}_j}{T} 
   - \mathcal K^{ij} \partial_i  \frac{\mu_j}{T}  -  \mu^d_i \Big( \frac{\lambda_i}{T} + \partial_j  \frac{ F_{ij}}{T} \Big) -\mu^d    \partial_i \partial_j  \frac{F_{ij}}{T}  \geq 0\,.
\end{split} \ee
Before classifying the complete set of constitutive relations, let us notice that the constitutive relations for $\mathcal J^i$ and $\mu_d^i$ contain the following terms
\be \begin{split}
\mathcal J^i &= \beta \Big( \frac{\mu_i}{T} - \partial_i \frac{\mu}{T}\Big) + \dots \\
\mu_i^d &= \gamma \Big( \frac{\lambda_i}{T} + \partial_j  \frac{F_{ij}}{T} \Big)+ \dots
\end{split}
\ee 
where $\beta\,, \gamma \geq 0$ and the dots represents other contributions that are not relevant to the forthcoming argument. We thus see that the equations for the gapped degrees of freedom $v^s_i$ and $\pi_i$ take the form of the relaxation equations 
\be \begin{split}
\partial_t \pi_i + \beta \chi_{\pi} \pi_i + \dots &= 0 \,, \\
\partial_t v^s_i + \gamma \chi_v v_i^s + \dots &= 0 \,. \\
\end{split}
\ee 
Therefore, there are again two possibilities. The first possibility correspond to the pure hydrodynamic regime where the transport coefficients are order one quantities $\beta\,, \gamma \sim \mathcal{O}(1)$. In this regime, one can set $\partial_t \pi_i + \beta \chi_{\pi} \pi_i \approx \beta \chi_{\pi} \pi_i$ and $\partial_t v^s_i + \gamma \chi_v v_i^s \approx \gamma \chi_v v_i^s$ effectively integrating out the gapped degrees of freedom in analogy to the case of diffusion discussed in \ref{sec:dipoleSubdiffusion2}.

When the transport coefficients are made perturbatively small, however, such that $\beta\,, \gamma \sim \partial_t$ the approximation is no longer valid as the two contributions are of the same order and the gapped degrees of freedom undergo independent dynamics. In this regime, the associated relaxation times are large and quantities $\pi_i$ and $v^s_i$ are ``almost" conserved. A comparable scenario, referred to as the \textit{spin dynamical regime}, has been explored in the context of spin hydrodynamics in \cite{Hongo:2021ona}.

While the latter possibility is of considerable interest, in the rest of this work, we will focus solely on the former, pure hydrodynamic regime, setting $\partial_t \pi_i=\partial_t v^i_s=0$ from now onwards. After integrating out the gapped degrees of freedom, we end up with four dynamical equations 
\be \label{eq:pureHydroEqs}
\begin{split}
    \partial_t n + \partial_i J^i &=\mathcal{O}(\nabla^{n+2}) \,, \\
    \partial_t \tilde p_i + \partial_j \tilde T^{ji} &=\mathcal{O}(\nabla^{n+2})\,, \\
    \partial_t \epsilon + \partial_i \epsilon^i &=\mathcal{O}(\nabla^{n+2}) \,, \\
    \partial_t \theta - \Big( \mu +\mu_d\Big)&=\mathcal{O}(\nabla^n)\,.  \\
\end{split}
\ee 
Supplemented with the two constraints, namely $J^i = -\partial_j K^{ji}$ and $\psi_i=\partial_i \theta$. The entropy production \eqref{eq:linearEntropyProduction} evaluated on-shell with respect to the constraints is then expressed as 
\be \begin{split}\label{eq:entropyProductionPureHydro}
\Delta = -\partial_j \Big( K^{ji} \frac{\mu_i}{T} - K^{ji} \partial_i \frac{\mu}{T} \Big) +  \mathcal E^i \partial_i \frac{ 1}{T} -\mathcal{\tilde{T}}^{ij} \partial_i \frac{\Tilde{V}_j}{T} 
   - \mathcal K^{ij} \partial_i  \partial_j \frac{\mu}{T}  -\mu^d    \partial_i \partial_j  \frac{F_{ij}}{T}  \geq 0\,.
\end{split} \ee
Therefore, after absorbing the total divergence term into the definition of the entropy current, we finally arrive at the desired form of the second law 
\be \begin{split}\label{eq:desired}
 \mathcal E^i \partial_i \frac{1}{T}  -\mathcal{\tilde{T}}^{ij} \partial_i \frac{\Tilde{V}_j}{T} 
   - \mathcal K^{ij} \partial_i  \partial_j \frac{\mu}{T}  -\mu^d    \partial_i \partial_j  \frac{F_{ij}}{T}  \geq 0\,.
\end{split} \ee
The dissipative currents will be constructed out of the 'non-hydrostatic' data whose classification is provided in the Table \ref{tab:Table3}.
\begin{table}[t]
    \centering
   \begin{tabular}{|l|c|c|c|}
   \hline
& First order & Second order & Third order  \\
 \hline
   Scalars & $\partial_i \frac{\tilde V_i}{T}$ &$\partial_i \partial_j \frac{F_{ij}}{T}\,, \partial^2 \mu\,, \partial^2 \frac{1}{T} $& $\partial^2 \partial_i \frac{\tilde V_i}{T}$\\
  Vectors   & $\partial_i \frac{1}{T}$ & $\partial_i \partial_j \frac{\tilde V_j}{T}$ &$\partial^2 \partial_i \frac{1}{T}$\\
      Tensors & $\partial_{\langle i} \frac{\tilde V_{j\rangle}}{T}$ &$\partial_{\langle i} \partial_{j \rangle} \frac{\mu}{T}\,, \partial_i \partial_j \frac{1}{T}$&$\partial^2 \partial_{\langle i} \frac{\tilde V_{j\rangle}}{T}\,, \partial_i \partial_j \partial_k \frac{\tilde V_k}{T}$\\
  \hline
 \end{tabular} 
    \caption{Non-hydrostatic data classification for fracton superfluids in the regime of pure (linearized) hydrodynamics.}
    \label{tab:Table3}
\end{table}
We are now able to identify the constitutive relations order by order in the derivative expansion. 

For the first order hydrodynamics, corresponding to the truncation of the entropy production \eqref{eq:desired} at the second order in derivatives i.e. $\partial_\mu S^{\mu} = \mathcal{O}(\nabla) \mathcal{O}(\nabla)$ we find
\be 
\begin{split}\label{eq:1order}
\mathcal E_{(1)}^i &=  \alpha \partial_i \frac{1}{T}\,, \\
T_{(1)}^{ij} &= -\zeta \partial_k \frac{\tilde V_k}{T} \delta_{ij} - \eta \partial_{\langle i} \frac{\tilde V_{j \rangle}}{T}
\end{split}
\ee 
with $\alpha\,, \zeta\,, \eta \geq0$.

Moving to second-order hydrodynamics, we observe that truncating entropy production at the third order, such that $\Delta \sim \mathcal{O}(\nabla^2)\mathcal{O}(\nabla)$, does not impose inequality constraints on transport coefficients. This is because such terms can be combined into squares, as follows: $\Delta \sim \Big(\mathcal{O}(\nabla)+\mathcal{O}(\nabla^2)\Big)^2 + \mathcal{O}(\nabla^4)$. Therefore, after completing squares, the positivity of entropy production is ensured as long as the $\mathcal{O}(\nabla)\mathcal{O}(\nabla)$ coefficients are positive, and any mistake made during the completion of a square is absorbed to higher order. 

Nevertheless, there are still possible (nondissipative) corrections to the currents that one can add at this order in the derivative expansion, namely
\be 
\begin{split}\label{eq:2order}
\mathcal E_{(2)}^i &= c_1 \partial_i \partial_j  \frac{\tilde V_j}{T}\,, \\
T_{(2)}^{ij} &= c_1 \partial^2 \frac{1}{T} \delta_{ij} +c_2 \partial^2 \frac{\mu}{T} \delta_{ij} +c_3 \partial_j \partial_k \frac{F_{jk}}{T}\delta_{ij}\,, \\
\mathcal{K}_{(1)}^{ij} &= c_2 \partial_k \frac{\tilde V_k}{T} \delta_{ij}\,, \\
-\mu_{(1)}^d &= c_3 \partial_i \frac{\tilde V_i}{T}
\end{split}
\ee 
where we have already imposed Onsager reciprocity relations.

Finally, we consider third order hydrodynamics contributing to the entropy production truncated at the fourth order. At this order we are able to construct squares of the form $\mathcal{O}(\nabla^2)\mathcal{O}(\nabla^2)$ and therefore we expect to establish some inequality type constraints. Indeed, the most general constitutive relations for third order hydrodynamics are 
\be 
\begin{split}\label{eq:3order}
\mathcal E_{(3)}^i &=-\bar \alpha \partial^2 \partial_i \frac{1}{T}+a_1  \partial^2 \partial_i \frac{\mu}{T}+a_2 \partial_i \partial_j \partial_k \frac{F_{jk}}{T}\,, \\
T_{(3)}^{ij} &= \bar \zeta_1 \partial^2 \partial_k \frac{\tilde V_k}{T} \delta_{ij}+ \bar \zeta_2 \partial_i \partial_j \partial_k \frac{\tilde V_k}{T} + \bar \eta \partial^2 \partial_{\langle i} \frac{\tilde V_{j \rangle}}{T} \,, \\
\mathcal{K}_{(2)}^{ij} &= -\sigma_{||} \partial^2 \frac{\mu}{T} \delta_{ij} - \sigma_{\perp} \partial_{\langle i} \frac{\partial_{j \rangle} \mu}{T} +a_1 \partial^2 \frac{1}{T}\delta_{ij}+a_3\partial_k \partial_l \frac{F_{kl}}{T} \delta_{ij}\,, \\
-\mu_{(2)}^d &=  \xi \partial_i \partial_j \frac{F_{ij}}{T}+a_2 \partial^2 \frac{1}{T} + a_3 \partial^2 \frac{\mu}{T} \,.
\end{split}
\ee 
The second law of thermodynamics is then satisfied provided that the dissipative transport coefficients satisfy $\bar \alpha\,, \bar \zeta_1\,, \bar \zeta_2\,, \bar\eta\,, \sigma_{||}\,, \sigma_{\perp}\,, \xi \geq 0$ and $\bar \alpha \sigma_{||} \geq  a^2_1$.
\paragraph{Dispersion relations.}
In this section we compute the dispersion relations $\omega(k)$ of the hydrodynamic modes up to the $\sim k^4$ contributions associated with the third order hydrodynamics. We confirm that the dispersion relations are organized systematically in a derivative expansion and that the constraints from the second law of thermodynamics and thermodynamic stability of the equilibrium state guarantee the linear stability of the modes.

Plugging the linearized constitutitve relations Eqs. \eqref{eq:0order}, \eqref{eq:1order}, \eqref{eq:2order} and \eqref{eq:3order} into the equations of motion \eqref{eq:pureHydroEqs} we arrive at the following set of dynamical equations for the third order hydrodynamics
\be \begin{split}
 \label{eq:Linearizedeoms}
\partial_t n + \partial_i \partial_j F_{ij} -\frac{c_2}{T_0}  \partial^2 \partial_i \tilde V_i +\sigma \partial^4 \frac{\mu}{T}-a_1 \partial^4 \frac{1}{T} -\frac{a_3}{T_0} \partial^2 \partial_i \partial_j F_{ij}&= 0\,, \\
\partial_t  \Tilde p_i + s_0 \partial_i  T - \frac{\hat \eta}{2T_0} \partial^2 \Tilde{V}_i - \frac{1}{T_0} \Big( \hat \zeta   + \hat \eta \frac{d-2}{2d}  \Big) \partial_i \partial_j \Tilde{V}_j +c_1 \partial_i \partial^2 \frac{1}{T} + c_2 \partial_i \partial^2 \frac{\mu}{T} + \frac{c_3}{T_0} \partial_i \partial_j \partial_k F_{jk} &=0 \\
\partial_t \epsilon + s_0 T_0 \partial_i \tilde V_i + \mu_0 \partial_i \partial_j F_{ij}  + \hat \alpha \partial^2 \frac{1}{T} + \frac{c_1}{T_0} \partial^2 \partial_i \tilde V_i+ a_1 \partial^4 \frac{\mu}{T} + \frac{a_2}{T_0} \partial^2 \partial_i \partial_j  F_{ij}  &= 0\,, \\
\partial_t \theta- \mu + \frac{c_3}{T_0} \partial_i \tilde V_i +\frac{\xi}{T_0}\partial_i \partial_j F_{ij} +\frac{a_2}{T_0}\partial^2 \frac{1}{T} + a_3 \partial^2 \frac{\mu}{T}&=0\,.
\end{split}
\ee
where we have defined the hatted coefficients representing corrected first order coefficients as follows
\be \begin{split}
\hat \eta &= \eta - \bar \eta \partial^2 \,, \\
\hat \zeta &= \zeta -  \big(\bar\zeta_1 +\bar\zeta_2\big)\partial^2 \,, \\
\hat \alpha &= \alpha-\bar\alpha \partial^2\,.
\end{split}
\ee 
Using the form of the entropy density \eqref{eq:entropyDensity} evaluated on the constraint $\psi_i=\partial_i\theta$ such that $\xi_{ij} = \partial_i \partial_j \theta$ we obtain the following thermodynamic identities  
\be \begin{split}
F_{ij} &= \lambda^{ijkl} \partial_k \partial_l \theta \,, \\
\delta \frac{1}{T} &=\frac{1}{T_0} \Big(   \chi_{ne} \delta n-\chi_{ee} \delta \epsilon\Big)\,, \\
\delta \frac{\mu}{T} &=\frac{1}{T_0} \Big(   \chi_{nn} \delta n-\chi_{ne} \delta \epsilon\Big)\,, \\
\tilde V_i &= \chi_p \tilde p_i\,, \\
 \delta T &=  T_0 \chi_{ee} \delta e - T_0 \chi_{ne} \delta n \,, \\
 \delta \mu &=  \chi_e \delta e +  \chi_n \delta n\,.
\end{split}
\ee
After expressing the variations of thermodynamic variables in terms of hydrodynamic densities using the above identities, we arrive at a closed system of differential equations for the evolution of the hydrodynamic variables
\be \begin{split}
 \label{eq:eomsFinal}
\partial_t \delta n + \lambda\partial^4 \theta -c_2\frac{\chi_p}{T_0}\partial^2 \partial_i \tilde p_i +  \tilde  \sigma \partial^4 \delta n+\tilde a_1 \partial^4 \delta \epsilon - \lambda \frac{a_3}{T_0} \partial^6 \theta &= 0\,, \\
\partial_t \tilde p_i +\hat\chi_{ee}\partial_i \delta \epsilon -\hat\chi_{ne} \partial_i \delta n - \frac{\chi_p }{2 T_0} \hat \eta  \partial^2 \Tilde{p}_i - \frac{\chi_p}{T_0} \Big( \hat \zeta   + \hat \eta \frac{d-2}{2d}  \Big) \partial_i \partial_j \Tilde{p}_j  + c_3 \frac{\lambda}{T_0} \partial^4 \partial_i\theta&= 0\,, \\
\partial_t \delta \epsilon +  \tilde \chi_p 
\partial_i \tilde p_i +\hat \lambda \partial^4 \theta + \hat \alpha_n  \partial^2 \delta n-\hat \alpha_e \partial^2 \delta \epsilon + c_1 \frac{\chi_p}{T_0} \partial^2 \partial_i \tilde p_i  &= 0\,, \\
\partial_t \theta- \hat \chi_n \delta n - \hat \chi_e \delta \epsilon +c_3 \frac{\chi_p}{T_0} \partial_i \tilde p_i +\frac{\lambda}{T_0} \xi \partial^4 \theta&=0\,.
\end{split}
\ee
For brevity of presentation, we have introduced the following notation
\be \begin{split}
\tilde \sigma &= \frac{1}{T_0}\Big(\sigma \chi_{nn}-a_1 \chi_{ne} \Big) \,, \\
\tilde a_1 &= \frac{1}{T_0}\Big(a_1 \chi_{ee}-\sigma \chi_{ne} \Big) \,, \\
\hat \lambda &= \lambda \Big( \mu_0 +\frac{a_2}{T_0} \partial^2 \Big)\,, \\
\tilde a_3 &= \frac{1}{T_0} \Big( a_3 \chi_{nn}-a_2 \chi_{ne} \Big) \,, \\
\tilde a_2 &= \frac{1}{T_0} \Big( a_2 \chi_{ee}-a_3 \chi_{ne} \Big) \,, \\
\hat \chi_{ee} &= s_0 T_0 \chi_{ee}-\frac{1}{T_0}\Big(c_1 \chi_{ee}+c_2 \chi_{ne}\Big)\partial^2 \,, \\
\hat \chi_{ne} &= s_0 T_0 \chi_{ne} -\frac{1}{T_0} \Big(c_1 \chi_{ne} +c_2 \chi_{nn}\Big)\partial^2\,, \\
\hat \alpha_n &= \frac{1}{T_0}\Big(  \chi_{nn} a_1\partial^2 + \chi_{ne} \hat \alpha \Big)\,, \\
\hat \alpha_e &= \frac{1}{T_0}\Big(\chi_{ne} a_1 \partial^2 + \chi_{ee} \hat \alpha   \Big)\,, \\
\hat \chi_n &=\chi_n-\tilde a_3\partial^2\,, \\
\hat \chi_e &= \chi_e-\tilde a_2\partial^2\,.
\end{split}
\ee 
Solving the eigenvalue problem, we verify the existence of two ``sound" modes in the longitudinal sector and one shear mode in the transverse sector with the exact formulas for the dispersion relations given by
 \be\begin{split}\label{eq:modesFull}
\omega_{\text{sound}} &= \pm\left( v_s  - \tilde v_s k^2\right)k- i\left( \Gamma + \tilde \Gamma k^2\right)k^2+\mathcal{O}(k^5) \,, \\
\omega_{\text{fracton}} &= \pm\left( v_m k + \tilde v_m k^3\right)k -i\Omega k^4 + \mathcal{O}(k^5)\,, \\
\omega_{\text{shear}} &= -i \left[\eta \frac{\chi_p}{2T_0} +\bar \eta \frac{\chi_p}{2T_0} k^2\right]k^2 +\mathcal{O}(k^5)\,.
\end{split}
\ee  
We have used tilded constants to denote the subleading corrections to the velocities and attenuation constants of the modes. 
\begin{table}[t]
    \centering
   \begin{tabular}{|l|c|c|c|c|}
   \hline
 & Zeroth order & First order & Second order & Third order  \\
 \hline
   $\omega_{\text{sound}}$ & $\pm v_s k$ & $-i\Gamma k^2$ & $\mp\tilde v_s k^3$& $-i\tilde \Gamma k^4$ \\
  $\omega_{\text{fracton}}$  & - & $\pm v_m k^2$ & - & $\pm \tilde v_m k^4-i\Omega k^4$ \\
      $\omega_{\text{shear}}$ & - & $-i\eta \frac{\chi_p}{2T_0}k^2$ & - & $-i\bar \eta \frac{\chi_p}{2T_0}k^4$\\
  \hline
 \end{tabular} 
    \caption{Contribution of momentum as polynomial expressions to the hydrodynamic modes in fracton superfluids, determined at a specific order of the derivative expansion.}
    \label{tab:Table4}
\end{table}
The explicit expressions for the constants are listed below
\be \begin{split}
v_s&=s_0 T_0 \sqrt{ \chi_{ee}  \chi_p }\,, \\
 v_m &= \sqrt{\frac{\lambda}{\chi_{ee}}\Big(\chi_{ee}\chi_{nn}-\chi^2_{ne} \Big)}\,, \\
\Gamma &= \frac{1}{2T_0 } \Big(  \big( \zeta + \eta \frac{d-1}{2d}\big) \chi_p + \alpha \chi_{ee} \Big) \,, \\
\tilde v_s &= \frac{1}{2 v_s} \Big[\Big(\Gamma - \frac{\alpha \chi_{ee}}{T_0}\Big)^2 -\frac{\lambda}{\chi_{ee}}\Big(\mu_0 \chi_{ee}-\chi_{ne} \Big)^2 \Big]\,, \\
\tilde \Gamma &= \bar \alpha \frac{\chi_{ee}}{2 T_0}+\Big(\sigma_{||}+\sigma_{\perp}\frac{d-1}{d} \Big) \frac{  \chi _{ne}^2}{2 T_0 \chi _{ee}}+\Big(\bar \zeta_1 +\bar \zeta_2+\bar \eta \frac{d-1}{d} \Big)\frac{\chi_p}{2T_0} -a_1 \frac{ \chi_{ne}}{T_0} \\
&-\Big(\zeta+\eta \frac{d-1}{d} \Big) \frac{  \chi_p \lambda  \left(\chi_{ne}-\mu _0 \chi_{ee}\right)^2}{2T_0 v^2_s \chi_{ee} }\,, \\ 
\Omega &= \xi \frac{\lambda}{2T_0}+\Big(\sigma_{||}+\sigma_{\perp}\frac{d-1}{d} \Big)\frac{\chi_{nn}\chi_{ee}-\chi^2_{ne}}{2T_0\chi_{ee}}+\Big(\zeta+\eta \frac{d-1}{d} \Big) \frac{  \chi_p \lambda  \left(\chi_{ne}-\mu _0 \chi_{ee}\right)^2}{2T_0 v^2_s \chi_{ee} }\,, \\
\tilde v_m &=  v_m \Big(\frac{a_3}{T_0} - \frac{  \lambda  \left(\chi_{ne}-\mu _0 \chi_{ee}\right)^2}{2 v^2_s \chi_{ee} } \Big)\,.
\end{split}
\ee 
It is straightforward to verify that the dispersion relations satisfy the consistency conditions \eqref{eq:conditions}. From the Table \ref{tab:Table4} it follows that the dispersion relations are organized consistently with respect to the derivative counting scheme in the sense that the transport coefficients corresponding to the $n$-th order hydrodynamics contribute to the $k^{n+1}$ order in the power series expansion of the dispersion relations $\omega(k)$ (consistent expansion). Furthermore, the conditions for thermodynamic stability Eq. \eqref{eq:stability} and second law of thermodynamics\footnote{These are the constraints on transport coefficients listed below Eqs. \eqref{eq:1order} and \eqref{eq:3order}.} guarantee that the imaginary part of the modes is negative so that the modes are linearly stable throughout the free parameter space (stable modes). We can also see that the only transport coefficients that contribute to the attenuation of the modes are the ones that produce entropy. In particular, the nondissipative coefficients $c_1, c_2$ and $c_3$ do not contribute to the attenuation of the modes (dissipation-entropy correspondence).

%%%%%%%%

\section*{Conclusion and outlook}\addcontentsline{toc}{section}{Conclusion and outlook}
In this thesis, we have established that the distinction between uniform and nonuniform symmetries provides a physically meaningful classification for the construction of low-energy effective field theories of many-body systems.

The main takeaway is that, in contrast to uniform symmetries, nonuniform symmetries do not give rise to additional independent gapless modes, but instead impose kinematic constraints on the effective theory. These constraints restrict the dynamics of the low-energy degrees of freedom and lead to qualitatively modified behavior, including the softening of NG modes and slowdown of hydrodynamic transport. In turn, a wide range of exotic collective phenomena in condensed matter systems can be traced back to their origin in nonuniform symmetries.

In particular, we have shown that the dynamics of Tkachenko waves, which encode the incompressibility of vortex lattices at low energies, emerge as a consequence of spontaneously broken nonuniform symmetries. In doing so, we have constructed an effective field theory for quantum vortex lattices directly from symmetry principles. Similarly, the unconventional hydrodynamics of fracton phases can be traced back to the presence of nonuniform dipole symmetry, which constrains the dynamics of elementary charges and manifests, at the collective level, as a slowdown of charge transport. These fracton fluids do not obey the ordinary Navier--Stokes equations, but instead satisfy modified hydrodynamic equations reflecting the constraints imposed by nonuniform dipole symmetry. To elucidate these constraints, we developed a hydrodynamic theory for systems with such symmetries by adopting the modern perspective of hydrodynamics as an effective theory. Underlying these results is an algebraic criterion for counting massless NG modes and gapless hydrodynamic excitations, determined entirely by the action of translations on the symmetry algebra of the broken charges. This criterion provides a systematic way to identify the low-energy degrees of freedom under minimal assumptions and generalizes existing counting theorems to the case of nonuniform symmetries.

Looking ahead, a natural direction for future work is to extend the algebraic counting rule developed here beyond the enumeration of gapless modes, and to extract information about their dispersion relations. We expect that this information is similarly encoded in the algebraic structure of the $\lambda_\mu$ matrices, in particular in their nilpotency properties. If viable, this would allow one to determine both the number of gapless modes and the scaling of their dispersion relations directly from the algebra.

Moreover, it would be interesting to apply the nonuniform perspective in a nonequilibrium effective field theory setting \cite{Liu:2018kfw}. In this framework, hydrodynamic modes can themselves be understood as Goldstones of  strong-to-weak spontaneous symmetry breaking \cite{Huang:2024rml}. This formalism has been successfully applied to fracton fluids, where it correctly reproduces their hydrodynamic description \cite{Glorioso_2023}. A natural extension would be to apply this framework to vortex lattices at finite temperature, where a normal component and dissipative effects are present, thereby generalizing the zero temperature theory constructed in this thesis to a full-fledged nonequilibrium effective field theory. More generally, it would be interesting to investigate whether a unified framework can be constructed in which both NG modes and hydrodynamic degrees of freedom are treated on the same footing, with their nonuniform symmetries incorporated systematically within the Schwinger--Keldysh effective field theory formalism.

Another direction for future work is to extend the nonuniform classification to higher-form symmetries \cite{Gaiotto:2014kfa}. Such symmetries arise in a variety of physical systems, particularly in those hosting topological defects, including the low-energy theory of elasticity \cite{Armas:2019sbe,Hirono:2022dci,Armas:2023tyx}. It would be interesting to explore whether this perspective can be employed to understand dissipative dynamics of elastic defects in terms of nonuniform higher-form magnetohydrodynamics, with topological defects identified as gauge charges.

Finally, while the present thesis has focused primarily on continuum systems, an important open direction is to extend the nonuniform perspective to lattice models. Particularly interesting are subsystem symmetries, which act on submanifolds of the lattice and lead to extensive ground-state degeneracy and constrained dynamics of elementary excitations. It would be interesting to investigate whether the algebraic framework developed here can be adapted to this setting, and in particular whether it can provide a classification of quantum phases based on the structure of nonuniform symmetries and their possible breaking patterns. This perspective may offer a systematic understanding of the spectrum of excitations and degeneracy, and in particular shed light on the origin of robustness in lattice models relevant for quantum memory.

%%%%%%%%

%\section{Charged fluids}\label{sec:chargedFluids}


% Provide your bibliography here. You have two options:

\appendix

\section{Coset construction for nonuniform symmetries}
\textit{This review is adapted from Ref.~\cite{Glodkowski:2025krf}. } \newline

\noindent From the perspective of a modern effective field theorist, the low-energy effective action of a given physical system consists of all operators consistent with the system's symmetries, each accompanied by a phenomenological coefficient. Identifying the complete set of such operators is a nontrivial task, especially when some of the symmetries are \textit{spontaneously broken}, giving rise to Goldstone fields that transform nonlinearly under the action of the broken symmetries.\footnote{Note that virtually all condensed matter systems spontaneously break at least some of their symmetries \cite{Nicolis:2015sra}.} The coset construction method \cite{Callan:1969sn,Coleman:1969sm,Volkov:1973vd,Ivanov:1975zq} (see also \cite{Penco:2020kvy,Naegels:2021ivf,Brauner:2024juy} for a modern introduction to the formalism) offers a systematic procedure for identifying the covariant structures constructed from Goldstone fields, with the symmetry breaking pattern as the only input.

In this appendix, we briefly review the coset formalism as adapted to the case of \emph{nonuniform symmetries}. When such symmetries are spontaneously broken, some of the associated Goldstone fields can be systematically eliminated by imposing the so-called inverse Higgs constraints, leading to an effective description in terms of a reduced set of degrees of freedom. We discuss the implementation of such constraints and their possible physical interpretations.

\subsection{Introduction to the coset formalism}\label{sec:coset}
We now outline the general procedure, starting with the identification of the symmetry generators and the parametrization of the coset space. Consider the symmetry breaking pattern $G \rightarrow H$ where the ground state spontaneously breaks the global symmetry group $G$ down to the subgroup $H$. We denote the generators of the symmetry algebra as follows
\begin{align}
\text{unbroken translations:} \quad & \bar P_\alpha \,, \\
\text{other unbroken generators:} \quad & T_A  \,, \\
\text{broken generators:} \quad & X_a \,.
\end{align}
In general, the set of broken generators $\{ X_a \}$ may contain both uniform and nonuniform generators.

Since $H$ leaves the ground state invariant, the set of physically inequivalent ground states (the vacuum manifold) is in one-to-one correspondence with the coset space $G/H$. A generic element of the coset space $U \in G/H$ can be parametrized as follows
\be \label{eq:coset}
    U(x, \pi) = e^{x^\alpha \bar P_\alpha} e^{ \pi^a(x) X_a},
\ee
where $x^\alpha$ represent the spacetime coordinates and $\pi^a(x)$ are the Goldstone fields. In the coset parametrization \eqref{eq:coset} we have included the factor $e^{x^\alpha \bar P_\alpha}$, which ensures that the Goldstone fields transform homogeneously under spacetime translations \cite{Delacretaz:2014oxa,Brauner:2024juy}. In the language of Sec.~\ref{sec:nonuniformGoldstone}, this choice corresponds to working in a \emph{uniform basis} for the Goldstone fields, in which the inhomogeneous part induced by nonuniform symmetries is factored out.

Under a global transformation $g \in G$, the coset representative transforms as
\begin{equation}\label{eq:transformation}
    g\, U(x, \pi) = U(x^\prime, \pi^\prime)\, h(x, \pi, g)\,, 
\end{equation}
where $h(x, \pi, g) \in H$ is a compensating transformation that is chosen to preserve the coset parametrization \eqref{eq:coset}. Using the symmetry algebra, one can invert the relation \eqref{eq:transformation} to extract the transformed spacetime coordinates $x^\prime$ and Goldstone fields $\pi^\prime$. Importantly, the Goldstone fields transform nonlinearly under the broken symmetries, making it cumbersome to construct invariant combinations directly from their transformation laws.

Instead of working with explicit transformation laws, one identifies covariant building blocks by evaluating the Maurer--Cartan form $\boldsymbol{\omega} \equiv U^{-1} d U$, which takes values in the Lie algebra. It can be decomposed as follows 
\be
\boldsymbol{\omega} = e^\alpha_\mu  \big( \bar P_\alpha + D_\alpha \pi^a X_a + A^B_\alpha T_B \big) dx^\mu\,,
\ee 
where we have introduced coefficients $e^\alpha_\mu$, $D_\alpha \pi^a$ and $A^B_\alpha$, which depend on the Goldstone fields and can be read-off directly from the Maurer-Cartan form. Their transformation properties under $g \in G$ follow from Eq. \eqref{eq:transformation}. In particular, we find  
\cite{Ivanov:1975zq,Penco:2020kvy}
\be 
\begin{split}\label{eq:structures}
    e^\alpha_\mu &\rightarrow
    [\rho(h)]^\alpha_\beta \,  e^\beta_\mu \,, \\
        D_\alpha \pi^a &\rightarrow  \, [\rho(h)]^a_b D_\beta \pi^b [\rho^{-1}(h)]^\beta_\alpha \,, \\
           e^\alpha_\mu  A^B_\alpha T_B &\rightarrow h \left( e^\alpha_\mu  A^B_\alpha T_B \right) h^{-1} + h \partial_\mu h^{-1}\,, \\
\end{split}
\ee 
where $[\rho(h)]^\alpha_\beta$ and $[\rho(h)]^a_b$ are different linear representations of the compensating element $h \equiv h(x, \pi, g) \in H$. 

The coefficients $D_\alpha \pi^a$ are called \textit{covariant derivatives} since they transform covariantly under the action of $G$ in a linear representation of $H$. One can then easily construct $G$--invariant combinations of the Goldstone fields by forming $H$--invariant tensor contractions of these covariant derivatives. 

Similarly, the object $e^\alpha_\mu$ is referred to as the \emph{coset vielbein}. It plays a role analogous to the standard vielbein in gravity, and, in particular, defines the covariant volume element $d^d x \det{e}$ \cite{Ivanov:1975zq,Penco:2020kvy}. Finally, $e^\alpha_\mu  A^B_\alpha T_B$ transforms as an $H$-connection, allowing us to define an $H$--covariant derivative
\begin{equation}
   \nabla^H_\alpha \equiv  E^\mu_\alpha \partial_\mu +  A_\alpha^B T_B\,,
\end{equation}
where $E^\mu_\alpha$ is an inverse to $e^\alpha_\mu$.

Putting the ingredients together, we can write down a $G$--invariant action as
\be 
S[\pi^a] = \int  d^d x \det{e} \, L(D_\alpha \pi^a, \nabla^H_\alpha)\,,
\ee 
where the free indices in $L(D_\alpha \pi^a, \nabla^H_\mu)$ must be contracted in the $H$--invariant fashion. Once the Maurer-Cartan form is evaluated using the symmetry algebra, the building blocks \eqref{eq:structures} can be read-off and the effective action constructed.

\subsection{Inverse Higgs mechanism}\label{sec:inverseHiggsMechanism}
In systems with spontaneously broken nonuniform symmetries, not all broken generators correspond to independent low-energy fluctuations. Within the coset construction, this reduction can be systematically implemented via the inverse Higgs mechanism \cite{Ivanov:1975zq}, which provides a prescription for eliminating inessential Goldstone fields in a manner consistent with the symmetry algebra, prior to specifying the effective action.

The inverse Higgs mechanism can be summarized as follows. Suppose the symmetry algebra contains a commutation relation of the form
\be 
[\bar P_\alpha, X^\prime] \sim X\,,
\ee 
where $\bar P_\alpha$ is an unbroken translation generator (in the $\alpha$-th direction), and $X$ and $X^\prime$ are broken generators. Furthermore, we assume that $X$ and $X^\prime$ do not transform into one another under the action of the unbroken symmetry group; in other words, they do not belong to the same irreducible representation of the unbroken subgroup $H$. This assumption fails, for example, if $X$ and $X^\prime$ are components of a vector that mixes under unbroken spatial rotations. 

The above conditions are sufficient to ensure that the covariant derivative $D_\alpha \pi$, associated with the broken generator $X$, contains a term linear in the Goldstone field $\pi^\prime$ corresponding to the generator $X^\prime$. One can then consistently impose the constraint $D_\alpha \pi = 0$. This results in an algebraic relation between the Goldstone fields $\pi^\prime$ and $\pi$, allowing for the elimination of $\pi^\prime$ in terms of derivatives of $\pi$. In this way, the Goldstone field $\pi^\prime$ can be systematically removed from the effective theory.

The physical meaning of such an elimination can be understood from two complementary points of view. The first interpretation, due to Low and Manohar \cite{PhysRevLett.88.101602}, is based on the observation that two linearly independent broken generators satisfying $[\bar P_\alpha, X'] \sim X$ do not necessarily generate independent massless fluctuations of the order parameter. In such cases, the elimination can be understood simply as a convenient gauge fixing, which removes the redundant Goldstones (see also \cite{Nicolis:2013sga}).

Depending on the details of the symmetry breaking mechanism\footnote{More precisely, on the representation furnished by the order parameter.}, however, such an interpretation is not always accurate. This brings us to the second interpretation of the inverse Higgs mechanism--namely, that it can be understood as integrating out physical massive modes to obtain an effective low-energy theory
\cite{Brauner:2014aha}. 

In Sec.~\ref{sec:vortex}, we encounter both interpretations in the context of vortex crystals and elucidate their physical origin.

\section{Matrix representation of the Aristotelian group}\label{app:aristotelian}
A convenient affine representation of $G_{\mathrm{sp}}$ on $\mathcal M$ is obtained by embedding 
$\mathcal M$ into $\mathbb R^4$ via
\begin{equation}
(t,x^i)\ \mapsto\ X=
\begin{pmatrix}
x^1\\
x^2\\
t\\
1
\end{pmatrix}\,.
\end{equation}
 In this setup, the physical spacetime is identified with the affine hyperplane
\begin{equation}
\mathcal M \simeq \left\{ X \in \mathbb R^4 \mid X^4 = 1 \right\}\,.
\end{equation}
The representation of $G_{\mathrm{sp}}$ on $\mathbb R^4$ is a subgroup of 
$\mathrm{GL}(4,\mathbb R)$ consisting of matrices of the form
\begin{equation}
M(R,\mathbf c,b)=
\begin{pmatrix}
R & \mathbf 0 & \mathbf c \\
\mathbf 0^{\mathsf T} & 1 & b \\
\mathbf 0^{\mathsf T} & 0 & 1
\end{pmatrix},
\end{equation}
where $R\in \mathrm{SO}(2)$, $\mathbf c\in \mathbb R^2$, and $b\in\mathbb R$. The group multiplication is
\begin{equation}
M(R,\mathbf c,b)\,M(R',\mathbf c',b') 
= M\big(RR',\, R\mathbf c' + \mathbf c,\, b + b'\big)\,,
\end{equation}
whereas the linear action on $\mathcal M$ is obtained by left multiplication
\begin{equation}
X \mapsto M X,
\qquad
\begin{pmatrix}
\mathbf x\\
t\\
1
\end{pmatrix}
\mapsto
\begin{pmatrix}
R\mathbf x + \mathbf c\\
t + b\\
1
\end{pmatrix}\,.
\end{equation}
%%%%%%
%%%%%%%

\section{Hydrodynamic analysis of dipole-conserving fluids}
\subsection{Entropy current analysis}
\label{sec:nextToLeading}
In this Appendix we show the algebraic manipulations that were used when going from Eq.  \eqref{eq:entprodc0} to Eq. \eqref{eq:master} explicitly. 

We start by rearranging Eq. \eqref{eq:entprodc0} as a total derivative minus the extra terms
\begin{equation}\label{eq:step1}
\begin{split}
0 &\leq \partial_i \Big( S^i - \frac{1}{T} \mathcal{E}^i 
    + \frac{\tilde{\mu}}{T}  \partial_j  J^{ij}  +\frac{V_j}{T} T^{ij} \Big) \\
    &+ \mathcal{E}^i \partial_i \frac{1}{T} - \partial_j  J^{ij} \partial_i \frac{\tilde{\mu}}{T} -  T^{ji} \partial_i \frac{V_j}{T} \,.
    \end{split}
\end{equation}
Driven by intuition from the conventional hydrodynamics, we now incorporate pressure into our construction by adding zero $0 = -\partial_i (P \frac{V_i}{T}) + P \partial_i \frac{V_i}{T}+\frac{V_i}{T} \partial_i P $ such that \eqref{eq:step1} becomes
\begin{equation}\label{eq:step2}
\begin{split}
0 &\leq  \partial_i \Big( S^i - \frac{1}{T} \mathcal{E}^i - P \frac{V_i}{T}
    + \frac{\tilde{\mu}}{T}  \partial_j  J^{ij}  +\frac{V_j}{T} T^{ij} \Big) \\
    &+ \mathcal{E}^i \partial_i \frac{1}{T} - \partial_j  J^{ij} \partial_i \frac{\tilde{\mu}}{T} +  (P\delta_{ij} - T^{ji}) \partial_i \frac{V_j}{T}  + \frac{V_i}{T} \partial_i P \,.
    \end{split}
\end{equation}
Using the definition of pressure Eq. \eqref{eq:pressureDef} it is possible to evaluate the gradient 
\begin{equation}\label{eq:pres}
    \partial_i P = n \partial_i \mu +  s \partial_i T - F_{jk} \partial_i V_{jk} \,.
\end{equation}
We substitute $\mu = \tilde{\mu} + n^{-1} V_i p_i$ and rewrite the last term in Eq. \eqref{eq:pres} obtaining
\begin{equation}
    \partial_i P = n \partial_i \tilde{\mu} + s\partial_i T + n \partial_i (\frac{V_j p_j}{n})  - F_{jk} \partial_i \partial_j \frac{p_k}{n} \,.
\end{equation}
Let us now express the last term as a total derivative  
\begin{equation}
\begin{split}
    \partial_i P &= n \partial_i \tilde{\mu} + s\partial_i T + n \partial_i (\frac{V_j p_j}{n}) \\& + \partial_j F_{jk} \partial_i \frac{p_k}{n} - \partial_j (F_{jk} \partial_i \frac{p_k}{n}) \,.
    \end{split}
\end{equation}
Using the definition of velocity $V_i = - n^{-1} \partial_j F_{ji}$ and relabelling the dummy indices $k \leftrightarrow j$ in the second-to-last term we find
\begin{equation}\label{eq:pressuregrad}
    \partial_i P = n \partial_i \tilde{\mu} + s\partial_i T + p_j \partial_i V_j - \partial_j (F_{jk} \partial_i \frac{p_k}{n}) \,.
\end{equation}
Thus, the last term in Eq. \eqref{eq:step2} can be written as follows 
\begin{equation}
\begin{split}
        \frac{V_i}{T} \partial_i P &=  \frac{V_i}{T} \Big (n \partial_i \tilde{\mu} + s\partial_i T + p_j \partial_i V_j - \partial_j (F_{jk} \partial_i \frac{p_k}{n}) \Big) \\
        &= nV_i \partial_i \frac{\tilde{\mu}}{T}  -\Big( \tilde{\mu} n V_i  + TsV_i  + V_j p_j V_i \Big) \partial_i \frac{1}{T} \\
        &+  p_j V_i   \partial_i  \frac{V_j}{T} +  F_{jk} \partial_i \frac{p_k}{n} \partial_j \frac{V_i}{T}- \partial_j \Big( \frac{V_i}{T}  F_{jk} \partial_i \frac{p_k}{n}       \Big) \,.
\end{split}
\end{equation}
Relabelling the dummy indices $i \leftrightarrow  j$ in the last two terms and using the definition of pressure Eq. \eqref{eq:pressureDef} we arrive at
\begin{equation}\label{eq:pressureGradient}
\begin{split}
        \frac{V_i}{T} \partial_i P  & = nV_i \partial_i \frac{\tilde{\mu}}{T} - \Big( P + \epsilon \Big) V_i \partial_i \frac{1}{T}  \\
        &+  (p_j V_i +  F_{ik} \partial_j \frac{p_k}{n})  \partial_i \frac{V_j}{T} - \partial_i \Big( \frac{V_j}{T}  F_{ik} \partial_j \frac{p_k}{n}       \Big) \,.
\end{split}
\end{equation}
Plugging \eqref{eq:pressureGradient} back into \eqref{eq:step2} we obtain an expression that closely resembles Eq. \eqref{eq:master}
\begin{equation}
\begin{gathered}
     \partial_i \Big( S^i - \frac{1}{T} \mathcal{E}^i - P \frac{V_i}{T}
    + \frac{\tilde{\mu}}{T}  \partial_j J^{ij} + \frac{V_j}{T} T^{ij} - \frac{V_j}{T}  F_{ik} \partial_j \frac{p_k}{n} \Big) + \Big( \mathcal{E}^i  - (p+\epsilon) V_i \Big) \partial_i (\frac{1}{T}) \\
    + (nV_i - \partial_j J^{ij}) \partial_i (\frac{\tilde{\mu}}{T})  +\Big(P \delta_{ij} + V_i p_j + F_{ik} \partial_j \frac{p_k}{n}  - T^{ij} \Big)\partial_i (\frac{V_j}{T} ) \geq 0 \,.
\end{gathered}
 \label{eq:master2}
\end{equation}

It is then straightforward to reach \eqref{eq:master} by adding another zero
\begin{equation}
\begin{split}
    0 &=- \partial_i \Big( \frac{V_j}{T} \partial_k \big(F_{ij} \frac{p_k}{n} - F_{kj} \frac{p_i}{n}\big) \Big) \\&+  \partial_k \big(F_{ij} \frac{p_k}{n} - F_{kj} \frac{p_i}{n}\big) \Big) \partial_i (\frac{V_j}{T}) \,.
    \end{split}
\end{equation}
This final step is necessary in order to obtain a stress tensor that is manifestly symmetric under the exchange of indices. 

\subsection{Fluid data classification}\label{sec:dissipativeCorrections}
We provide a classification of the independent fluid variables organized according to their transformations under rotations. To this aim we decompose the symmetric tensor $V_{ij}$ as follows 
\begin{equation}
    V_{ij} = \sigma^{ij} + \frac{\delta_{ij}}{d}\theta 
\end{equation}
where we have defined a transverse tensor $\sigma^{ij}$ and a scalar $\theta$ satisfying 
\begin{equation}
    \sigma^{ij} = \partial_{\langle i} \frac{p_{j\rangle}}{n}\,, \hspace{5px}\theta=\partial_{ i} \frac{p_{i}}{n} \,.
\end{equation}
One may then construct independent structures order by order in the gradient expansion according to the power counting scheme established in \ref{sec:grad}. In table \ref{tab:1} we present a list of the onshell independent linear terms up to the first order in the derivative expansion.
\begin{table}[t]
\centering
\caption{Classification of the on-shell independent linear data up to first order in the derivative expansion.}
\label{tab:1}

\begin{tabular}{||c|c|c|c||}
\hline
Order & Scalars & Vectors & Tensors \\
\hline\hline

$\mathcal{O}(0)$
&
\makecell{$n,\ \epsilon,\ \theta$}
&
\makecell{
$\partial_i n,\ \partial_i \epsilon,$\\
$\partial_j \sigma^{ij},\ \partial_i \theta$
}
&
\makecell{$\delta_{ij},\ \sigma^{ij}$}
\\
\hline

$\mathcal{O}(1)$
&
\makecell{
$\nabla^2 n,\ \nabla^2 \epsilon,$\\
$\nabla^2 \theta$
}
&
\makecell{
$\partial_i \nabla^2 n,\ \partial_i \nabla^2 \epsilon,$\\
$\nabla^2 \partial_j \sigma^{ij},\ \partial_i \nabla^2 \theta$
}
&
\makecell{
$\partial_{\langle i}\partial_{j\rangle} n,$\\
$\partial_{\langle i}\partial_{j\rangle} \epsilon,$\\
$\partial_{\langle i}\partial_{j\rangle} \theta,$\\
$\nabla^2 \sigma^{ij}$
}
\\
\hline

\end{tabular}
\end{table}

\subsection{Constitutive relations at first order}
Let us now consider the most general form of the first order (linearized around the equilibrium state Eq. \eqref{eq:equilibriumH}) corrections to the currents, written on the basis of the derivative expansion. 
These are given by (see Table \ref{tab:1})
\begin{equation}
\begin{split}\label{eq:macro}
    J_{1}^{ij} &=\Big(j_{n_1}\nabla^2 n + j_{e_1}\nabla^2\epsilon + j_{v_1} \nabla^2\theta  \Big)\delta^{ij}+ j_{v_2} \nabla^2\sigma^{ij}\\
& + j_{n_2} \partial_{\langle i}\partial_{j\rangle}n + j_{e_2}\partial_{\langle i}\partial_{j\rangle}\epsilon + j_{v_3}\partial_{\langle i}\partial_{j\rangle}\theta \,, \\[2.5pt]
     T_{1}^{ij} &=\Big(t_{n_1}\nabla^2 n + t_{e_1}\nabla^2\epsilon + t_{v_1} \nabla^2\theta  \Big)\delta^{ij}+ t_{v_2} \nabla^2\sigma^{ij}\\
& + t_{n_2} \partial_{\langle i}\partial_{j\rangle}n + t_{e_2}\partial_{\langle i}\partial_{j\rangle}\epsilon + t_{v_3}\partial_{\langle i}\partial_{j\rangle}\theta  \,,  \\[2.5pt]
    \mathcal{E}^i_{1} &=e_{n}\partial_i \nabla^2 n + e_{e}\partial_i \nabla^2\epsilon + e_{\theta} \partial_i \nabla^2\theta  + e_{\sigma} \nabla^2\partial_j \sigma^{ij}\,.\\
\end{split}
\end{equation}
In writing Eq. \eqref{eq:macro} we have introduced a new set of phenomenological parameters. These can be related to the dissipative transport coefficients presented in Eq. \eqref{eq:macroscopic}. 

To this aim, we express $V_i$ in terms of the variables $\theta$ and $\sigma^{ij}$ using Eq. \eqref{eq:therm1}
\begin{equation}
    F_{ij} = f_{||} \theta \delta_{ij} + f_{\perp} \sigma_{ij} \rightarrow V_i = -n^{-1}_0 \Big( f_{||} \partial_i \theta + f_{\perp} \partial_j \sigma_{ij} \Big) \,.
\end{equation}
Thus
\begin{equation}
\begin{split}
    \partial_i V_i &= -n^{-1}_0 \Big( f_{||} + f_{\perp}\frac{d-1}{d}  \Big) \nabla^2 \theta\,, \\
    \partial_{\langle i} V_{j\rangle} &= -n^{-1}_0 \Big( f_{||} \partial_{\langle i} \partial_{j \rangle} \theta + f_{\perp}  \nabla^2\sigma_{ij} - f_{\perp}\frac{d-1}{d^2} \nabla^2 \theta \delta_{ij} \Big) \,, \\
    \nabla^2 V_i &= -n^{-1}_0 \Big( f_{||} \partial_i \nabla^2 \theta+ f_{\perp} \nabla^2 \partial_j \sigma^{ij} \Big) \,, \\
    \partial_i \partial_j V_j &= -n^{-1}_0 \Big( f_{||} + f_{\perp}\frac{d-1}{d}  \Big) \partial_i \nabla^2 \theta \,.
    \end{split}
\end{equation}
Hence we see that 
\begin{equation}\label{eq:transport0}
    \begin{split}
        j_{v_1} &= -T^{-1}_0 n^{-1}_0 \Big[  \Big( f_{||} + f_{\perp}\frac{d-1}{d}  \Big) \gamma_{||} - f_{\perp}\frac{d-1}{d^2} \gamma_{\perp} \Big]\,, \\[1pt]
        t_{v_1} &= T^{-1}_0 n^{-1}_0 \Big[  \Big( f_{||} + f_{\perp}\frac{d-1}{d}  \Big) \zeta - f_{\perp}\frac{d-1}{d^2} \eta \Big]\,, \\[1pt]
           e_{\theta} & = T^{-1}_0 n^{-1}_0 \Big[ \big(f_{||} + f_{\perp}\frac{d-1}{d}\big)\big( \alpha_{||} + \alpha_{\perp} \frac{d-1}{d}\big) + \alpha_{\perp} f_{||} \Big] \,, \\[1pt]
        j_{v_2} &= -T^{-1}_0 n^{-1}_0 f_{\perp} \gamma_{\perp} \,, \hspace{10px} t_{v_2} = T^{-1}_0 n^{-1}_0 f_{\perp} \eta  \,, \\[1pt]
        j_{v_3} &= -T^{-1}_0 n^{-1}_0 f_{||} \gamma_{\perp} \,, \hspace{10px} t_{v_3} = T^{-1}_0 n^{-1}_0 f_{||} \eta \,. \\[1pt]
        e_{\sigma} & = T^{-1}_0 n^{-1}_0 f_{\perp} \alpha_{\perp} \,.
    \end{split}
\end{equation}
Even though $j_{v_2}(t_{v_2})$ and $j_{v_3}(t_{v_3})$ are \textit{a priori} independent parameters, they are in fact related via $\frac{j_{v_2}}{ j_{v_3}} = \frac{t_{v_2}}{ t_{v_3}} = \frac{f_{\perp}}{f_{||}}$ by the requirement of the non-negative entropy production Eq. \eqref{eq:diss2}.

Now, we re-express the terms in Eq. \eqref{eq:macroscopic} involving $\delta \frac{1}{T}$ and $\delta \frac{\mu}{T}$ in terms of the variables $\delta \epsilon$ and $\delta n$. Using the thermodynamic relations Eqs. \eqref{eq:therm1} we can identify 
\begin{equation}\label{eq:transport2}
\begin{split}
    j_{n_1} &=  s_{n\epsilon} \beta_{||} -s_{nn} \sigma_{||}\,, \hspace{10px} j_{n_2} =  s_{n\epsilon} \beta_{\perp} -s_{nn} \sigma_{\perp}\,, \\
        j_{e_1} &=  s_{\epsilon \epsilon} \beta_{||} -s_{n\epsilon} \sigma_{||}\,, \hspace{10px} j_{e_2} =  s_{\epsilon \epsilon} \beta_{\perp} -s_{n\epsilon} \sigma_{\perp}\,, \\
            t_{n_1} &=  s_{nn} \gamma_{||} - s_{n\epsilon} \alpha_{||} \,, \hspace{10px} t_{n_2} =  s_{nn} \gamma_{\perp} - s_{n\epsilon} \alpha_{\perp} \,, \\
        t_{e_1} &= s_{n\epsilon} \gamma_{||} - s_{\epsilon \epsilon} \alpha_{||} \,, \hspace{10px} t_{e_2} =  s_{n\epsilon} \gamma_{\perp} - s_{\epsilon \epsilon} \alpha_{\perp} \,. \\
        e_n & = s_{nn} \big( \beta_{||} + \frac{d-1}{d} \beta_{\perp} \big)  - s_{n \epsilon}  \big( \kappa_{||} + \frac{d-1}{d} \kappa_{\perp} \big) \,, \\
        e_{e} & = s_{ee} \big( \kappa_{||} + \frac{d-1}{d} \kappa_{\perp} \big) - s_{n \epsilon} \big( \beta_{||} + \frac{d-1}{d} \beta_{\perp} \big)\,.
    \end{split}
\end{equation}

\subsection{Linearized equations of motion}\label{sec:linearizedeoms}
In this Appendix, we derive the linearized equations of motion with the first order corrections (Eqs. \eqref{eq:eoms1}). This is most easily done in the basis used in Eqs. \eqref{eq:macro}, which are, of course, completely equivalent to Eqs. \eqref{eq:macroscopic} provided that the transport coefficients are identified according to Eqs. \eqref{eq:transport0} and \eqref{eq:transport2}.

It is then straightforward to compute the relevant gradients of the dissipative currents 
\begin{equation}
\begin{split}\label{eq:currentsGradients}
    \partial_i \partial_j J_{1}^{ij} &= \Big( j_{n_1} +\frac{d-1}{d} j_{n_2} \Big) \nabla^4 n + \Big( j_{e_1} +\frac{d-1}{d} j_{e_2} \Big) \nabla^4 \epsilon  \\
    & + \Big( j_{v_1} + \frac{d-1}{d} \big( j_{v_2}+j_{v_3} \big) \Big) \nabla^4 \theta\,, \\[1pt]
     \partial_j T_{1}^{ij} &=\Big(t_{n_1} +  \frac{d-1}{d} t_{n_2} \Big) \partial_i \nabla^2 n + \Big(t_{e_1} +  \frac{d-1}{d} t_{e_2} \Big) \partial_i \nabla^2 \epsilon   \\ 
     &+\Big(t_{v_1} +  \frac{d-2}{d} t_{v_2} + \frac{d-1}{d} t_{v_3}  \Big) \partial_i \nabla^2 \theta  + n_0^{-1} t_{v_2} \nabla^4 \delta p_i \,, \\[1pt]
    \partial_i \mathcal{E}^i_{1} &=e_{n} \nabla^4 n + e_{e} \nabla^4 \epsilon + \Big( e_{\theta} + \frac{d-1}{d}  e_{\sigma} \Big)  \nabla^4 \theta\,.\\
\end{split}
\end{equation}
Thus, we see that the linearized equations of motion up to first order in the derivative expansion are given by Eqs. \eqref{eq:eoms1} with
\begin{equation}\label{eq:eomsDefs}
    \begin{split}
        j_n &= j_{n_1} +\frac{d-1}{d} j_{n_2}\,, \quad
        j_e = j_{e_1} +\frac{d-1}{d} j_{e_2}\,,\\
        j_v &= j_{v_1} + \frac{d-1}{d} \big( j_{v_2}+j_{v_3} \big)\,, \quad  t_{v_{\perp}} =n^{-1}_0 t_{v_2}\,, \\
         t_{v_{||}} &= t_{v_1} +  \frac{d-2}{d} t_{v_2} + \frac{d-1}{d} t_{v_3} \,, \\
         t_n &= t_{n_1} +\frac{d-1}{d} t_{n_2}\,, \quad
        t_e = t_{e_1} +\frac{d-1}{d} t_{e_2}\,,\\
          e_n &= e_{n_1} +\frac{d-1}{d} e_{n_2}\,, \quad
        e_e = e_{e_1} +\frac{d-1}{d} e_{e_2}\,,\\
        e_v &= e_{\theta} + \frac{d-1}{d}  e_{\sigma}\,.
    \end{split}
\end{equation}
Going to Fourier space, we find that the shear mode picks up a subdiffusive contribution Eq. \eqref{eq:shear} while the dispersion relations of the longitudinal modes are now given by the roots of the modified polynomial Eq. \eqref{eq:polynomial1} where
\begin{equation}\label{eq:bsDef}
    \begin{split}
        b_0 & = \alpha  s_{\epsilon \epsilon}\Big(j_n+t_v\Big)  + \bar{f}\Big( t_n+ n^{-1}_0 (p_0 +\epsilon_0) t_e \Big)\,,  \\
    & +T_0 \Big(e_v P_{\epsilon} +  j_v P_n\Big) - \alpha j_e s_{ne} \\[2.5pt]
    b_1 & = \bar{f}  P_{\epsilon} T_0\Big(e_n -n^{-1}_0 (p_0 +\epsilon_0) j_n \Big)  \\
    &- \bar{f}  P_n T_0 \Big(e_e - n^{-1}_0 (p_0 +\epsilon_0)j_e \Big)\\
    & - \alpha s_{n \epsilon} \Big(\bar{f} t_e + j_v P_{\epsilon} T_0\Big) + \alpha s_{\epsilon \epsilon} \Big(\bar{f} t_n + j_v P_n T_0\Big)\,, \\[2.5pt]
        b_2 &= e_e + j_n + t_v \,.
    \end{split}
\end{equation}

% Use your bibtex library, formatted by the SciPost style file.
\bibliography{bibliography.bib}

%%%%%%%%%% END TODO: BIBLIOGRAPHY

\end{document}